\documentclass[12pt]{article}
\pdfoutput=1
\usepackage{jheppub}
\usepackage[utf8x]{inputenc}
\usepackage{amssymb}
\usepackage{amsmath}
\usepackage{amsbsy}    
\usepackage{graphicx}
\usepackage{bm}

\def\bea{\begin{eqnarray}}
\def\eea{\end{eqnarray}}

\newcommand{\bt}{{\tilde{\b}}}
\newcommand{\st}{{\tilde{\s}}}

\newcommand{\CB}{{\cal B}}

\newcommand{\CF}{{\cal F}}

\newcommand{\CN}{{\cal N}}

\newcommand{\CT}{{\cal T}}

\newcommand{\CW}{{\cal W}}

\newcommand{\CZ}{{\cal Z}}

\def\M{{\mathfrak M}}

\def\IZ{{\mathbb Z}}

\def\Gb{\Gamma_b}

\newcommand{\Qt}{\tilde{Q}}
\def\a{\alpha}
\def\b{\beta}
\def\s{\sigma}

\def\N{{\cal N}}

\newcommand{\Ups}{\Upsilon}
\newcommand{\Gm}{\Gamma}
\newcommand{\sfr}{S^4}

\newcommand{\rpf}{\mathbb{RP}^4}
\newcommand{\rpt}{\mathbb{RP}^2}
\newcommand{\intz}{\mathbb{Z}}
\newcommand{\ntwo}{\mathcal{N} = 2}
\newcommand{\sq}{\mathcal{Q}}
\newcommand{\sqb}{\mathcal{Q}_B}
\newcommand{\sqh}{\widehat{\mathcal{Q}}}
\newcommand{\ckrd}{\mathrm{Coker}\, D_{10}}
\newcommand{\krd}{\mathrm{Ker}\, {D_{10}}}
\newcommand{\zins}{Z_\mathrm{inst}}
\def\at{{\tilde \a}}
\def\Gb{\Gamma_b}
\def\pt{\tilde p}

\usepackage{tikz}
\usetikzlibrary{decorations.markings,arrows,backgrounds,patterns,decorations.pathreplacing}
\tikzset{>=latex}
\tikzset{every picture/.append style={scale=1.1}}
\tikzset{flavor/.style={draw}}
\tikzset{gauge/.style={draw,circle,inner sep=2pt}}

\newcommand{\be}{\begin{equation}}
\newcommand{\ee}{\end{equation}}
\def\ba{\begin{aligned}}
\def\ea{\end{aligned}}
\def\ben{\begin{eqnarray}\displaystyle}
\def\een{\end{eqnarray}}

\preprint{SISSA  49/2017/MATE-FISI}

\title{$\mathcal{N}=2$ gauge theories on unoriented/open four-manifolds and their AGT counterparts}

\author[1]{Aditya Bawane,}
\author[2]{Sergio Benvenuti,}
\author[2]{Giulio Bonelli,}
\author[3]{\\Nouman Muteeb}
\author[2]{and Alessandro Tanzini}

\affiliation[1]{Department of Physics, Sogang University, Seoul 121-742, Korea.} 
\affiliation[2]{International School of Advanced Studies (SISSA), 
via Bonomea 265, 34136 Trieste, Italy \& INFN, Sezione di Trieste.}
\affiliation[3]{International Center for Theoretical Physics (ICTP), Strada Costiera 11, 34151 Trieste, Italy.}
\emailAdd{aditya.bawane@gmail.com,benve79@gmail.com,\\bonelli@sissa.it,nouman01e72@yahoo.com,tanzini@sissa.it}

\abstract
{We compute the exact path integral of $\mathcal{N}=2$ supersymmetric gauge theories with general gauge group on $\mathbb{RP}^4$ and a $\mathbb{Z}_2$-quotient of the hemi-$S^4$. By specializing to $SU(2)$ superconformal quivers, we show that these, together with hemi-$S^4$ partition functions, compute Liouville correlators on unoriented/open Riemann surfaces. We perform explicit checks for Riemann surfaces obtained as $\mathbb{Z}_2$ quotients of the sphere and the torus. We also discuss the coupled $3d\!-\!4d$ systems associated to Liouville amplitudes with boundary punctures.}

\begin{document}
\maketitle

\section{Introduction and summary of results}

Equivariant localization provided new tools to explore supersymmetric gauge theories on various space-time geometries, unveiling rich algebraic and geometric structures
of their moduli spaces of vacua. This also had a deep impact on various branches of mathematics. In particular, $\mathcal{N}=2$ supersymmetric gauge theories
in four dimensions were analysed via these techniques starting from \cite{Nekrasov:2002qd}. See the review \cite{Pestun:2016zxk} for a recent overview on the subject.
In this paper we apply equivariant localization to compute the supersymmetric partition function
and BPS correlators of gauge theories with general gauge group on {\it unoriented} and {\it open} four manifolds. 
More precisely, we investigate gauge theories on $\mathbb{Z}_2$-quotients of the four-sphere, namely $\rpf$ and the hemi-sphere $HS^4$, and on a $\mathbb{Z}_2$-quotient of the $HS^4$, 
whose boundary is the Lens space $L(2,1)=\mathbb{RP}^3$. 
The one-loop gauge theory computation is performed by a direct analysis of harmonic modes expansion around the fixed locus of the equivariant action, by generalizing the techniques applied in \cite{Gava:2016oep} in the $HS^4$ case for $b=1$.

We also study the three-dimensional gauge theories living on the boundaries of the four-manifolds, finding coupled 3d/4d systems which describe defects in the four-dimensional gauge theories. 

An immediate application of our results is obtained by specializing to $SU(2)$ quiver gauge theories, which realize an 
extension of AGT correspondence \cite{Alday:2009aq} to unoriented/open Liouville correlators. AGT correspondence
naturally arises from the study of two M5-branes compactified on product manifolds $M_4\times C_{g,n}$, where the latter is a Riemann surface of genus $g$ and $n$ regular punctures.  
The gauge group $SU(2)^{3g-3+n}$ and field content 
of the four-dimensional $\mathcal{N}=2$ supersymmetric theory are specified by the geometry of $C_{g,n}$ \cite{Gaiotto:2009we}. In particular,
for genus $g=0$ one gets superconformal linear quivers with $n-3$ nodes and for $g=1$ circular quivers with $n$ nodes \cite{Witten:1997sc}.
Liouville theory correlators on $C_{g,n}$ are reproduced by the supersymmetric partition function of the corresponding gauge theory on $M_4=S^4$.
This correspondence can be generalized by considering supersymmetric quotients of the M5-brane theory. 
Orientable quotients of $S^4$ by a finite group $\Gamma\in SU(2)$, resulting in four-manifolds locally described by ALE spaces
$\mathbb{C}^2/\Gamma$, were first considered in \cite{Belavin:2011pp,Nishioka:2011jk,BMT1,Belavin:2011sw}, were it was proposed that the corresponding
two-dimensional CFT becomes a parafermionic theory. In particular, for $\Gamma=\mathbb{Z}_2$ the gauge theory reproduces the correlators of  $\mathcal{N}=1$ Super-Liouville  theory \cite{BMT2}. 

The $\mathbb{Z}_2$-quotients we consider in this paper result instead in 
open and/or unoriented four-manifolds. Since the $(2,0)$ six-dimensional superconformal theory expected to describe the dynamics of
M5-branes is chiral, the four-dimensional $\mathbb{Z}_2$-action has to be accompanied by a suitable involution of $C_{g,n}$
producing open/unoriented Riemann surfaces. 
For Riemann surfaces of arbitrary topology we expect a superconformal quiver $SU(2)$ gauge theory to arise whose Coulomb moduli space is described by 
the moduli space of complex structures of open/unoriented Riemann surfaces. Since these surfaces can be always constructed as the $\mathbb{Z}_2$-quotient of closed Riemann surfaces,
we expect the corresponding gauge theory to be obtained as a suitable quotient of the quiver gauge theory associated to the oriented double.

For $g=0$ the quotients result in the two-disk $\mathcal{D}$
or the real projective plane $\rpt$.
Liouville amplitudes on the latter are compared with the gauge theory partition function on $\rpf$. Concerning the disk, one can 
choose two types of boundary conditions, namely FZZT \cite{Fateev:2000ik,Teschner:2000md} or ZZ \cite{Zamolodchikov:2001ah}, whose gauge theory counterpart are Dirichlet
or Neumann boundary conditions on the $S^3$ boundary of the $HS^4$. 
For $g=1$, $\mathbb{Z}_2$ involutions give rise to annulus $\mathcal{A}$, Moebius strip $\mathcal{M}$ and Klein bottle $\mathcal{K}$. From the M5-branes
perspective, it is natural to expect these to correspond to suitable quotients of circular quivers. When the resulting open/unoriented Riemann surface
has boundaries, the relevant FZZT/ZZ boundary conditions are implemented by imposing Dirichlet/Neumann boundary conditions on the fields associated to the $SU(2)$ nodes
of the gauge theory quiver.  
    
Notice that
one can consider both bulk and boundary insertions. In particular, boundary insertions are necessary to describe the strongly coupled frame
of gauge theory, arising when bulk insertions are separated very far away. Indeed, in this case 
bulk insertions can approach a boundary making an open Liouville channel open up. 

We consider the basic building blocks, i.e. the bulk-to-boundary two-point function and the boundary three-point function. We interpret them as the partition functions, respectively, of $3$ and $4$ free massless $4d$ hypermultiplets, with a specific $3d$ gauge theory living on the boundary. The whole $3d-4d$ system preserves $4$ supercharges.

Our results connecting unoriented/boundary Liouville correlators with $\mathcal{N}=2$ gauge theories on quotients of $S^4$ can be summarized as follows.
\begin{itemize}
\item
Crosscap with two bulk insertions versus partitition function of $SU(2)$, $N_f=4$ gauge theory on $\mathbb{RP}^4$,  in the topological sector of gauge connections with trivial holonomy 
\be 
\langle V_{\alpha_1} (q) V_{\alpha_2}(0)\rangle_{\mathbb{RP}^2} = Z_{N_f=4}^{\mathbb{RP}^4} (\mu_1=\alpha_1 -\frac{Q}{2}, \mu_1=\alpha_2 -\frac{Q}{2} ; q,\epsilon_1,\epsilon_2)\, .
\ee
\item
Disk with two bulk insertions and FZZT boundary conditions with $s=0$ versus $HS^4$,  $SU(2)$, $N_f=4$ with Dirichlet boundary conditions
\be 
\langle V_{\alpha_1} (q) V_{\alpha_2}(0)\rangle_{\mathcal{D}}^{{\rm FZZT},\, s=0} = Z_{N_f=4}^{HS^4, \, {\rm Dir}} (\mu_1=\alpha_1 -\frac{Q}{2}, \mu_1=\alpha_2 -\frac{Q}{2} ; q,\epsilon_1,\epsilon_2)
\ee
\item
Disk with two bulk insertions and ZZ boundary conditions versus $HS^4$,  $SU(2)$, $N_f=4$ with Neumann boundary conditions
\be 
\langle V_{\alpha_1} (q) V_{\alpha_2}(0)\rangle_{\mathcal{D}}^{{\rm ZZ},\, (s_1=s_2=1)} = Z_{N_f=4}^{HS^4, \, {\rm Neu}} (\mu_1=\alpha_1 -\frac{Q}{2}, \mu_1=\alpha_2 -\frac{Q}{2} ; q,\epsilon_1,\epsilon_2) \, .
\ee
\item
Klein bottle amplitude versus $\mathbb{Z}_2$-quotient of $SU(2)\times SU(2)$ circular quiver on $\mathbb{RP}^4$ in the topological sector of gauge connections with trivial holonomy
\be
\langle {\bf 1}
 \rangle_{\mathcal{K}} = Z^{\mathbb{RP}^4}_{[SU(2)\times SU(2)]^{\mathbb{Z}_2}} (q,\epsilon_1,\epsilon_2) \, 
\ee
The identity operator insertion on the torus in Liouville CFT corresponds to $\mathcal{N}=2^*$ with mass $\mu=1$ for $b=1$. This applies also for the  
cases below until \eqref{moe}.
\item
Annulus with FZZT boundary conditions versus $\mathbb{Z}_2$-quotient of $SU(2)\times SU(2)$ circular quiver on $HS^4$ with Dirichlet boundary conditions
\be
\langle {\bf 1}
 \rangle_{\mathcal{A}}^{{\rm FZZT},\, s_1=s_2=0} = Z^{{HS}^4, \, {\rm Dir}}_{[SU(2)\times SU(2)]^{\mathbb{Z}_2}} (q,\epsilon_1,\epsilon_2)\, .
\ee
\item
Annulus with ZZ boundary conditions versus $\mathbb{Z}_2$-quotient of $SU(2)\times SU(2)$ circular quiver on $HS^4$ with Neumann boundary conditions
\be
\langle {\bf 1}
\rangle_{\mathcal{A}}^{{\rm ZZ},\, (r_1,r_2)=(s_1,s_2)=(1,1)} = Z^{{HS}^4, \, {\rm Neu}}_{[SU(2)\times SU(2)]^{\mathbb{Z}_2}} (q,\epsilon_1,\epsilon_2)\, .
\ee
\item Annulus with (FZZT,ZZ) boundary conditions at the two ends as a circular quiver quotients on $HS^4$ with Dirichlet/Neumann boundary conditions on the two nodes
\be
\langle {\bf 1}
\rangle_{\mathcal{A}}^{{\rm FZZT/ZZ},\, (s=0/(s_1,s_2)=(1,1))} = Z^{{HS}^4, \, {\rm Dir/Neu}}_{[SU(2)\times SU(2)]^{\mathbb{Z}_2}} (q,\epsilon_1,\epsilon_2)\, .
\ee
\item
Moebius strip with FZZT/ZZ boundary condition vs. $\mathbb{Z}_2$ quotient of $SU(2)\times SU(2)$ circular quiver on $HS^4/\mathbb{Z}_2$ with Dirichlet/Neumann boundary conditions 
\be\label{moe}
\langle {\bf 1}
\rangle_{\mathcal{M}}^{{\rm FZZT/ZZ}} = Z^{{HS}^4/\mathbb{Z}_2, \, {\rm Dir/Neu}}_{[SU(2)\times SU(2)]^{\mathbb{Z}_2}} (q,\epsilon_1,\epsilon_2)\, .
\ee
were, as above,  $s=0$ for FZZT and $(s_1,s_2)=(1,1)$ for ZZ. The non-orientability of the Moebius surface is compensated by an half-integral Chern-Simons level
on the boundary $\mathbb{RP}^3$ theory.
\item
Disk bulk/boundary two point function with FZZT boundary conditions versus $3d$ $U(1)$ $\CN=4$ with $N_f=2$ (called $TSU(2)$) coupled to $3$ $4d$ free hypers
\be
\langle \Psi_\beta(1) V_\alpha(0)\rangle^{\rm FZZT}_{{\mathcal D}, \s}=\mathcal{Z}[\mathcal{B}_{TSU(2)}](\beta,\alpha;\s)
\ee
\item
Disk three point boundary function with FZZT boundary conditions versus $3d$ $\mathcal{N}=2$ $U(1)$ with $N_f=4$ and monopole superpotential coupled to $4$ $4d$ free hypers
\be
\langle \Psi_{\beta_1}(0) \Psi_{\beta_2}(1)\Psi_{\beta_3}(\infty) \rangle^{\rm FZZT}_{{\mathcal D}, \s_1,\s_2,\s_3}=\CZ[\CB_{U(1), N_f=4}](\beta_1,\beta_2,\beta_3;\s_1,\s_2,\s_3)
\ee
\end{itemize}
We remark that the FZZT amplitudes with $s=mb+nb^{-1}$ and ZZ amplitudes with $(s_1,s_2)=(m,n)$ are reproduced from the respective cases described above
via Wilson line insertions.
 
The structure of the paper is as follows. In Section 2 we describe the equivariant localization formulae  for general gauge theories on $\rpf$ and $HS^4$.
In Section 3 we specialize to $SU(2)$ quivers and perform the comparison with Liouville amplitudes on open/unoriented Riemann surfaces with bulk insertions, while Section 4
is devoted to the boundary insertions case and the corresponding three dimensional gauge theory sectors. Section 5 contains some open questions and few technical Appendices complete the presentation.

\vspace{.5 cm}

While we were typing this paper, \cite{LeFloch:2017lbt} appeared which has partial overlap with our Section 3.

\section{Localising gauge theories on $\mathbb{RP}^4$ and $HS^4$ }

The calculation of partition functions and BPS correlators of supersymmetric gauge theories on $\mathbb{RP}^4$ and $HS^4$ via localization can be obtained by realizing those manifolds as $\mathbb{Z}_2$-quotients of $S^4$. Henceforth, we first review the results for the $S^4$ partition function
in order to set-up the notation for the subsequent computations.

It is well known that $\mathcal{N}=2$ supersymmetry can be realised on a four-manifold by turning on some background fields, which can be either seen
as coming from a rigid limit of supergravity backgrounds \cite{Gupta:2012cy,Hama:2012bg,Klare:2013dka} or as auxiliary fields needed to enforce the
rigid $\mathcal{N}=2$ supersymmetry algebra \cite{Bawane:2014uka}. Global supersymmetry transformations are generated by suitable Killing spinors satisfying the
following equations
 \begin{eqnarray}\label{eq:mainKE}
 D_m\xi_A + T^{kl}\sigma_{kl}\sigma_m\bar\xi_A &=& -i\sigma_m\bar\xi_A^\prime \nonumber\\
 D_m\bar\xi_A + \bar T^{kl}\bar\sigma_{kl}\bar\sigma_m\xi_A &=& -i\bar\sigma_m\xi_A^\prime
\end{eqnarray}
with $\xi_A^\prime = \frac{i}{4} \bar\sigma_n D^n \xi_A$ and $\bar\xi_A^\prime=\frac{i}{4} \sigma_n D^n \bar \xi_A$, and the auxiliary equations:
\begin{eqnarray}\label{eq:auxKE}
 \sigma^m\bar\sigma^nD_mD_n\xi_A + 4D_lT_{mn}\sigma^{mn}\sigma^l\bar\xi_A &=& M\xi_A \nonumber\\
\bar\sigma^m\sigma^nD_mD_n\bar\xi_A + 4D_l\bar T_{mn}\bar\sigma^{mn}\bar\sigma^l\xi_A &=& M\bar\xi_A \ \ ,
\end{eqnarray}
where $T^{kl}$, $\bar T^{kl}$ and $M$ are anti-self-dual, self-dual and scalar background fields respectively. We denote by
$A$, $B$,... the $SU(2)_\mathcal{R}$ doublet indices\footnote{$SU(2)_\mathcal{R}$ denotes the $R$-symmetry group, whereas $SU(2)_R$ will be used to denote the right-handed generators of the Lie algebra of the isometry group on $\sfr$.}. The covariant derivative acting on the spinor and its derivative is given by 
\begin{align*}
 D_m\xi_A &\equiv \partial_m\xi_A +\frac{1}{4}\Omega_m^{ab}\sigma_{ab}\xi_A + i\xi_BV_{m\phantom{B}A}^{\phantom{m}B}\\
 D_mD_n\xi_A &\equiv  \partial_mD_n\xi_A +\frac{1}{4}\Omega_m^{ab}\sigma_{ab}D_n\xi_A - \Gamma^p_{\phantom{p}mn}D_p\xi_A+ iD_n\xi_BV_{m\phantom{B}A}^{\phantom{m}B},
\end{align*}
where $V_{m\phantom{B}A}^{\phantom{m}B}$ is an  $SU(2)_\mathcal{R}$ gauge field, $\Omega_m^{ab}$ are components of the spin connection one-form and $\Gamma^p_{\phantom{p}mn}$ are the Christoffel symbols\footnote{Indices from the middle of the Latin alphabet $(l,m,n,p...)$ are curved and those from the beginning $(a,b,c,..)$ are flat.}. Similar formulae hold for the right handed spinors 
\footnote{The representation we work with is given by
\begin{equation}
 \begin{array}{lll}
  \sigma^a = -i\tau^a, & \bar\sigma^a = i\tau^a, & (a=1,2,3)\\
  \sigma^4 = 1, &\bar\sigma^4 = 1, &
 \end{array}
\end{equation}
where $\tau^a$ are the Pauli matrices.}
with $\bar\sigma_{ab}$ instead of $\sigma_{ab}$.

Let us review the solution of the above equations on round $S^4$.
We use the following angular coordinates
\begin{equation}\label{eq:CoordsS4}
\begin{aligned}
 x_0 &= r \cos\rho \\
  x_1 &= r \sin\rho\cos\left( \frac{\theta}{2}\right)\cos\left(\frac{\psi+\phi}{2}\right) \\
    x_2 &= r \sin\rho\cos\left( \frac{\theta}{2}\right)\sin\left(\frac{\psi+\phi}{2}\right) \\
      x_3 &= r \sin\rho\sin\left( \frac{\theta}{2}\right)\cos\left(\frac{\psi-\phi}{2}\right) \\
        x_4 &= r \sin\rho\sin\left( \frac{\theta}{2}\right)\sin\left(\frac{\psi-\phi}{2}\right)
\end{aligned}
\end{equation}
where $0\leq\rho\leq\pi$, $0\leq\theta\leq\pi$,  $0\leq\phi\leq2\pi$,  $0\leq\psi\leq4\pi$, and $\psi \sim \psi+4\pi$. The metric is given by
\begin{equation*}
 ds^2 = d\rho^2+\frac{1}{4}\sin^2\rho\,\big(d\theta^2+\sin\theta^2d\phi^2+(d\psi+\cos\theta d\phi)^2\big)
\end{equation*}
Using the above data, one may verify that a solution of the main Killing spinor equation Eq. \ref{eq:mainKE} is
\begin{equation}\label{eq:Kspinors}
\xi_{\alpha A} =  \frac{1}{\sqrt{2}} \cos \left(\frac{\rho}{2}\right) \,\delta_{\alpha A}
\ \ \ , \ \ \ 
\bar\xi^{\dot\alpha}_A =  \frac{1}{\sqrt{2}} \sin \left(\frac{\rho}{2}\right)\, \delta^{\dot\alpha}_A
\end{equation}
corresponding to the background fields:
\begin{equation}\label{bg}
 V_{m\phantom{B}A}^{\phantom{m}B} = 0, \, T^{kl} =0, \, \bar T^{kl} =0.
\end{equation}
The auxiliary equation Eq. \ref{eq:auxKE} is automatically satisfied with $M = -R/3 = -4/r^2$, where $R$ is the Ricci scalar and $r$ is the radius of the four-sphere. We set $r=1$ here onwards.

We start by considering pure $\mathcal{N}=2$ Super Yang-Mills theory, whose vector multiplet consists of a pair of scalars $(\phi,\bar\phi)$, a gauge field $A_m$, gauginos $(\lambda_{\alpha A},\bar\lambda_{\dot\alpha}^A)$, and a triplet of auxiliary fields $D_{AB}=D_{BA}$, all transforming in the adjoint representation of the gauge group $G$. The details of the supersymmetry transformation and gauge fixing terms are relegated to the Appendix \ref{app:susy}. The general $\ntwo$ supersymmetric Yang-Mills action is given by
\begin{equation}\label{eq:YMAction}
\begin{split}
 \mathcal{L}_\mathrm{YM} = \mathrm{Tr}\Big[&\tfrac{1}{2}F_{mn}F^{mn} + 16 F_{mn}(\bar\phi T^{mn} +\phi \bar T^{mn})
 \\&- 4D_m\bar\phi D^m\phi + 2M\bar\phi\phi+ 64\bar\phi^2T_{mn}T^{mn}+64\phi^2\bar T_{mn}\bar T^{mn} 
 \\&- 2i\lambda^A\sigma^mD_m\bar\lambda_A -2\lambda^A[\bar\phi,\lambda_A] + 2\bar\lambda^A[\phi,\bar\lambda_A] 
 \\&+ 4[\phi,\bar\phi]^2 - \tfrac{1}{2}D^{AB}D_{AB}\Big]
\end{split}
\end{equation}
which can be specified to $S^4$ by fixing the background fields as in \eqref{bg}.

The saddle-point about which one considers fluctuations is found by requiring that $\sqh \, \mathrm{fermions} = 0$, where $\sqh = \sq + \sqb$, which are the supersymmetry differential (Eqs. \ref{eq:SUSYVect}, \ref{eq:SUSYGhost}) and the BRST differential (Eqs. \ref{eq:BRSTGhost}, \ref{eq:BRSTVect}) respectively. The saddle-point turns out to be \cite{Hama:2012bg}
\begin{equation*}
 A_m=0,\hspace{0.2cm} \phi_2 =0, \hspace{0.2cm}\phi_1 = a_0, \hspace{0.2cm} D_{AB} = -ia_0w_{AB},
\end{equation*}
where $a_0$ is a Lie algebra-valued constant,
$\phi_1 \equiv i(\phi + \bar\phi)$, $\phi_2 \equiv \phi - \bar\phi$ and 
$w_{AB} $ a bilinear in the $\xi_A$s.

The localizing action is given by the $\sqh$-transform of
\begin{equation*}
V_\mathrm{vec}=\text{Tr}\left[(\sqh\lambda_{\alpha A})^{\dagger}\lambda_{\alpha A}+(\sqh\bar{\lambda}^{\dot{\alpha}}_A)^{\dagger}\bar{\lambda}^{\dot{\alpha}}_A\right] + \textrm{gauge-fixing terms},
\end{equation*}
which in terms of the ``cohomological variables''
\begin{align*}
 \Psi &= \sq\phi_2\nonumber,\\
 \Psi_m &= \sq A_m=\sqh A_m - D_m c\nonumber,\\
 \chi_a &= \xi^B\lambda_A\left(\sigma^a\right)^A_{\phantom{A}B},
\end{align*}
can be written as
\begin{equation}\label{eq:LocAct}
 V_\mathrm{vec}=\text{Tr}\left[(\sqh\Psi)^{\dagger}\Psi + (\sqh\Psi^m)^{\dagger}\Psi_m + (\sqh\chi_a)^{\dagger}\chi_a\right] + \textrm{gauge-fixing terms}.
\end{equation}
The gauge-fixing terms are:
\begin{equation*}
 \text{Tr}\left[\bar c G+\bar c B_0+ c\bar a_0\right].
\end{equation*}
We choose the gauge-fixing function $G= \partial_nA^n$. What is relevant for the localization procedure is only the quadratic truncation of the localizing action 
$\sqh V_\mathrm{vec}$, which we schematically write as
\begin{equation}\label{ciccio}
 \sqh V_\mathrm{vec}|_\mathrm{quad}= (\textrm{fermions}) ^{\dagger}K_\textrm{fermion}(\textrm{fermions})+(\textrm{bosons}) ^{\dagger}K_\textrm{boson}(\textrm{bosons}).
\end{equation}
Upon integration over the fluctuations, one finds the one-loop determinants
\begin{equation}
 Z^\textrm{vec}_\textrm{1-loop} = \sqrt{\frac{\det K_\textrm{fermion}}{\det K_\textrm{boson}}}.
\end{equation}
The calculation of the one-loop determinants can be simplified by introducing
\begin{equation*}
 X\equiv(\phi_2,A_m;\bar a_0,B_0),\hspace{0.5cm} \Xi \equiv (\chi_a,\bar c, c).
\end{equation*}
in terms of which \eqref{ciccio} can be written as
\begin{equation*}
 V_\mathrm{vec}|_\mathrm{quad}=\left(\sqh X, \,\Xi \right)\left(\begin{array}{cc}
                                                         D_{00} & D_{01}\\
                                                         D_{10} & D_{11}
                                                        \end{array}\right)
                                                        \left(\begin{array}{c}
                                                         X\\\sqh\Xi
                                                        \end{array}\right)  \ \ .
\end{equation*}
Correspondingly, we get
\begin{equation}\label{eq:Detratio}
 \frac{\det K_\textrm{fermion}}{\det K_\textrm{boson}} = \frac{\det_\Xi \sqh^2}{\det_X \sqh^2}= \frac{\det_{\ckrd} \sqh^2}{\det_{\krd} \sqh^2} \ \ .
\end{equation}
Rather than computing the ratio of determinants above via index theorem, as done in \cite{Pestun:2007rz} and \cite{Hama:2012bg}, we now proceed to explicitly determine the spaces $\ckrd$ and $\krd$, and compute the corresponding $\sqh^2$ eigenvalues. As we will see in the following, this method can be easily extended to $\mathbb{Z}_2$ quotients of the four-sphere calculating the relevant ratio of determinants for the $\mathbb{RP}^4$ and $HS^4$ cases.

The part of $V_\mathrm{vec}|_\mathrm{quad}$ that yields $D_{10}$ is:
\begin{equation}\label{eq:D10}
\sum_{a=1}^3\chi_a({\sqh\chi_a})^\dagger+cD_n(\sqh\Psi^n)^\dagger+\bar c\partial_n A^n+ \bar cB_0+c\bar a_0
\end{equation}
from which the terms containing $\phi_1$ and $D_{AB}$ may be dropped since they do not contribute to $D_{10}$.

\subsection{Gauge theories on ${\mathbb R}{\mathbb P}^4$}

In this Section we proceed to the computation of the partition function and Wilson loops of $\mathcal{N}=2$ supersymmetric gauge theories with compact semi-simple gauge group $G$  on $\mathbb{RP}^4$. We first discuss
the one-loop determinants calculation. The instanton contribution will be discussed in the next subsection \ref{inst-rp4}.
As anticipated in the previous Section, we compute one loop determinants for supersymmetry multiplets by solving the kernel and cokernel partial differential equations(PDEs) corresponding to $D_{10}$ operator. The  $D_{10}$ operator is read from the fermionic functional 
$V$ used to localize the physical actions $S$ for gauge and hyper multiplets. The PDEs are solved by diagonalizing them in the basis provided by $SO(4)\sim SU(2)_L\times SU(2)_R$ harmonics, which are discussed in Appendix \ref{app:harmonics}. 
It turns out that the PDEs reduce to ordinary differential equations in variable $r$ and as explained in the appendix \ref{app:KE}, each ODE can be expressed as linear combinations of the generators of $SU(2)_L$. This has the important consequence that $SU(2)_R$ commutes with kernel and cokernel differential equations and the solutions arrange themselves in $SU(2)_R$ multiplets. Let us denote by $X_0$ the set of dynamical fields in the kernel, and by $X_1$ the fields in the cokernel of $D_{10}$. In our computation $\phi_1,D_{AB}$ only contribute classically and so we set them to zero in the one-loop computation. We denote by $a=1,2,3$ the three directions in the tangent space of $S^3$ and use the following combinations of gauge fields $A_a$ belonging to 
$X_0$ in this basis.
\bea
A_+\equiv A_1+i A_2,\quad A_-\equiv A_1-i A_2\quad {\rm and} \quad A_3 \, .
\eea
The fermions $\chi_a$ are correspondingly denoted as
\bea
\chi_+\equiv\chi_1+i \chi_2,\quad \chi_-\equiv\chi_1-i \chi_2\quad {\rm and}   \quad \chi_3 \, .
\eea
In this basis the differential equations appear simpler. The computation can be further simplified by noting that, since $D_{10}$ commutes with $\hat{Q}^2$, it closes on the fields of same $\hat{Q}^2$ eigenvalues both in $X_0$ and $X_1$. Since the kernel and cokernel differential equations can be written only in term of $SU(2)_L$ generators, $SU(2)_R$ commutes with them and therefore all the fields in $X_0$ carry same $SU(2)_R$ charge equal to $q_R$ and those in $X_1$ will carry $-q_R$. Solutions of kernel and cokernel equations are organized in $SU(2)_R$ multiplets each of dimension $2j_R+1$ with the following possible values of $j_R$
\bea
j_R=j_L,j_L+1,j_L-1,
\eea
as determined by the $SO(4)$ harmonics.

The strategy to compute  the one-loop determinants
on $\mathbb{RP}^4\simeq \mathbb{S}^4/\mathbb{Z}_2$, is to take the solutions for kernel and cokernel equations on $\mathbb{S}^4$ and then apply the  antipodal $\mathbb{Z}_2$ projection on this solution set. This is done both for vector multiplet and hyper multiplet and then the invariant modes are combined to get the total one loop determinant.\\
 The analysis of ODEs can be performed in the simpler case of $U(1)$ gauge group, the generalization to general gauge group $G$, being obtained by multiplying the vector multiplet index by the factor $\sum_{\alpha \in Roots}e^{i \alpha \cdot a}$ in the adjoint representation, and the hyper multiplet in the representation $R$ of G by the factor  
 $\sum_{\rho \in R}e^{i \rho \cdot a}$. Finally, by using the eigenvalues of $\hat{Q}^2$ operator for the kernel and cokernel zero modes of $D_{10}$ we calculate the one-loop factor
\bea\label{eq:z1loop}
Z_{1-loop}=(\frac{\textbf{det}_{CokerD_{10}}\hat{\textbf{Q}}^2}{\textbf{det}_{KerD_{10}}\hat{\textbf{Q}}^2})^{\frac{1}{2}} \, .
\eea
We will work with the following killing spinor  on the covering space $S^4$
\bea
\xi=\left(
\begin{array}{cc}
 \frac{\cos \left(\frac{r}{2}\right)}{\sqrt{2}} & 0 \\
 0 & \frac{\cos \left(\frac{r}{2}\right)}{\sqrt{2}} \\
 \frac{i \sin \left(\frac{r}{2}\right)}{\sqrt{2}} & 0 \\
 0 & -\frac{i \sin \left(\frac{r}{2}\right)}{\sqrt{2}} \\
\end{array}
\right)
\eea
for the metric 
\bea
ds^2=g_{\mu\nu}dx^{\mu}dx^{\nu}=dr^2+\frac{f(r)^2}{4}\big(d\theta^2+\sin\theta^2d\phi^2+(d\psi+\cos\theta d\phi)^2\big) \, .
\eea
We use Hopf vibration coordinates
\bea
z_1= \sin(r)\sin(\frac{\theta}{2})e^{i\frac{(\psi-\phi)}{2}},\quad z_2=\sin(r)\cos(\frac{\theta}{2})e^{i\frac{(\psi+\phi)}{2}},\quad t=\cos(r)
\eea
with $0\le\theta\le\pi$, $0\le\phi\le2\pi$ and $0\le\psi\le4\pi$.
In this coordinates, the antipodal $\mathbb{Z}_2$ action reads
\bea
r\to\pi-r,\quad \psi\to\psi+2\pi
\eea
and correspondingly
\bea
z_1\to -z_1,\quad z_2\to -z_2,\quad t\to -t \, .
\eea
On the killing spinor we have
\bea
\mathbb{Z}_2:\xi\to \left(
\begin{array}{cc}
 \frac{\sin \left(\frac{r}{2}\right)}{\sqrt{2}} & 0 \\
 0 & \frac{\sin \left(\frac{r}{2}\right)}{\sqrt{2}} \\
 \frac{i \cos \left(\frac{r}{2}\right)}{\sqrt{2}} & 0 \\
 0 & -\frac{i \cos \left(\frac{r}{2}\right)}{\sqrt{2}} \\
\end{array}
\right)
\eea
so that this is not preserved under the $\mathbb{Z}_2$ projection. By using the two component notation  
\bea
\xi_A=\left(
\begin{array}{cc}
 \frac{\cos \left(\frac{r}{2}\right)}{\sqrt{2}} & 0 \\
 0 & \frac{\cos \left(\frac{r}{2}\right)}{\sqrt{2}} \\
\end{array}
\right),\quad
 \bar{\xi}_A=\left(
\begin{array}{cc}
 \frac{i \sin \left(\frac{r}{2}\right)}{\sqrt{2}} & 0 \\
 0 & -\frac{i \sin \left(\frac{r}{2}\right)}{\sqrt{2}} \\
\end{array}
\right)
\eea
we have indeed
\bea
\mathbb{Z}_2:\xi_A\to -i \bar{\xi}_B \sigma_{3A}^B,\quad \bar{\xi}_A\to i {\xi}_B \sigma_{3A}^B \, .
\eea
It is easy to check that anyway the localising action $\hat{Q}_{\xi}V$ remains invariant provided that we  choose the following transformation properties of dynamical and ghost fields
\bea
&\chi_3\to -\chi_3,\quad \chi_+\to \chi_+,\quad \chi_-\to \chi_- ,\quad c\to c,\quad \bar{c}\to \bar{c}\nonumber\\
&A_{\psi}\to A_{\psi},\quad A_{\theta}\to A_{\theta},\quad A_{\phi}\to A_{\phi},\quad A_{r}\to -A_{r},\quad \phi_1\to \phi_1,\quad \phi_2\to -\phi_2\nonumber\\
\eea 
The projection for fermions can also be written in terms of $\lambda$s as 
\bea
\mathbb{Z}_2:\lambda_A\to -i \bar{\lambda}_B\sigma_{3A}^B,\quad \bar{\lambda}_A\to i {\lambda}_B\sigma_{3A}^B
\eea
and these projections are consistent with supersymmetry transformations.

\subsection*{$\mathbb{Z}_2$-projection on the vector multiplet}
On $\mathbb{RP}^4$ spinors form $Pin_+$ representations depending on which connected component of $O(4)$, the tangent space group, one is in. There are two $Pin_+$ structures on $\mathbb{RP}^4$ one with monodromy $+1$ as on moves along the orientation reversing loop and the second with monodromy $-1$. On $S^4$, the double cover of $\mathbb{RP}^4$, these two types of spinors translate into two parity conditions
\bea
P \Psi=\pm \Psi
\eea
These  parity projections have to be consistent with ${\mathcal N}=2$ supersymmetry.

It is interesting to remark that\footnote{When all fields are in the adjoint representation, the gauge group is actually $SU(2)/\mathbb{Z}_2\sim SO(3)$.} since $\pi_1(SO(3))=\mathbb{Z}_2$ and $\pi_3(SO(3))=\mathbb{Z}$, an $SO(3)$ bundle on a general 4-manifold $M$ is characterized by two topological invariants. One is the instanton number $k$ and the other is the non-abelian magnetic flux or an element of $H^2(M,\mathbb{Z}_2)$ called the second Stieffel-Whitney class of the $SO(3)$ gauge bundle. If we represent the second Stieffel-Whitney class as $u=\omega_2(SO(3))$, then these two topological invariants are related by 
\bea\label{eq:ku}
k=-\frac{u\cdot u}{4}\quad mod\quad 1.
\eea
see e.g. \cite{Vafa:1994tf} for more explanations. The element of $H^2(\mathbb{RP}^4,\mathbb{Z}_2)$ with $u=0$ corresponds to the untwisted sector. of perturbative and non-perturbative parts of the partition function.It is identical to the contribution of an $SU(2)$ gauge bundle. However for $u\ne 0\in H^2(\mathbb{RP}^4,\mathbb{Z}_2)$ one get the contribution of the twisted sector.
This operation is carried out on the one-loop part by shifting the product over the modes by $\frac{1}{2}$ and in the non-perturbative  sector by shifting $k$ by the relation (\ref{eq:ku}).

Let us start by discussing the trivial holonomy sector.
As shown in the Appendix \ref{app:KE}, for the vector multiplet the solution set of the kernel equations is  empty. We have the following projection condition for cokernel fields
\bea\label{eq:untwisted1}
\chi_3\to -\chi_3,\quad \chi_+\to \chi_+,\quad \chi_-\to \chi_- ,\quad c\to c,\quad \bar{c}\to \bar{c}\nonumber\\
\eea
The $\hat{Q}^2$ eigenvalue of the fields $\chi_3,c$ and $\bar{c}$ is $n+ia\cdot \alpha$ with multiplicity $n+1$, and that of $\chi_+$,$\chi_-$ is $n+ia\cdot \alpha$ with multiplicity $n-1$. 
It is important to observe that the cokernel equations are coupled and the system for $\chi_3,c,\bar{c}$ admits a unique solution up to a constant. Therefore they count as a single mode in the harmonic expansion.
Thus according to (\ref{eq:untwisted1}) , $\chi_3$ give contribution from odd modes only and $\chi_+$ and $\chi_-$ will give contribution just from even modes. Explicitely for $\chi_3$
\bea
\mathbb{Z}_2:\prod_{\alpha\in\Delta}  \prod_{n\ge1}  (n+i a\cdot \alpha)^{n+1}= \prod_{\alpha\in\Delta} \prod_{n\ge1}(2n-1+i a\cdot \alpha)^{2n-1+1}(2n+i a\cdot \alpha)^{2n+1})\nonumber\\\Longrightarrow \prod_{\alpha\in\Delta}\prod_{n\ge0} (2n-1+i a\cdot \alpha)^{2n-1+1}\nonumber\\
\eea
and for $\chi_+,\chi_-$
\bea
\mathbb{Z}_2:  \prod_{\alpha\in\Delta}\prod_{n\ge1}  (n+i a\cdot \alpha)^{n-1}=\prod_{\alpha\in\Delta} \prod_{n\ge1}(2n-1+i a\cdot \alpha)^{2n-1-1}(2n+i a\cdot \alpha)^{2n-1})\nonumber\\\Longrightarrow\prod_{\alpha\in\Delta} \prod_{n\ge1} (2n+i a\cdot \alpha)^{2n-1}\nonumber\\
\eea

The unregularized product can be rewritten as 
\begin{equation}
\prod_{\alpha\in\Delta_+}  \prod_{n\geq 1}(2n + ia\cdot \alpha)^{2n-1}(2n-1 + ia\cdot \alpha)^{2n-1+1}(2n -
ia\cdot \alpha)^{2n-1}(2n-1 - ia\cdot \alpha)^{2n-1+1}
\end{equation}
whose $\zeta$-function regularised form is 
\begin{equation}\label{vec-rp4}
Z_{1-loop}^{{\rm vec},\mathbb{RP}^4}=
\prod_{\alpha\in\Delta_+}\frac{\Ups(ia\cdot \alpha)}{ia\cdot \alpha}\Gm(1+ia\cdot \alpha)^2\cosh^2\frac{\pi a\cdot \alpha}{2}
\end{equation}

For the non trivial holonomy sector,
a similar analysis yields the following expression for the vector multiplet contribution to the one-loop determinant 
\begin{equation}
\prod_{\alpha\in\Delta_+}\frac{\Ups(ia\cdot \alpha)}{ia\cdot \alpha}\Gm(1+ia\cdot \alpha)^2\sinh^2\frac{\pi a\cdot \alpha}{2}
\end{equation}

\subsection*{ $\mathbb{Z}_2$-projection on hypermultiplets}

Under the $\mathbb{Z}_2$ action the fields in matter hypermultiplet belonging to the kernel and Cokernel of $D_{10}^{hyper}$ transform as follows
\bea\label{eq:hyperproj}
q_{AB}:&q_{11}\to q_{11},\quad q_{12}\to q_{12},\quad
q_{21}\to -q_{21},\quad q_{22}\to -q_{22},\nonumber\\
\Sigma_{AB}:&\Sigma_{11}\to \Sigma_{11},\quad \Sigma_{12}\to \Sigma_{12},\quad
\Sigma_{21}\to -\Sigma_{21},\quad \Sigma_{22}\to -\Sigma_{22}.\nonumber\\
\eea
As shown in the Appendix \ref{app:KE},  for the hypermultiplet the solution set of cokernel PDEs is empty and only the kernel fields contribute. For completeness we give the mode expansion of kernel fields $q_A$ in a general representation ${\bf R}$
\bea
&&q_{11}:\quad  \prod_{\rho\in \bf{R}} \prod_{k\ge 1}(k+\mu+i a\cdot\rho)^{k},\nonumber\\&
&q_{12}:\quad  \prod_{\rho\in \bf{R}} \prod_{k\ge 1}(k-\mu+i a\cdot\rho)^{k},\nonumber\\&
&q_{21}:\quad  \prod_{\rho\in \bf{R}} \prod_{k\ge 1}(-k+\mu+i a\cdot\rho)^{k},\nonumber\\&
&q_{22}:\quad  \prod_{\rho\in \bf{R}} \prod_{k\ge 1}(-k-\mu+i a\cdot\rho)^{k} \, ,
\eea
whose $\mathbb{Z}_2$ projection under (\ref{eq:hyperproj}) gives
\bea
&&q_{11}:\quad  \prod_{\rho\in \bf{R}} \prod_{k\ge 1}(2k+\mu+i a\cdot\rho)^{2k-1},\nonumber\\&
&q_{12}:\quad  \prod_{\rho\in \bf{R}} \prod_{k\ge 1}(2k-\mu+i a\cdot\rho)^{2k-1},\nonumber\\&
&q_{21}:\quad  \prod_{\rho\in \bf{R}} \prod_{k\ge 1}(-2k+1+\mu+i a\cdot\rho)^{2k},\nonumber\\&
&q_{22}:\quad  \prod_{\rho\in \bf{R}} \prod_{k\ge 1}(-2k+1-\mu+i a\cdot\rho)^{2k}.
\eea
where $\mu$ is the mass parameter. The one-loop determinant turns out to be
\bea
\bigg( \prod_{\rho\in \bf{R}} \prod_{k\ge 1}(2k+\mu+i a\cdot\rho)^{2k}(2k-\mu+i a\cdot\rho)^{2k}(-2k+1+\mu+i a\cdot\rho)^{2k-1}\nonumber\\ \times(-2k+1-\mu+i a\cdot\rho)^{2k-1}\bigg)^{\frac{1}{2}}\nonumber\\
=\bigg( \prod_{\rho\in \bf{R}} \prod_{k\ge 1}(2k+\mu+i a\cdot\rho)^{2k}(2k-\mu+i a\cdot\rho)^{2k}(2k-1-\mu-i a\cdot\rho)^{2k-1}\nonumber\\ \times (2k-1+\mu-i a\cdot\rho)^{2k-1}\bigg)^{\frac{1}{2}}
\eea
which can be further simplified for a real representation to
\bea
\bigg( \prod_{\rho\in \bf{R}} \prod_{k\ge 1}(2k+\mu+i a\cdot\rho)^{2k}(2k-\mu+i a\cdot\rho)^{2k}(2k-1-\mu+i a\cdot\rho)^{2k-1}\nonumber\\ \times (2k-1+\mu+i a\cdot\rho)^{2k-1}\bigg)^{\frac{1}{2}}\nonumber\\
=\bigg( \prod_{\rho\in \bf{R}} \prod_{k\ge 1}(k+\mu+i a\cdot\rho)^{k}(k-\mu+i a\cdot\rho)^{k}\bigg)^{\frac{1}{2}}
\eea
The regularised form of the above is given by
\bea
Z_{1-loop}^{{\rm hyper},\mathbb{RP}^4}&=&
\bigg(\prod_{\rho\in \bf{R}} G(1+\mu+i a\cdot\rho)G(1-\mu+i a\cdot\rho)\bigg)^{\frac{1}{2}}
\eea
In the massless case, $\bf{\mu}=0$
this is further simplified to
\bea
Z_{1-loop,\mu=0}^{{\rm hyper},\mathbb{RP}^4}&=&\bigg(\prod_{\rho\in \bf{R}} G(1+i a\cdot\rho)G(1-i a\cdot\rho)\bigg)^{\frac{1}{2}}\nonumber\\
&=&\bigg(\prod_{\rho\in \bf{R}} H(i a\cdot\rho)\bigg)^{\frac{1}{2}}
\eea
Combining the above with the vector multiplet contribution \eqref{vec-rp4}, the one loop partition function of ${\cal N}=2$ with a massless hypermultiplet in a real representation reads
\bea
Z_{1-loop,\mu=0}^{{\bf R},\mathbb{RP}^4}=\prod_{\alpha\in\Delta_+} \frac{\Ups(ia\cdot \alpha)}{ia\cdot \alpha}\Gm(1+ia\cdot \alpha)^2\cosh^2\frac{\pi a\cdot \alpha}{2}
{(\prod_{\rho\in \bf{R}} H(i a\cdot\rho))^{-\frac{1}{2}}}
\eea
A notable case we will need in the following is that of ${\mathcal N}=2^*$ theory with
$\bf{\mu}=1$ (namely, the maximally supersymmetric case for the unsquashed case $b=1$)
whose partition function reads
\bea
Z_{1-loop,\mu=1}^{Adj,\mathbb{RP}^4}&=&\bigg(\prod_{\alpha\in \Delta} G(2+i a\cdot \alpha)G(i a\cdot \alpha)\bigg)^{\frac{1}{2}}\nonumber\\
&=&\bigg(\prod_{\alpha\in \Delta} \Gamma(1+ia\cdot \alpha)G(1+i a\cdot \alpha)\frac{G(1-i a\cdot \alpha)}{\Gamma(-i a\cdot \alpha)}\bigg)^{\frac{1}{2}}\nonumber\\
&=&\bigg(\prod_{\alpha\in \Delta} \frac{\Gamma(1+ia\cdot \alpha)}{\Gamma(-i a\cdot \alpha)}H(i a\cdot \alpha)\bigg)^{\frac{1}{2}}\nonumber\\
&=&\bigg(\prod_{\alpha\in \Delta}H(1+i a\cdot \alpha)\bigg)^{\frac{1}{2}}=\bigg(\prod_{\alpha\in \Delta}\Upsilon(i a\cdot \alpha)\bigg)^{\frac{1}{2}}
\eea
Combining it with the vector multiplet contribution one finally gets
\bea
Z_{1-loop,\mu=1}^{{\mathcal N}=2^*,\mathbb{RP}^4}= \prod_{\alpha\in\Delta_+}\frac{\Ups(ia\cdot \alpha)}{ia\cdot \alpha}\Gm(1+ia\cdot \alpha)^2\cosh^2\frac{\pi a\cdot \alpha}{2}
{\bigg(\prod_{\alpha\in \Delta}\Upsilon(i a\cdot \alpha)\bigg)^{-\frac{1}{2}}
}
\eea
Let us present the one-loop contribution for $\mathcal{N}=4$ gauge theory
\bea
Z_{1-loop}^{\mathcal{N}=4, \mathbb{RP}^4}&=&\prod_{\alpha\in\Delta_+} \frac{\Ups(ia\cdot \alpha)}{ia\cdot \alpha}\Gm(1+ia\cdot \alpha)^2\cosh^2\frac{\pi a\cdot \alpha}{2} \frac{1}{\bigg(\prod_{\alpha\in \Delta}H(i a\cdot \alpha)\bigg)^{\frac{1}{2}}}\nonumber\\
&=&\prod_{\alpha\in\Delta_+} \frac{\Ups(ia\cdot \alpha)}{ia\cdot \alpha}\Gm(1+ia\cdot \alpha)^2\cosh^2\frac{\pi a\cdot \alpha}{2} \bigg(\prod_{\alpha\in \Delta}\frac{\Gamma(-ia\cdot \alpha)}{\Gamma(1+ia\cdot \alpha)(H(1+i a\cdot \alpha))}\bigg)^{\frac{1}{2}}\nonumber\\
&=&\prod_{\alpha\in\Delta_+} \frac{\Ups(ia\cdot \alpha)}{ia\cdot \alpha}\Gm(1+ia\cdot \alpha)^2\cosh^2\frac{\pi a\cdot \alpha}{2} \bigg(\prod_{\alpha\in \Delta}\frac{\Gamma(-ia\cdot \alpha)}{\Gamma(1+ia\cdot \alpha)(\Upsilon(i a\cdot \alpha))}\bigg)^{\frac{1}{2}}\nonumber\\
\eea

Finally, 
for the system of four hypermultiplets we denote the mass parameters as $\mu_i,i=1,...,4$. These mass terms in the Lagrangian are generated by gauging a $U(1)^4$ subgroup of the commutant of $SU(2)$ in the flavor symmetry group $SO(8)$. By using the above results for the massive hypermultiplet in the fundamental representation and 
in order to implement the $\mathbb{Z}_2$-projection in the matter sector, we make the following identification of mass parameters
\bea
\mu_1=\mu_3,\quad \mu_2=\mu_4.
\eea
As a consequence under the $\mathbb{Z}_2$ projection the hypers of masses $\mu_1,\mu_3$  give rise to two half hypers which under the identification $\mu_1=\mu_3$ combine to give one hypermultiplet. Similarly other two hypers of masses $\mu_2,\mu_4$ under $\mathbb{Z}_2$ projection and identification $\mu_2=\mu_4$ give rise to a second hypermultiplet.
Explicity to compute the hypermultiplet contributions, we have to determine the transformation of the component fields of the hypermultiplet under the $\mathbb{Z}_2$ action. We first note that the Killing spinors $(\xi,\bar{\xi})$ and $(\check{\xi},\bar{\check{\xi}})$ transform as
\bea
\xi_{1\alpha}\to -i \bar{\xi}_{1\dot{\alpha}},\quad \xi_{2\alpha}\to i \bar{\xi}_{2\dot{\alpha}}\nonumber\\
\check{\xi}_{1\alpha}\to i \bar{\check{\xi}}_{1\dot{\alpha}},\quad \check{\xi}_{2\alpha}\to -i \bar{\check{\xi}}_{2\dot{\alpha}}
\eea
Consistency with supersymmetry transformations given in appendix (\ref{app:susy}) require us to choose the transformation of the scalar and fermions in the hypermultiplet as
\bea
q_1\to q_1,\quad q_2\to-q_2,\nonumber\\
\psi^{\alpha}\to i \bar{\psi}^{\dot{\alpha}},\quad \bar{\psi}^{\dot{\alpha}}\to- i \psi^{\alpha},
\eea
Projecting out the Fourier components which do not obey this transformation, we get the following contribution to the one-loop part
\bea
Z_{4\, {\rm half-hypers}}= \prod_{\pm}\prod_{i=1,2}\Gamma_2(1\pm i a\pm\mu_i)
\eea
It is interesting to note that this contribution of hypers is identical to that on the Hemi-$S^4$.
Finally the full expression for the one-loop part is the following 
\bea \label{Znf4}
Z^{N_f=4,\,\mathbb{RP}_4}_{1-loop}=\prod_{\alpha\in\Delta_+}\frac{\Ups(ia\cdot \alpha)}{ia\cdot \alpha}\Gm(1+ia\cdot \alpha)^2\cosh^2\frac{\pi a\cdot \alpha}{2} \prod_{\pm}\prod_{i=1,2}\Gamma_2(1\pm i a\pm\mu_i)
\eea

\subsubsection{Instanton contribution}
\label{inst-rp4}

Let us now discuss the instanton contribution to the supersymmetric partition function on $\mathbb{RP}^4$.
It is convenient to work on its double cover $S^4$, with a suitable antipodal identification. For example in stereographic coordinates $(\frac{4 X_i}{X_iX_i+4},\frac{X_iX_i-4}{X_iX_i+4})$ with $X_i\in \mathbb{R}^4$ the antipodal identification is
\bea
X_i\to -\frac{4X_i}{X_iX_i}
\eea
where summation over $i=1,..,4$ is understood.
On $S^4$ the contribution to the supersymmetric partition function comes only from point-like instantons and anti instantons at the South pole and North pole respectively. Under the antipodal identification instantons with charge $k$ is mapped to anti-instantons with charge $-k$. In other words, being $ \mathbb{RP}^4$  non-orientable, as an instanton is moved along the orientation reversing path, it comes back to the original position as an anti-instanton due to non-trivial monodromy. As a result the $\theta$-term 
\bea
S_{\theta}=\frac{\theta}{32 \pi^2}\int d^4x \epsilon_{\mu\nu\rho\sigma}F^{\mu\nu}F^{\rho\sigma}
\eea
which is added to the SYM action to give the vacuum a topological charge, flips it sign under antipodal map and hence is ill defined. However for  $\theta=0,\pi$ the integral remains invariant. For $\theta=0$ this is obvious, and for $\theta=\pi$ due to $2\pi$ periodicity. 
For $\theta=0$ the  coupling constant is $\tau=\frac{4\pi i}{g^2}$ and, due to the antipodal identification of North and South poles, the non-perturbative contribution is just one factor of Nekrasov instanton partition function with the above value of $\tau$.
The other case $\theta=\pi$ will be discussed later in the case corresponding to the Moebius strip.


\subsection{Gauge theory on $HS^4$}
\label{sec:hemisphere}

The explicit computation of the supersymmetric partition function on $HS^4$ was performed in \cite{Gava:2016oep}. Here for completeness we recall their results and also compute new cases
which are needed for the comparison with Liouville theory. The possible choices of supersymmetric boundary conditions are Dirichlet and Neumann. 

\subsection*{Dirichlet Boundary Condition for the vector multiplet}\label{sec:OneLoopDir}

The Dirichlet boundary conditions on the boundary $S^3$ of the hemisphere for the fermionic fields $\Xi$ are given by
\begin{equation*}
 \begin{array}{ll}
  \chi_\pm(\rho)|_{\rho=\pi/2}=\chi_0,& c|_{\rho=\pi/2}=0, \\
  \chi_3|_{\rho=\pi/2}=0,& \bar c|_{\rho=\pi/2}=0,
 \end{array}
\end{equation*}
where $\chi_0$ is a constant. The solutions of the cokernel equations, are displayed in Appendix \ref{app:KE}. The ones compatible with the above boundary conditions are those corresponding to $|q_L| = j_L + 1$. Therefore, the multiplicity of an eigenvalue $n+ia\cdot \alpha$ is $|n|-1$, as opposed to $2|n|$ for the spherical case. The corresponding unregularized product is given by: 
\begin{equation}\label{eq:unregDir}
  \prod_{\alpha \in \Delta_+}\prod_{n\geq 1}(n + ia\cdot \alpha)^{n-1}(n - ia\cdot \alpha)^{n-1}
\end{equation}
which, upon regularization and choosing the gauge group to be $SU(2)$ becomes
\begin{equation}\label{Zdir}
 Z_\mathrm{Dir,\, 1-loop}^\mathrm{vec}(a) = \Ups(2ia)\frac{\Gm(1 + 2ia)^2}{ia}
\end{equation}
Let us consider Wilson loops in the boundary of the hemisphere. Solely for this section, we will use the coordinates $(\rho,\vartheta,\varphi,\chi)$, where $\varphi = \frac{\psi+\phi}{2}$, $\chi = \frac{\psi-\phi}{2}$, and where $\varphi$ and $\chi$ are periodic with period $2\pi$. 
 There are two classes of closed loops in the boundary $S^3$, one winding around $\varphi$, and another winding around $\chi$. Following \cite{Hama:2012bg}, the circles supporting the Wilson loops are:
\begin{equation*}
\begin{aligned}
 S^1_\varphi(\rho=\tfrac{\pi}{2})&:(x_0,x_1,x_2,x_3,x_4)=(0,\epsilon_1\cos\varphi,\epsilon_1\sin\varphi,0,0),\\
 S^1_\chi(\rho=\tfrac{\pi}{2})&:(x_0,x_1,x_2,x_3,x_4)=(0,0,0,\epsilon_2\cos\chi,\epsilon_2\sin\chi).
 \end{aligned}
\end{equation*}
The supersymmetric Wilson loops are
\begin{equation*}
\begin{aligned}
 W_\varphi(R)&\equiv \mathrm{Tr}_R\mathrm{P}\exp i\int_{S^1_\varphi\left(\rho=\tfrac{\pi}{2}\right)}d\varphi(A_\varphi + i\epsilon_1\phi_1),\\
 W_\chi(R)&\equiv \mathrm{Tr}_R\mathrm{P}\exp i\int_{S^1_\chi\left(\rho=\tfrac{\pi}{2}\right)}d\chi(A_\chi + i\epsilon_2\phi_1).
 \end{aligned}
\end{equation*}
These have the following vevs:
\begin{equation*}
\begin{aligned}
W_\varphi(R)&=\mathrm{Tr}_R\mathrm{P}\exp\left(-2\pi b\hat a \right)\\
W_\chi(R)&= \mathrm{Tr}_R\mathrm{P}\exp\left(-2\pi b^{-1}\hat a \right).
 \end{aligned}
\end{equation*}
where $\hat a = \sqrt{\epsilon_1\epsilon_2}a$. To compute their expectation values, one simply inserts these classical expressions in the integral expression for the partition function.
One could also consider a more general Wilson loop that winds $m$-times around $\varphi$ and $n$-times around $\chi$ in the fundamental representation. 
These play a relevant r\^ole for the matching between gauge theory and Liouville amplitudes.

\subsection*{Neumann Boundary Conditions for the vector multiplet}\label{sec:OneLoopNeu}
Neumann boundary conditions are:
\begin{equation*}
 \chi_\pm(\rho)|_{\rho=\pi/2}=0.
\end{equation*}
The only non-trivial solutions to the cokernel equations compatible with this boundary condition correspond to $|q_L| = j_L + 1$. The multiplicity of a solution with eigenvalue $n+ia\cdot \alpha$ is $|n|+1$. The one-loop partition function is
\begin{equation}\label{eq:unregNeu}
  Z_\mathrm{Neu}^\mathrm{vec}(a) = \prod_{\alpha \in \Delta_+}\prod_{n\geq 1}(n + ia\cdot \alpha)^{n+1}(n - ia\cdot \alpha)^{n+1}
\end{equation}
which, upon regularization and choosing the gauge group to be $SU(2)$, gives
\begin{equation*}
\Ups(2ia)\frac{1}{\Gm(1 + 2ia)^2(ia)}
\end{equation*}
The partition function is obtained by integrating the above over the Coulomb branch parameter. Analogous to the case of a sphere, we must include the factor arising from the Vandermonde determinant:
\begin{equation}\label{Zvec}
  Z_\textrm{1-loop, Neu}^\mathrm{vec}(a) = \Ups(2ia)\frac{ia}{\Gm(1 + 2ia)^2}.
\end{equation}

\subsection*{Dirichlet boundary conditions for the  hyper-multiplet}

Let us recall that the solution set of cokernel PDEs is empty in the hypermultiplet case. Therefore only the mode contribution of  the  kernel fields is relevant and is given by
\bea
&&q_{11}:\quad  \prod_{\rho\in \bf{R}} \prod_{k\ge 1}(-k+\mu+i a\cdot\rho)^{k},\nonumber\\&
&q_{12}:\quad  \prod_{\rho\in \bf{R}} \prod_{k\ge 1}(-k-\mu+i a\cdot\rho)^{k},\nonumber\\&
&q_{21}:\quad  \prod_{\rho\in \bf{R}} \prod_{k\ge 1}(k+\mu+i a\cdot\rho)^{k},\nonumber\\&
&q_{22}:\quad  \prod_{\rho\in \bf{R}} \prod_{k\ge 1}(k-\mu+i a\cdot\rho)^{k}.
\eea
where $\bf{R}$ is the matter representation of the gauge group and $\mu$ is mass parameter.
Dirichlet boundary conditions amount to
\bea
\text{q}_{12}(\theta ,\frac{\pi}{2})&=&0,\quad  \text{q}_{21}(\theta ,\frac{\pi}{2})=0,\quad  \Sigma_{11}(\theta ,\frac{\pi}{2})=0\quad  \Sigma_{22}(\theta ,\frac{\pi}{2})=0,\nonumber\\
\partial_{\theta}\text{q}_{12}(\theta ,\frac{\pi}{2})&=&0,\quad \partial_{\theta}\text{q}_{21}(\theta ,\frac{\pi}{2})=0,\quad\partial_{\theta}\Sigma_{11}(\theta ,\frac{\pi}{2})=0,\quad \partial_{\theta}\Sigma_{22}(\theta ,\frac{\pi}{2})=0\nonumber\\
\eea
and 
\bea
\partial_r\text{q}_{11}(\theta ,\frac{\pi}{2})=0,\quad \partial_r\text{q}_{22}(\theta ,\frac{\pi}{2})=0,\quad \partial_r\Sigma_{12}(\theta ,\frac{\pi}{2})=0,\quad \partial_r\Sigma_{21}(\theta ,\frac{\pi}{2})=0
\eea
This translates into the following expression for the one-loop determinant 
\bea
\left(  \prod_{\rho\in \bf{R}} \prod_{k\ge 1}(k-\mu-i a\cdot\rho)^{k} (k+\mu+i a\cdot\rho)^{k}\right)^{\frac{1}{2}}\nonumber\\
\eea
and after regularization to
\bea\label{Zhyp}
Z^{{\rm hyper}}_{{\rm Dir, 1-loop}}&=& \left(  \prod_{\rho\in \bf{R}}G(1-\mu-i a\cdot\rho)G(1+\mu+i a\cdot\rho)\right)^{\frac{1}{2}}
\eea
For the the $\mathcal{N}=2^*$ theory at $\mu=1$ 
\bea
Z^{\mathcal{N}=2^*,\, hyper}_{{\rm Dir,1-loop}}&=& \left(   \prod_{\alpha\in \Delta} G(-i a\cdot \alpha)G(2+i a\cdot \alpha)\right)^{\frac{1}{2}}\nonumber\\
&=& \left(  \prod_{\alpha\in \Delta} \frac{G(1-i a\cdot \alpha)G(1+i a\cdot \alpha)\Gamma(1+ia\cdot \alpha)}{\Gamma(-ia\cdot \alpha)}\right)^{\frac{1}{2}}\nonumber\\
&=&\left(   \prod_{\alpha\in \Delta} \frac{\Gamma(1+ia\cdot \alpha)H(i a\cdot \alpha)}{\Gamma(-ia\cdot \alpha)}\right)^{\frac{1}{2}}\nonumber\\
\eea
Combining it with the vector one loop part one gets
\bea
Z_{{\rm Dir, 1-loop}}^{\mathcal{N}=2^*, \mu=1} &=&(\prod_{\alpha\in \Delta} H(i a\cdot \alpha)\frac{1}{i a\cdot \alpha \sinh(i \pi a\cdot \alpha) })^{\frac{1}{2}}(\prod_{\alpha\in \Delta}\frac{\Gamma(-ia\cdot \alpha)}{ \Gamma(1+ia\cdot \alpha)H(i a\cdot \alpha)})^{\frac{1}{2}}\nonumber\\
&=&(\prod_{\alpha\in \Delta} \frac{\Gamma(ia\cdot \alpha)}{i a\cdot \alpha })^{\frac{1}{2}}(\prod_{\alpha\in \Delta}\Gamma(-ia\cdot \alpha))^{\frac{1}{2}}\nonumber\\
&=&(\prod_{\alpha\in \Delta} \frac{\Gamma(ia\cdot \alpha)\Gamma(-ia\cdot \alpha)}{i a\cdot \alpha })^{\frac{1}{2}}
\eea
Finally, the one loop contribution of $\mathcal{N}=4$ theory with Dirichlet BCs  can be obtained by combining the one loop part of $\mathcal{N}=2$ vector multiplet with that of $\mathcal{N}=2$ hypermultiplet in the adjoint representation
\bea
Z^{{\rm vec}}_{{\rm Dir, 1-loop}}&=& \prod_{\alpha\in\Delta_+}G(1+i a\cdot \alpha)G(1-i a\cdot \alpha)\frac{ 1}{a\cdot \alpha\sinh(\pi a\cdot \alpha)}\nonumber\\
&=& \prod_{\alpha\in\Delta_+} H(i a\cdot \alpha)\frac{1}{a\cdot \alpha \sinh( \pi a\cdot \alpha) }
\eea
\bea
Z^{{\rm hyper}}_{{\rm Dir, 1-loop}}=\bigg(\prod_{\alpha\in\Delta}\frac{1}{H(i a\cdot \alpha)}\bigg)^{\frac{1}{2}}
\eea
Combining them one get the perturbative part of $\mathcal{N}=4$ vector multiplet with Dirichlet BCs on $HS^4$
\bea
Z^{\mathcal{N}=4}_{{\rm Dir, 1-loop}}&=& \prod_{\alpha\in\Delta_+} H(i a\cdot \alpha)\frac{1}{a\cdot \alpha \sinh( \pi a\cdot \alpha) }\times \bigg(\prod_{\alpha\in\Delta}\frac{1}{H(i a\cdot \alpha)}\bigg)^{\frac{1}{2}}
\eea
From the identity $G(1+i a\cdot \alpha)G(1-i a\cdot \alpha)=H(i a\cdot \alpha)$, it is clear that $H(i a\cdot \alpha)=H(-i a\cdot \alpha)$. Therefore one gets
\bea
Z^{\mathcal{N}=4}_{{\rm Dir, 1-loop}}&=&\prod_{\alpha\in\Delta_+} H(i a\cdot \alpha)\frac{1}{a\cdot \alpha\sinh( \pi a\cdot \alpha) }\times \frac{1}{H(i a\cdot \alpha)}\nonumber\\
&=&\prod_{\alpha\in\Delta_+}\frac{1}{a\cdot \alpha\sinh( \pi a\cdot \alpha) }
\eea

\subsection*{Neumann boundary conditions for the  hyper-multiplet}

Neumann boundary conditions imply
\bea
\text{q}_{11}(\theta ,\frac{\pi}{2})&=&0,\quad  \text{q}_{22}(\theta ,\frac{\pi}{2})=0,\quad  \Sigma_{12}(\theta ,\frac{\pi}{2})=0\quad  \Sigma_{21}(\theta ,\frac{\pi}{2})=0,\nonumber\\
\partial_{\theta}\text{q}_{11}(\theta ,\frac{\pi}{2})&=&0,\quad \partial_{\theta}\text{q}_{22}(\theta ,\frac{\pi}{2})=0,\quad\partial_{\theta}\Sigma_{12}(\theta ,\frac{\pi}{2})=0,\quad \partial_{\theta}\Sigma_{21}(\theta ,\frac{\pi}{2})=0\nonumber\\
\eea
and 
\bea
\partial_r\text{q}_{12}(\theta ,\frac{\pi}{2})=0,\quad \partial_r\text{q}_{22}(\theta ,\frac{\pi}{2}),\quad \partial_r\Sigma_{11}(\theta ,\frac{\pi}{2})=0,\quad \partial_r\Sigma_{22}(\theta ,\frac{\pi}{2})=0
\eea
It is clear from the mode decomposition that we will get the following result 
\bea \label{Zhypneu}
\left(  \prod_{\rho\in \bf{R}} \prod_{k\ge 1}(k-\mu-i a\cdot\rho)^{k} (k+\mu+i a\cdot\rho)^{k}\right)^{\frac{1}{2}}\nonumber\\
= \left( \prod_{\rho\in \bf{R}} \prod_{k\ge 1}G(1+\mu+i a\cdot\rho)G(1-\mu-i a\cdot\rho)\right)^{\frac{1}{2}}
\eea
This is identical to the previous case which for $\mu=1$ leads to 
\be 
Z^{{\rm hyper}, \mu=1}_{{\rm Neu, 1-loop}}=\left( \prod_{\alpha\in \bf{R}} \frac{\Gamma(1+ia\cdot \alpha)H(i a\cdot \alpha)}{\Gamma(-ia\cdot \alpha)}\right)^{\frac{1}{2}}
\ee
Combining it with the vector one loop part one gets
\bea
Z_{{\rm Neu, 1-loop}}^{\mathcal{N}=2^*,\, \mu=1}&=&(\prod_{\alpha\in R} H(i a\cdot \alpha)a\cdot \alpha \sinh(i \pi a\cdot \alpha))^{\frac{1}{2}}(  \prod_{\alpha\in \bf{R}} \frac{\Gamma(-ia\cdot \alpha)}{\Gamma(1+ia\cdot \alpha)H(i a\cdot \alpha)})^{\frac{1}{2}}\nonumber\\
&=&(\prod_{\alpha\in R} a\cdot \alpha \sinh(\pi a\cdot \alpha))^{\frac{1}{2}}(  \prod_{\alpha\in \bf{R}} \frac{\Gamma(1+ia\cdot \alpha)\Gamma(-ia\cdot \alpha)}{(\Gamma(1+ia\cdot \alpha))^2})^{\frac{1}{2}}\nonumber\\
&=&\left(\prod_{\alpha\in R} \frac{a\cdot \alpha}{\Gamma(1+ia\cdot \alpha)^2}\right)^{\frac{1}{2}}
\eea
The perturbative part of $\mathcal{N}=4$ vector multiplet with Neumann BCs is given by
\bea
Z^{\mathcal{N}=4}_{{\rm Neu, 1-loop}}
&=& \prod_{\alpha\in\Delta_+} H(i a\cdot \alpha)a\cdot \alpha\sinh( \pi a\cdot \alpha)\times \bigg(\prod_{\alpha\in\Delta}\frac{1}{H(i a\cdot \alpha)}\bigg)^{\frac{1}{2}}\nonumber\\
&=&\prod_{\alpha\in\Delta_+}a\cdot \alpha\sinh( \pi a\cdot \alpha)\, .
\eea

\subsubsection{Instanton contribution}

In the computation of the partition function on $S^4$ the non-perturbative contributions arise only from point like instantons and anti-instantons located at the fixed points of the $U(1)$ isometry which is used to localize the path integral, namely the South and North poles of $S^4$. 
If a $\mathbb{Z}_2$-involution acts on $S^4$ with the $S^3$ at the equator as its fixed locus, then, with either supersymmetric Dirichlet or Neumann BCs, the non-perturbative contribution will be just one factor of $Z^k_{inst}\equiv Z^k_{Nekrasov}$. Since $\pi_3(S^3)=\mathbb{Z}$, the instanton sector is characterized by an integer $k$ 
equal the  winding mode at the equator. Let us remind for completeness the expression of the instanton partition function for $\mathcal{N}=2^*$ theory 
\bea
Z^k_{inst,\mathcal {N}=2^*}=q^{|\vec{Y}|}\prod_{\alpha,\beta=1}^N\prod_{s\in Y_{\alpha}}\frac{(E_{\alpha\beta}-\tilde{m})(\epsilon_1+\epsilon_2-E_{\alpha\beta}-\tilde{m})}{E_{\alpha\beta}(\epsilon_1+\epsilon_2-E_{\alpha\beta})}
\eea
where ${Y_{\alpha}},\alpha=1...N$ is the set of Young diagrams $N-$tuples, $|\vec{Y}|=\sum_{\alpha}Y_{\alpha}=k$ the instanton number, the equivariant hypermutiplet mass $\tilde{m}$ is related to the physics mass $m$ as $\tilde{m}=m+\frac{\epsilon_1+\epsilon_2}{2}$ and $E_{\alpha\beta}(s)=(-h_{Y_{\beta}}\epsilon_1+(v_{Y_{\alpha}}+1)\epsilon_2)+a_{\beta}-a_{\alpha}$ with $a_{\alpha},a_{\beta}$ elements of the cartan of the gauge  group.  
Note that for the special value $\tilde{m}=0$ the ratio in $Z^k_{inst,N=2^*}$ cancels to one and therefore the sum over the Young diagrams produces $1/\eta(\tau)$ factors
\cite{Bruzzo:2002xf}. This will turn out to be useful in the comparison with Liouville theory amplitudes. to this end, it is important to underline that 
the insertion of Liouville identity operator on the torus corresponds to $\mathcal{N}=2^*$ theory on $\sfr$, where the mass of the hypermultiplet is $i/r$ \cite{Okuda:2010ke}. In our
nomralization this is the case $\mu=1$. 


\section{Comparison with Liouville theory}

Let us start by briefly reviewing basic relevant facts about AGT correspondence \cite{Alday:2009aq}. The ${\cal N}=2$ $SU(2)$ gauge theory can be formulated on the squashed four sphere
\begin{equation*}
 \sfr_{\epsilon_1,\epsilon_2}\equiv\left\{(x_0,...,x_4)| x_0^2 + \epsilon_1^2(x_1^2 +x_2^2 )+ \epsilon_2^2(x_3^2 +x_4^2 )=1\right\}.
\end{equation*}
and its spectral content can be put in correspondence with 
the geometry of an auxiliary Riemann surface. 
Liouville theory correlators on such a  surface can be put in correspondence with supersymmetric gauge theory partition function.
In particular, the four points correlator on the Riemann sphere corresponds to the partition function of the $\mathcal{N}_f=4$ gauge theory on the squashed four sphere.
\begin{equation}
 Z(\vec{m};\tau;\epsilon_1,\epsilon_2) \propto \langle e^{2\alpha_4\phi(\infty)}e^{2\alpha_3\phi(1)}e^{2\alpha_2\phi(q)}e^{2\alpha_1\phi(0)}\rangle_b,
\end{equation}
where $\vec{m}$ encodes the mass parameters $m_1,...,m_4$, and $b$ is the parameter that appears in the Liouville action
\begin{equation}\label{eq:LiouvilleAction}
 S_b = \frac{1}{4\pi}\int d^2z\left[\left(\partial_a\phi\right)^2 + 4\pi\mu e^{2b\phi}\right].
\end{equation}
The dictionary between the two theories has been tested to be
\begin{equation*}
 b = \sqrt\frac{\epsilon_1}{\epsilon_2},\hspace{0.5cm} q=e^{2\pi i \tau}, \hspace{0.5cm}\alpha_j = (Q/2)+im_j,
\end{equation*}
where $Q=b+b^{-1}$. 
%

The usual correspondence is based on closed Riemann surfaces, while here we discuss the case of open/unoriented Riemann surfaces.
These can be obtained as ${\mathbb Z}_2$ involutions of closed ones. It is then natural to consider gauge theories on ${\mathbb Z}_2$ involutions of
the four sphere. Indeed, AGT correspondence arises from M5-branes compactifications on the product of the four sphere and the Riemann surface itself.
Since the six-dimensional (2,0) theory describing the dynamics of  M5-branes is chiral, the ${\mathbb Z}_2$ involution has to act on both factors simultaneously
\cite{Tachikawa:unpub}.
There are different classes of ${\mathbb Z}_2$ actions, distinguished by their fixed point locus.
We will study the resulting quotients of the original theory in two cases, focusing on the $N_f=4$ and ${\cal N}=2^*$ theories, that is on the possible quotients 
of Riemann surfaces of genus zero and one.

Let us consider the ${\mathbb Z}_2$ quotients of the four punctured Riemann sphere. This can be a two-disk or an $\rpt$. 
In the unorientable case, the ${\mathbb Z}_2$ action is
the antipodal action on the four-punctured two sphere producing the $\rpt$ with two punctures, namely 
\begin{center}
\includegraphics[scale=0.5]{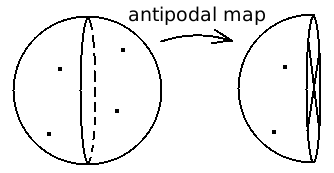}
\end{center}
where we have assumed that the punctures were aligned so as to make the antipodal identification possible.
On the four dimensional factor, the ${\mathbb Z}_2$ quotient is the $\rpf$ geometry.

In the orientable case the fixed locus of the involution is the equator and, depending on the location of the punctures with respect to it, one has three possible configurations:
\begin{itemize}
 \item Two points in the bulk, 
 \begin{center}
 \includegraphics[scale=0.5]{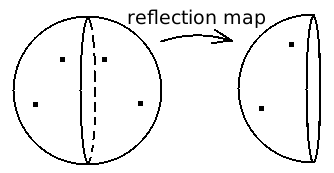}
  \end{center}
 \item Two points on the boundary and one point in the bulk, 
  \begin{center}
 \includegraphics[scale=0.5]{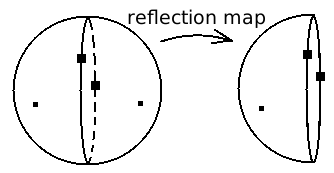}
  \end{center}
 \item Four points on the boundary
  \begin{center}
 \includegraphics[scale=0.5]{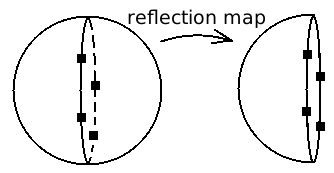}
  \end{center}

\end{itemize}
On the four dimensional factor, the ${\mathbb Z}_2$ quotient is the $HS^4$ geometry.

In the study of Liouville field theory on open surfaces, one has to choose the boundary conditions.
We are allowed a choice of two boundary conditions: $\partial_a \phi =0$ or $\phi \rightarrow \infty$ at the boundary of the disk. These correspond, respectively, to the FZZT \cite{Fateev:2000ik,Teschner:2000md} and the ZZ brane \cite{Zamolodchikov:2001ah}.

Let us first consider the FZZT boundary condition. As seen earlier in Eq.(\ref{eq:LiouvilleAction}),  the bulk Liouville interaction term is given by $\int_\Sigma\sqrt{g}\mu e^{2b\phi}$. In the presence of a boundary, one also has the boundary interaction term $\int_{\partial\Sigma}g^{1/4}\mu_B e^{b\phi}$. The constant $\mu_B$, often called the boundary cosmological constant, is unrestricted and parametrizes the family of conformally invariant boundary conditions.
One has therefore a scale invariant ratio $(\mu_B^2/\mu)$ on which the correlation functions depend. 
It is customary to trade the scale invariant ratio with a parameter $s$ as follows:
\begin{equation*}
\frac{\cosh^2 \pi b s}{\sin \pi b^2}=\frac{\mu_B^2}{\mu}.
\end{equation*}
Therefore the bulk one-point function with FZZT boundary condition will depend on the continuous parameter $s$.
For the ZZ boundary condition the bulk one-point function in this case depends instead on a choice of two positive integers, $m$ and $n$,
as explained in \cite{Nakayama:2004vk}.
As we will show in the following, these two classes of boundary conditions correspond respectively to Dirichlet and Neumann boundary conditions of the gauge theory on the $HS^4$.
In the rest of this section we discuss bulk punctures, the boundary ones being deferred to the subsequent Section 4.

In the following we will provide evidence of the correspondence described so far. Moreover, we will also treat $\mathbb{Z}_2$ involutions of genus one curves.
In this case the complex
double is a torus and the Klein bottle, annulus and  Moebius strip can be obtained by applying
different anti-holomorphic involutions, as show in the following figure
\begin{center}
\includegraphics[scale=0.5]{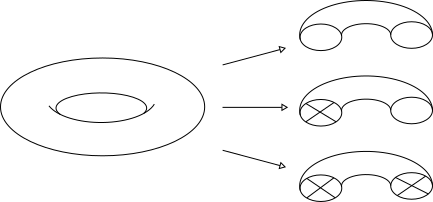} 
\end{center}

\noindent
The conformal families of tori admitting such involutions
are Lagrangian submanifolds in the Teichm\"uller space of the covering torus modded by the translations $\tau\to \tau+1$, with $\left\{\tau \in \mathbb{C}\, | \, {\rm Im}(\tau)>0, \, -\frac{1}{2}\le {\rm Re}(\tau) \le \frac{1}{2} \right\}$. These are vertical straight lines at
${\rm Re}(\tau)=0$
for the annulus and the Klein bottle while at ${\rm Re}(\tau)=\pm \frac{1}{2}$ for the Moebius strip. 
Since the double cover of these internal geometries of the M5 compactification is a torus, we expect that the corresponding gauge theory
to be a $\mathbb{Z}_2$-quotient of an appropriate circular quiver. This is indeed the case as we will show in detail in the following.
For the case of Moebius strip, we will actually follow an equivalent approach, by performing a $\mathbb{Z}_2$-quotient of the annulus amplitudes.
However, keeping in mind the torus double covering is useful to discuss the instanton sector of the corresponding gauge theory. In particular
this makes transparent the arising of a $\theta=\pi$ topological term in the four dimensional gauge theory action in this case.

\subsection{$\rpt$ with two punctures vs. $N_f=4$ on $\rpf$}

The Liouville two point correlator on $\rpt$ can be conveniently expressed in terms of the OPE as in 
the following diagram:
\begin{center}
\includegraphics[scale=0.5]{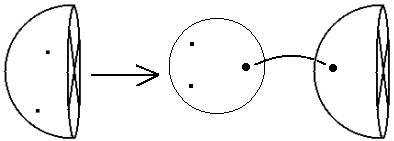}
\end{center}
The right hand side of the figure corresponds to the expression (up to the conformal block which is not relevant for the present discussion):
\begin{equation}\label{ciccio}
 C\left(\frac{Q}{2} + im_1,\frac{Q}{2} + im_2,\frac{Q}{2} -ia\right)\times \langle V_{Q/2 + ia} \rangle_{\rpt}\, ,
\end{equation}
where $C$ is the DOZZ three-point function and
the second factor is the crosscap wavefunction $\Psi_C(a)$ as in \cite{Hikida:2002bt}.

Let us unpack the product in Eq.(\ref{ciccio}) as
\bea\label{eq:ThreepointCrosscap}
&&C\left(\frac{Q}{2} + im_1,\frac{Q}{2} + im_2,\frac{Q}{2} -ia
\right)\Psi_C(a)= 
\prod_{\pm\pm\pm}\Gm_2\left(\pm ia \pm im_1 \pm im_2 +
\frac{Q}{2}\right)\times \nonumber\\
&\times&
\left(\Ups(2ia)\frac{\Gm(1 + 2iab)\Gm(1 + 2iab^{-1})\cosh(\pi ab)\cosh(\pi a b^{-1})}{ia} \right)\nonumber\\
&=_{b=1}&\prod_{\pm\pm\pm}\Gm_2\left(\pm ia \pm im_1 \pm im_2 +
1\right) \left(\Ups(2ia)\frac{\Gm(1 + 2ia)^2\cosh(\pi a
)^2}{ia} \right)\nonumber\\
\eea
The last expression makes precise contact with the one-loop partition function of $N_f=4$ on $\rpf$ as computed in Section 2, formula \ref{Znf4}.
In particular,
the first factor on the right corresponds to the hypermultiplet contribution and the second factor to one loop contribution of the vector multiplet in the 
trivial holonomy sector (namely, the projection on even modes).

\subsection{Disk with two bulk punctures vs. $N_f=4$ on $HS^4$}

As before, the Liouville amplitude is decomposed as:
\begin{center}
\includegraphics[scale=0.5]{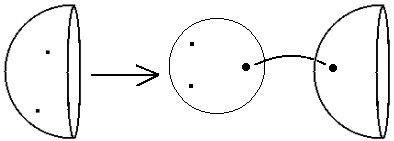}
\end{center}
giving
\begin{equation}\label{eq:DiskProduct}
 C\left(\frac{Q}{2} + im_1,\frac{Q}{2} + im_2,\frac{Q}{2} -ia\right)\times \langle V_{Q/2 + ia} \rangle_\mathrm{disk}.
\end{equation}
Before we can evaluate the second factor on the right above, we have to choose the boundary condition for the Liouville field on the disk, namely either FZZT or ZZ.

In the FZZT case, the disk one point function turns out to be 
\begin{equation}\label{eq:FZZT1}
  \langle V_{Q/2 + ia} \rangle_{\mathrm{FZZT},s}\simeq \Psi_s(a) \simeq  \frac{\Gm(1 + 2iba)\Gm(1 +
2iab^{-1})\cos(2\pi s a)}{-2i\pi a}.
\end{equation}
By writing the above product as
\begin{equation}\label{eq:FZZT2}
\begin{split}
 &\frac{1}{\prod_{\pm\pm\pm}\Gm_2\left(\pm ia \pm im_1 \pm im_2 +
\frac{Q}{2}\right)} \\ &\times \left(\Ups(2ia)\frac{\Gm(1 + 2iba)\Gm(1 +
2iab^{-1})\cos(2\pi s a)}{-2i\pi a}\right)
\end{split}
\end{equation}
one can easily find the dictionary with gauge theory. Indeed, for $s=0$, the above expression coincides with the one-loop partition function of $N_f=4$ on $HS^4$
with Dirichlet boundary conditions, which can be obtained from \ref{Zdir} and \ref{Zhyp}. The case $s=pb+rb^{-1}$ is related to the expectation value of a corresponding Wilson loop winding $(p,r)$ times the two circles at the
$S^3$ equator.

Now we turn to the ZZ boundary condition. Rather than a continuous parameter as in the previous case, the bulk one-point function now depends on two positive integers.
When both of them are equal to one 
we have:
\begin{equation}
  \langle V_{Q/2 + ia} \rangle_{\mathrm{ZZ},(1,1)}\simeq \Psi_{1,1}(a) \simeq  \frac{2i\pi a}{\Gm(1 - 2iba)\Gm(1 -2iab^{-1})}.
\end{equation}
The full Liouville theory amplitude reads 
\begin{equation}
  C\left(\frac{Q}{2} + im_1,\frac{Q}{2} + im_2,\frac{Q}{2} -ia\right)\times  \Psi_{1,1}(a).
\end{equation}
and is explicity given by
\begin{equation}\label{eq:ZZ2}
\begin{split}
C\left(\frac{Q}{2} + im_1,\frac{Q}{2} + im_2,\frac{Q}{2} -ia\right)\times  \Psi_{1,1}(a)\simeq \\
\prod_{\pm\pm\pm}\Gm_2\left(\pm ia \pm im_1 \pm im_2 +
\frac{Q}{2}\right) \left(\Ups(2ia)\frac{2i\pi a}{\Gm(1 - 2iba)\Gm(1 -2iab^{-1})}\right)
\end{split}
\end{equation}
This expression 
corresponds to the one-loop partition function of of $N_f=4$ on $HS^4$ with Neumann boundary conditions, see \ref{Zvec} and \ref{Zhypneu}.
Similarly to the FZZT case, for higher values of the integer parameters in the Liouville amplitude, one finds the corresponding Wilson loop expectation values.

\subsection{Klein bottle vs. quotiented circular quiver on $\rpf$ }

Let us now turn to the amplitudes obtained from quotients of the torus. We decompose along an intermediate channel according to the
following pictures
\begin{center}
\vspace{.5cm}
\includegraphics[scale=0.3]{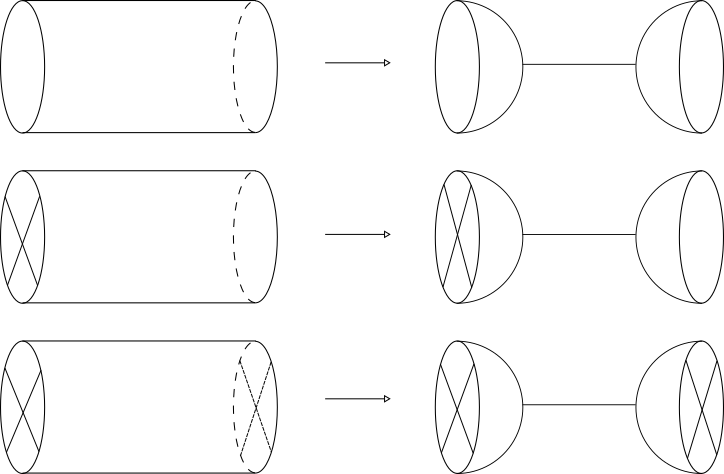} 
\vspace{.5cm}
\end{center}

\noindent
The Liouville amplitude for for Klein bottle ${\mathcal K}_2$ is given by \cite{Nakayama:2004vk}:
\begin{equation}
 Z_{\mathcal K} = \int_{-\infty}^{\infty} dP  \Psi_C(P) \Psi_C(-P)\frac{q^{P^2}}{\eta(\tau)}
\end{equation}
where
\begin{equation}
 \Psi_C(P) = \pmb{\mu}^{-iPb^{-1}} 2^{1/4}\frac{\Gm(1+2iPb)\Gm(1 + 2iPb^{-1})}{2\pi i P}\cosh(\pi Pb)\cosh(\pi P b^{-1})
\end{equation}
and $q = e^{-2\pi\tau}$. We have defined $\pmb{\mu} = \pi\mu\gamma(b^2)$ and we will use this throughout. Simplifying this expression we get
\begin{equation}\label{klein}
  Z_{{\mathcal K}_2}  = \frac{1}{2\sqrt{2}}\int_{-\infty}^{\infty} dP \frac{\cosh(\pi Pb)\cosh(\pi P b^{-1})}{\sinh(\pi Pb)\sinh(\pi P b^{-1})}\frac{q^{P^2}}{\eta(q)}
\end{equation}
where the identity
\begin{equation}
 \Gamma(1+x)\Gamma(1-x) = \frac{\pi x}{\sin(\pi x)}
\end{equation}
has been used. As discussed at the beginning of this Section, 
the  corresponding gauge theory turns is a circular quiver with two $SU(2)$ gauge nodes and two bifundamental hypermultiplets. The $\mathbb{Z}_2$ action  which defines 
$\mathbb{RP}^4$ acts on the gauge group $SU(2)\times SU(2)$ as an automorphism. Under this automorphism both the two $SU(2)$ nodes and bifiundamentals are swapped and one resulting in an $\mathcal{N}=2^*$ $SU(2)$ gauge theory. The resulting modes combine to give the correct expression for the Klein bottle partition function, as we will now explain.
 The Dynkin diagram associated to $SU(2)\times SU(2)$ is of type $D_2\simeq A_1\times A_1$. It is represented by two disconnected nodes, with each node corresponding to a simple root and moreover these two simple roots are orthogonal. The automorphism symmetry corresponds to switching the two nodes and is implemented by conjugating with a $O(4)$ matrix with determinant equal to $-1$. Since the nodes of our root system are disconnected, it can be consistently quotiented by this automorphism symmetry to yield the invariant part of it and we get
 \bea
 &D_2\to B_1,\nonumber\\
 &\quad SU(2)\times SU(2)\to SO(3)
 \eea

\begin{tikzpicture}
\node[gauge] (B) at (0,0) {SU(2)};
\node[gauge] (C) at (6,0) {SU(2)};
\draw[thick] (0.4,0.44)--(5.6,0.44) ;
\draw[thick] (0.4,-0.44)--(5.6,-0.44) ;
\node at (2.5,0.7) {bi-fundamental} ;
\node at (2.5,-0.7) {bi-fundamental} ;
\node at (3, -2) {$\Bigg\downarrow
 \mathbb{Z}_2$ projection} ;
\node[gauge] (B) at (2,-4) {SU(2)};
\draw[thick, ->] (2.5,-3.6) arc (150:-150:0.8) ;
\node at (5.5,-3.9) {adjoint-hyper} ;
\end{tikzpicture}\\

Consequently the one-loop part, by using the results of Section 2, is given by
\bea\label{eq:KB}
Z_{1-loop}&=&\frac{\Upsilon(-2 i a)\Gamma(1-2 i a)^2 \cosh(\pi a)^2}{ a H(1+2 i a)}\times \frac{\Upsilon(2 i a)\Gamma(1+2 i a)^2 \cosh(\pi a)^2}{ a H(1-2 i a)}\nonumber\\
&=&\frac{4\pi^2 a^2\cosh(\pi a)^4}{a^2\sinh(2\pi a)^2}\nonumber\\
&=&\frac{\pi^2\cosh(\pi a)^2}{\sinh(\pi a)^2}
\eea
where we used the identities $H(1-2 i a)=\Upsilon(-2i a)$. We remark that the above is the one-loop contribution of $\mathcal{N}=2^*$ theory with $\mu=1$, which corresponds
to the insertion of the identity operator in Liouville theory \cite{Okuda:2010ke}, thus perfectly matching our expectations.

\subsection*{Annulus vs. $\mathbb{Z}_2$-quotient of $SU(2)\times SU(2)$ circular quiver on $HS^4$}

Let us start by considering an annulus with full FZZT boundary conditions parametrised by $s_1$ and $s_2$ :
\begin{equation}
 Z_{s_1s_2} = \int_{-\infty}^{\infty} dP  \Psi_{s_1}(P) \Psi_{s_2}(-P)\frac{q^{P^2}}{\eta(\tau)}
\end{equation}
where $q = e^{2\pi i\tau}$ and
\begin{equation}
 \Psi_s(P) = 2^{-1/4}\pmb\mu^{-iPb^{-1}} \frac{\Gm(1+2iPb)\Gm(1 + 2iPb^{-1})}{-2\pi i P}\cos(2\pi s P).
\end{equation}
On simplifying:
\begin{equation}
 Z_{s_1s_2} = \frac{1}{\sqrt{2}}\int_{-\infty}^{\infty} dP \frac{\cos(2\pi s_1 P) \cos(2\pi s_2 P)}{\sinh(2b\pi P)\sinh(2b^{-1}\pi P)} \frac{q^{P^2}}{\eta(\tau)}.
\end{equation}
As we learnt in the disk case, the relevant boundary conditions on the gauge theory side are the Dirichlet ones.
 
Let us discuss first the case $s_1=s_2=0$. The $\mathbb{Z}_2$ quotient of the circular quiver gauge theory corresponding to FZZT boundary conditions on the two sides of the annulus amounts 
to impose SUSY Dirichlet boundary conditions on the two $SU(2)$ gauge nodes and on the two bifundamental hypermultiplets. Moreover, the two Coulomb branch parameters $a_1,a_2$ have to be identified under the $\mathbb{Z}_2$ action. The Dirichlet boundary conditions on the circular quiver gauge theory is imposed in the following way:\\
a) for one $SU(2)$ node the Dirichlet BCs are imposed on the positive roots, whereas for the other $SU(2)$ the boundary conditions are imposed on the negative roots. \\
b) for $SU(2)$ gauge group at the two nodes, each of the two bifundamentals transforms in the $(2,2)$ representation, so that imposing Dirichlet BCs on the bifundamental will result in a 'half' bifundamental.\\
c) identifying the two Coulomb branch parameters $a_1=a_2=a$ under the $\mathbb{Z}_2$ action  will imply that the two half nodes and two half bifundamentals combine to give $N=2^*$ $SU(2)$ theory with $\mu=1$.\\ 
The resulting one-loop partition function is therefore
\bea
Z^{1-loop}&=&\prod_{\alpha\in\Delta}\frac{H(i a\cdot \alpha)}{(H(i a\cdot \alpha)H(i a\cdot \alpha))^{\frac{1}{2}}}\frac{a\cdot\alpha}{(\sinh\pi(a\cdot \alpha))}\nonumber\\
&=&\prod_{\alpha\in\Delta}\frac{1}{(\sinh\pi(a\cdot \alpha))}\nonumber\\
&=&\frac{1}{(\sinh(2\pi a))^2} \, .
\eea
For the case of Dirichlet boundary conditions there are no overall integrals and hence no Vandermonde factor.
The general result for $s_i=m_ib+n_ib^{-1} \, , i=1,2$ is obtained by inserting Wilson loops in the supersymmetric path integral.

In the case of full ZZ boundary conditions the annulus Liouville amplitude is 
\begin{align}\label{benign}
   Z_{(m_1n_1),(m_2n_2)} &= \int_{-\infty}^{\infty} dP  \Psi_{(m_1n_1)}(P) \Psi_{(m_2n_2)}(-P)\frac{q^{P^2}}{\eta(\tau)} \nonumber\\
 &=2\sqrt{2} \int_{-\infty}^{\infty} dP \Bigg[ \sinh(2\pi bP)\sinh(2\pi b^{-1}P)\frac{q^{P^2}}{\eta(\tau)}\nonumber\\
&\times\left(\frac{\sinh(2\pi m_1 Pb^{-1})\sinh(2\pi m_2 Pb^{-1})\sinh(2\pi n_1 Pb)\sinh(2\pi n_2 Pb)}{\sinh^2(2\pi Pb^{-1})\sinh^2(2\pi Pb)}\right)\Bigg]
\end{align}
where
\begin{equation}
 \Psi_{(m,n)}(P) = \Psi_{(1,1)}(P)\frac{\sinh(2\pi m Pb^{-1})\sinh(2\pi n Pb)}{\sinh(2\pi Pb^{-1})\sinh(2\pi Pb)}
\end{equation}
and 
\begin{equation}
 \Psi_{(1,1)}(P) = 2^{3/4}(\pi\mu\gamma(b^2))^{-iPb^{-1}} \frac{2\pi i P}{\Gm(1-2iPb)\Gm(1 - 2iPb^{-1})}.
\end{equation}
This correlator for $m_1=m_2=n_1=n_2=1$ corresponds to a $\mathbb{Z}_2$ quotient of the circular quiver with Neumann BCs imposed in the following way:\\
a) for one $SU(2)$ node the Neumann BCs are imposed on the positive roots, whereas for the other $SU(2)$ the boundary conditions are imposed on the negative roots \\
b) for each of the $SU(2)$ nodes, the  two bifundamentals transform in the $(2,2)$ representation. Therefore, imposing Neumann BCs results in a 'half' bifundamental.\\
c) identifying the two Coulomb branch parameters $a_1=a_2=a$ under the $\mathbb{Z}_2$ action implies that the two 'half' nodes and two 'half' bifundamentals combine to give ${\cal N}=2^*$ $SU(2)$ theory with $\mu=1$.\\ 
This results in the following expression for the one-loop partition function
\bea
Z^{1-loop}&=&\prod_{\alpha\in\Delta}\frac{H(i a\cdot \alpha)}{(H(i a\cdot \alpha-1)H(i a\cdot \alpha+1))^{\frac{1}{2}}}\frac{(\sinh\pi(a\cdot \alpha))}{(a\cdot \alpha)}\nonumber\\
&=&\prod_{\alpha\in\Delta}\frac{1}{a\cdot \alpha}\times\frac{(\sinh\pi(a\cdot \alpha))}{(a\cdot \alpha)}\nonumber\\
&=&\frac{(\sinh(2\pi a))^2}{4 a^4}\, .
\eea
This accounts correctly for the Vandermonde factor in the integration measure associated to the two $SU(2)$ nodes with Neumann BCs, which gives a factor of $a^4$. The formula for general values of the integer parameters should correspond to the vev
of Wilson loops analogously to what discussed in previous cases.

A remark is in order about S-duality properties of the amplitude \eqref{benign}. Indeed by expressing this in terms of the modular-transformed parameter $\tau' = -1/\tau$ one obtains
a finite sum of characters of degenerate representations of the Virasoro algebra:
\begin{equation}
 Z_{(m\, n),(m'\, n')} = \sum_{k=0}^{\min(m,m')-1}\sum_{l=0}^{\min(n,n')-1}\chi_{m+m'-2k-1,n+n'-2l-1}(\tau')
\end{equation}
where
\begin{equation}
 \chi_{m,n}(\tau) = \frac{q^{-{mb^{-1}+nb}^2/4}-q^{-{mb^{-1}-nb}^2/4}}{\eta(q)}
\end{equation}
The above formulae claim for a simple gauge theory interpretation of this dual phase which it would be interesting to explore further.
%

In the mixed case FZZT/ZZ we finally have the following Liouville amplitude
\begin{align}
   Z_{s,(m,n)} &= \int_{-\infty}^{\infty} dP  \Psi_{s}(P) \Psi_{(m,n)}(-P)\frac{q^{P^2}}{\eta(\tau)} \nonumber\\
 &=\sqrt{2} \int_{-\infty}^{\infty} dP \cos(2\pi s P)\left(\frac{\sinh(2\pi m Pb^{-1} )\sinh(2\pi n Pb)}{\sinh(2\pi Pb^{-1})\sinh(2\pi Pb)}\right)\frac{q^{P^2}}{\eta(\tau)}\nonumber\\
\end{align}
Similarly to the previous two cases, the gauge theory counterpart of this amplitude is obtained by taking the $\mathbb{Z}_2$ quotient of circular quiver in the following way:\\
a) for one $SU(2)$ node the Neumann BCs are imposed on the positive roots, whereas for the other $SU(2)$ the Dirichlet boundary conditions are imposed on the negative roots. \\
b) on each bifundamental we have to impose Dirichlet BCs at one node and Neumann at the other. Since the representation $(2,2)$ is symmetric with respect to the exchange of the two nodes, it does not matter at which node we impose Dirichlet and on which node Neumann BCs. This set of BCs again gives two 'half' bifundamentals.\\
c)  identifying the two Coulomb branch parameters $a_1=a_2=a$ under the $\mathbb{Z}_2$ action implies that the two 'half' nodes and two  'half' bifundamentals combine to give ${\cal N}=2^*$ $SU(2)$ theory with $\mu=1$.\\ 
This results in the following expression one the one-loop partition function 
\bea
Z^{1-loop}&=&\prod_{\alpha\in\Delta}\frac{H(i a\cdot \alpha)}{(H(i a\cdot \alpha+\mu)H(i a\cdot \alpha-\mu))^{\frac{1}{2}}}\nonumber\\
&=& \prod_{\alpha\in\Delta}\frac{H(i a\cdot \alpha)}{(H(i a\cdot \alpha+1)H(i a\cdot \alpha-1))^{\frac{1}{2}}}\quad for\quad \mu=1\nonumber\\
&=&\prod_{\alpha\in\Delta} \frac{1}{i a\cdot \alpha}\nonumber\\
&=&\frac{1}{4 a^2}\quad
\eea
where we have used the identity $H(x+1)H(x-1)=H(x)^2 x^2$ in the last line.
Notice that the factor $\frac{1}{a^2}$ cancels with the Vandermonde determinant.\\

\subsection{The Moebius strip cases vs circular quiver on $HS^4/\mathbb{Z}_2$}

In this subsection we consider the Moebius strip by realising it as a $\mathbb{Z}_2$-quotient of the annulus, or, equivalently, as a $\mathbb{Z}_2\times \mathbb{Z}_2$
quotient of the circular quiver on $S^4$. More precisely,
we consider a circular quiver consisting of gauge group $SU(2)\times SU(2)$ with edges representing bifundamental hypermultiplets on $S^4$, and apply the following sequence of 
$\mathbb{Z}_2$ quotients.
The first $\mathbb{Z}_2$-quotient leads to $HS^4$ and corresponds to imposing supersymmetric boundary conditions, either Dirichlet or Neumann, on the matter content of the gauge theory. This $\mathbb{Z}_2$ action will generate two half-vector multiplets coupled through two 'half'-bifundamentals.
On top of the previous $\mathbb{Z}_2$ action, we apply orientation reversing or antipodal identification 
\begin{equation*}
 \rho\rightarrow \pi-\rho,\hspace{0.25cm}\theta\rightarrow\theta,\hspace{0.25cm}\psi\rightarrow\psi+2\pi,\hspace{0.25cm}\phi\rightarrow\phi.
\end{equation*}
Note the important fact that the killing spinor 
\bea
\xi=\left(
\begin{array}{cc}
 \frac{\cos \left(\frac{\rho}{2}\right)}{\sqrt{2}} & 0 \\
 0 & \frac{\cos \left(\frac{\rho}{2}\right)}{\sqrt{2}} \\
 \frac{i \sin \left(\frac{\rho}{2}\right)}{\sqrt{2}} & 0 \\
 0 & -\frac{i \sin \left(\frac{\rho}{2}\right)}{\sqrt{2}} \\
\end{array}
\right)
\eea
which is used to perform localization does not depend on $\psi,\theta,\phi$ and the action on  $\rho$  is the same as in the previous $\mathbb{Z}_2$ action. Therefore the second $\mathbb{Z}_2$  quotient  does not break supersymmetry and we can consistently apply it. As usual by now, the different FZZT/ZZ boundary conditions correspond to Dirichlet/Neumann boundary conditions
on the $\mathbb{RP}^3$ boundary.

For FZZT the Liouville amplitude is 
\begin{align*}
 Z^M_{s} &= \int_{-\infty}^{\infty}dP\Psi_C(P)\Psi_s(-P)\frac{q^{P^2}}{\eta(i\tau_c + \tfrac{1}{2})}\\
 &= -\int_{-\infty}^{\infty}dP\frac{\cos 2\pi s P}{4\sinh \pi bP\sinh \pi b^{-1}P}\frac{q^{P^2}}{\eta(i\tau_c + \tfrac{1}{2})}
\end{align*}
Acting with first $\mathbb{Z}_2$ on the circular quiver theory implies imposing Dirichlet boundary conditions on the two $SU(2)$ nodes and the adjoint hyper. As already stated before, the first 
$\mathbb{Z}_2$ action reduces round $S^4$ to a Hemi-$S^4$. 
The antipodal identification acts only on the boundary of the Hemi $-S^4$, and, as an important consequence, it acts only on the field modes at the boundary $\rho=\frac{\pi}{2}$. Since hypermultiplets have no boundary contribution, the antipodal indetification does not act on them. The final expression we get is therefore
\bea
Z_{half}^{1-loop}&=&\frac{ \prod_{\alpha\in\Delta}\prod_{n\ge 1}(n+i a.\alpha)^{n}}{( \prod_{\alpha\in\Delta}\prod_{n\ge1}(n+1+i a.\alpha)^{n}(-n+1+i a.\alpha)^{n})^{\frac{1}{2}}}\times\frac{1}{ \prod_{\alpha\in\Delta}\prod_{n\ge 1}(2n+i a.\alpha)}\nonumber\\
&=&\frac{\prod_{\alpha\in\Delta}G(1+ia.\alpha)}{(\prod_{\alpha\in\Delta}G(2+ia.\alpha)G(-ia.\alpha))^{\frac{1}{2}}}\Gamma(1+\frac{ia.\alpha}{2})
\nonumber\\
&=&\frac{\prod_{\alpha\in\Delta}G(1+ia.\alpha)}{(\prod_{\alpha\in\Delta}G(1+ia.\alpha)G(1-ia.\alpha))^{\frac{1}{2}}}(\frac{\Gamma(-ia.\alpha)}{\Gamma(1+ia.\alpha)})^{\frac{1}{2}}\Gamma(1+\frac{ia.\alpha}{2})\nonumber\\
&=&\prod_{\alpha}(\frac{\Gamma(-ia.\alpha)}{\Gamma(1+ia.\alpha)})^{\frac{1}{2}}
\Gamma(1+\frac{ia.\alpha}{2})
\eea
Combining it with the other half we will get
\bea
Z^{1-loop}&=&\prod_{\alpha}(\frac{\Gamma(-ia.\alpha)}{\Gamma(1+ia.\alpha)})^{\frac{1}{2}}
\Gamma(1+\frac{ia.\alpha}{2})
\prod_{\alpha}(\frac{\Gamma(ia.\alpha)}{\Gamma(1-ia.\alpha)})^{\frac{1}{2}}
\Gamma(1-\frac{ia.\alpha}{2})\nonumber\\
&=&\frac{1}{\sinh(\pi a)^2}
\eea

For ZZ boundary condition the Liouville amplitude reads
\begin{align*}
 Z^M_{m,n} &= \int_{-\infty}^{\infty}dP\Psi_C(P)\Psi_{m,n}(-P)\frac{q^{P^2}}{\eta(i\tau_c + \tfrac{1}{2})}\\
 &=-2\int_{-\infty}^{\infty}dP\cosh{\pi P b}\cosh{\pi P b^{-1}}\frac{\sinh{2\pi nP b}\sinh{2\pi mP b^{-1}}}{\sinh{2\pi P b}\sinh{2\pi P b^{-1}}}\frac{q^{P^2}}{\eta\left(i\tau_c + \tfrac{1}{2}\right)}\\
 &=-\frac{1}{2}\int_{-\infty}^{\infty}dP\frac{\sinh{2\pi nP b}\sinh{2\pi mP b^{-1}}}{\sinh{\pi P b}\sinh{\pi P b^{-1}}}\frac{q^{P^2}}{\eta\left(i\tau_c + \tfrac{1}{2}\right)}.
\end{align*}
Considerations similar to the FZZT case takes us to the following expression for Neumann boundary conditions
\bea
Z_{half}^{1-loop}&=&\frac{ \prod_{\alpha\in\Delta}\prod_{n\ge 1}(n+i a.\alpha)^{n}}{( \prod_{\alpha\in\Delta}\prod_{n\ge1}(n+1+i a.\alpha)^{n}(-n+1+i a.\alpha)^{n})^{\frac{1}{2}}}\times \prod_{\alpha\in\Delta}\prod_{n\ge 1}(2n-1+i a.\alpha)\nonumber\\
&=&\frac{\prod_{\alpha\in\Delta}G(1+ia.\alpha)}{(\prod_{\alpha\in\Delta}G(2+ia.\alpha)G(-ia.\alpha))^{\frac{1}{2}}}\frac{1}{\Gamma(\frac{1}{2}+\frac{ia.\alpha}{2})}
\nonumber\\
&=&\frac{\prod_{\alpha\in\Delta}G(1+ia.\alpha)}{(\prod_{\alpha\in\Delta}G(1+ia.\alpha)G(1-ia.\alpha))^{\frac{1}{2}}}(\frac{\Gamma(-ia.\alpha)}{\Gamma(1+ia.\alpha)})^{\frac{1}{2}}\frac{1}{\Gamma(\frac{1}{2}+\frac{ia.\alpha}{2})}\nonumber\\
&=&\prod_{\alpha}(\frac{\Gamma(-ia.\alpha)}{\Gamma(1+ia.\alpha)})^{\frac{1}{2}}
\frac{1}{\Gamma(\frac{1}{2}+\frac{ia.\alpha}{2})}
\eea
Combining it with the other half we will get
\bea
Z^{1-loop}&=&\prod_{\alpha}(\frac{\Gamma(-ia.\alpha)}{\Gamma(1+ia.\alpha)})^{\frac{1}{2}}
\frac{1}{\Gamma(\frac{1}{2}+\frac{ia.\alpha}{2})}
\prod_{\alpha}(\frac{\Gamma(ia.\alpha)}{\Gamma(1-ia.\alpha)})^{\frac{1}{2}}
\frac{1}{\Gamma(\frac{1}{2}+\frac{-ia.\alpha}{2})}\nonumber\\
&=&\frac{\cosh(\pi a)^2}{a^2}
\eea
which matches with $Z_{1,1}^M$. FZZT and ZZ amplitudes with more general values of the boundary parameters can be obtained as usual via Wilson loop insertions.

Let us now briefly discuss the instanton contribution for the comparison with the Moebius strip amplitudes.
In this case the gauge theory coupling constant  is $\tau=\frac{4\pi i}{g^2}+\frac{1}{2}=i\tau_c + \frac{1}{2}$, where the $\frac{1}{2}$ corresponds to turn on a half-integer Chern-Simons term on the three-sphere fixed under the involution. 
The simplest way to understand this relation is by noting that the Moebius strip can also be realized as a $\mathbb{Z}_2$ quotient of a torus, as we remarked at the beginning of this Section.
More precisely, the Moebius strip can be obtained by acting on a torus with the above complex modulus by the following involutions
\bea
z\to 1-\bar{z}+i\tau_2,\nonumber\\
\quad z\to -\bar{z},\quad z\to 2-\bar{z}
\eea
The second set of involutions has a fixed point set which defines the boundary of the Moebius strip. On the gauge theory side, this has the interesting interpretation of turning on a  
$\theta$-term with $\theta=\pi$. This can also be interpreted as a contribution from a Chern Simons term on the $\mathbb{RP}^3$ boundary.

\section{Coupled 3d/4d Gauge theories and boundary Liouville insertions}
In this section we consider Liouville theory on Riemann surfaces with punctures on the boundary. The two building blocks for all the possible amplitudes are given by 
\begin{itemize}
\item the disk with a boundary and bulk puncture.
\item the disk with $3$ boundary punctures.
\end{itemize}
Both of these cases can be obtained taking a $\IZ_2$ quotient of the $2$-sphere with $3$ punctures. In the first case one starts from the $2$-sphere with one puncture on the equator.
\begin{center}
\includegraphics[scale=0.8]{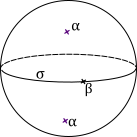} 
\end{center}
In the second case from the $2$-sphere with all three punctures on the equator.
\begin{center}
\includegraphics[scale=0.8]{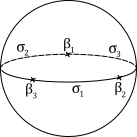} 
\end{center}
The $2$-sphere with $3$ punctures is AGT dual of the $4d$ $\CN=2$ theory of the $SU(2)^3$ tri-fundamental, that is $4$ free massless hypermultiplets.

The amplitude for these two building blocks are well known. In this section we interpret them as $3d$ boundary conditions for the theory of $4$ free massless hypermultiplets, following \cite{Dimofte:2012pd} (see also \cite{Dimofte:2011ju, Dimofte:2011py, Dimofte:2013lba}). It turns out that the original $SU(2)$ bulk symmetries are broken to $U(1)$ if the puncture is on the equator fixed by the $\IZ_2$, while if the puncture is in the bulk of the Riemann surface the global symmetry is still $SU(2)$. There are also purely $3d$ symmetries, whose fugacities are the cosmological constants $\s$ living on the boundary segments. Such purely $3d$ symmetries seems to always be  $SU(2)$, sometimes it's an apparently $U(1)_\s$ non-trivially enhanced to $SU(2)_\s$.

We then write the amplitudes for the boundary four point function and the "two boundary - one bulk" three point function. These can be thought of as the $\IZ_2$ quotient of the sphere with $4$ punctures, that is $\CN=2$ $SU(2)$ SQCD with $N_f=4$, so they represent boundary conditions for interacting $4d$ theories. These amplitude are written in terms of the two basic amplitudes, but they also contain instanton factors.

\subsection{"one bulk - one boundary" two point function: $\CN=4$ $U(1)$ with $2$ flavors coupled to $3$ hypers}
The Liouville amplitude for the disk with one bulk puncture (with fugacity $\a$) and one boundary puncture (with fugacity $\b$) 
\be
\langle \Psi_\beta(1) V_\alpha(0)\rangle^{\rm FZZT}_{{\mathcal D}, \s}=\mathcal{Z}[\mathcal{B}_{TSU(2)}](\beta,\alpha;\s)
\ee
can be taken from eq 5.39 of \cite{Nakayama:2004vk} and gives us the hemisphere partition function $\mathcal{Z}$ for the $3d-4d$ system:
\be \label{BB} \mathcal{Z}[\mathcal{B}_{TSU(2)}]=\frac{\Gb(Q/2 \pm 2 \at -\bt)\Gb(Q/2-\bt)^2}{\Gb(Q)\Gb(-2\bt)\Gb(\at)\Gb(Q-\at)} 
S_b(\frac{Q}{2}-\bt) \int_{-i\infty}^{+i\infty} \!\!\!e^{2 \pi \sigma x} S_b\!\left(\frac{Q}{4}\!\pm\!x\!\pm\!\at\!+\!\frac{\bt}{2}\right)\!dx \ee
where $F(x \pm y) =F(x+y)F(x-y)$ and we redefined $\at = -Q/2+\alpha$, $\bt = -Q/2+\b$. We applied the definition of $S_b(x)$
\be \frac{\Gamma_b(x)}{\Gamma_b(Q-x)}=S_b(x) \ee
One can interpret \ref{BB} as the partition function of a coupled $3d-4d$ system living on a half-$S^4$. $\Gb$ is the contribution of a half-hyper in the $4d$ bulk. $S_b(\frac{Q}{2}r+iy)$ is the contribution of $3d$ chiral multiplet living on the $S^3$ at the boundary, with R-charge $r$ and global-symmetry fugacity $y$.

Looking at \ref{BB}, it is easy to see that on the boundary there is $U(1)$ gauge theory with $4$ charged fields. $x$ is the fugacity for a $3d$ $U(1)$ gauge symmetry, with FI parameter $\sigma$, which is the fugacity of the topological symmetry.

The four $4d$ $\Gb$ fields are in the numerator of the prefactor in \ref{BB}:
\begin{itemize}
\item $\Gb(Q/2 \pm 2 \at -\bt)\Gb(Q-\bt)$ is a $4d$ half-hyper, $SU(2)_\at$-triplet, $X^{4d}_I$
\item $\Gb(Q/2-\bt)$ is a $4d$ half-hyper, $SU(2)_\at$-singlet, $Y^{4d}$
\item $S_b(Q/4 + x \pm \at +\bt/2)$ are $3d$ flavors $p_i$ of gauge charge $+1$
\item $S_b(Q/4 - x \mp \at +\bt/2)$ are $3d$ flavors $\pt_i$ of gauge charge $-1$
\item $S_b(Q/2-\bt)$ is a $3d$ gauge singlet $\Sigma^{3d}$
\end{itemize}

\begin{figure}\centering
\begin{tikzpicture}
\node[gauge] (B) at (0,0) {$U(1)^{3d}_x$};
\node[flavor] (C) at (4,0) {$SU(2)^{4d}_{\at}$};
\node[flavor] (D) at (7.6,0) {$U(1)^{4d}_\b$};
\draw[thick, ->] (B)--(C) ;
\draw[thick, ->] (C)--(B) ;
\draw[thick] (6.3,0)--(6.9,0) ;
\draw[thick, ->] (4.8,0.4) arc (150:-150:0.8) ;
\draw[thick, ->] (-0.6,0.4) arc (30:330:0.6) ;
\node at (-2,0) {$\Sigma^{3d}$} ;
\node at (2,-0.5) {$p_i$} ;
\node at (2, 0.5) {$\pt_i$} ;
\node at (6.2,0.9) {$X^{4d}$} ;
\end{tikzpicture}\caption{Quiver depiction of the $3d-4d$ system for the case of the "one bulk - one boundary" Liouville two point function.}
\end{figure}
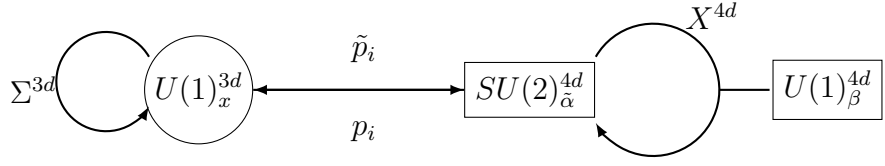

We only see $4$ of the $8$ half-hypers because this is a hemi-sphere partition function instead of a sphere partition function.\footnote{The $4d$ fields  are the field content of $SU(2)$ $\N=4$ SYM on a half-$S^4$, that is Liouville theory on a torus with one puncture, modded out by $\IZ_2$. On the full $S^4$ we would see $8$ half hypers, an the integral over the Cartan of $SU(2)_\at$ and the symmetry with fugacity $\beta$ would be $SU(2)$ instead of $U(1)$. We expect the $3d$ gauge theory to be related to the S-duality wall gauge theory of $4d$ $\CN=4$ SYM, that is $TSU(N)$, with $N=2$.}

In order to describe completely the boundary condition, we need to write the superpotential. We consider the most general gauge invariant terms which are also uncharged under the $\at$, $\bt$ and $\s$ fugacities, and also impose $SU(2)_\a$ non-Abelian global symmetry. This is given by
\be \label{WBB} \CW_{3d-4d} = \Sigma^{3d} (p_1 \pt_1 + p_2\pt_2) + \lambda \left(X^{4d}_{-1} p_1 \pt_2 + X^{4d}_{0} (p_1 \pt_1- p_2 \pt_2) +X^{4d}_{+1} p_2 \pt_1 \right)\ee
The $3d$ part of this boundary condition is precisely the matter content and superpotential of the $3d$ $\CN=4$ gauge theory $TSU(2)$, that is $U(1)$ with $2$ flavors.

The $SU(2)_\at$ doublets $p_i, \pt_j$ have scaling dimension $1/2$, the singlet $\Sigma^{3d}$ has scaling dimension $1$. The superpotential \ref{WBB} is scale invariant because the $4d$ fields and the $3d$ mesons have scaling dimension $1$. Notice that in this case we don't need the $3d-4d$ superpotential to fix the correct scaling dimensions of the $3d$ fields, these are fixed just by the $3d$ superpotential. The coupling in front of the term $\Sigma (p_1 \pt_1 + p_2\pt_2)$ is not marginal, it is needed to fix the scaling dimension of $\Sigma$. The coupling $\lambda$ is instead exactly marginal.

Another term preserving all the global symmetries is $Y^{4d} (p_1 \pt_1 + p_2\pt_2) $, but the $\CF$-terms of $\Sigma$ coming from \ref{WBB} are setting to zero the gauge invariant $(p_1 \pt_1 + p_2\pt_2)$ in the $3d$ chiral ring, so we don't expect the term $Y^{4d} (p_1 \pt_1 + p_2\pt_2) $ to be present. Therefore, the $3d-4d$ system considered here is a boundary condition for the $3$ $4d$ hypers $X^{4d}_I$.

The topological symmetry, with fugacity $\sigma$, is just a $3d$ symmetry, in the sense that no $4d$ fields are charged under it, and it's enhanced to $SU(2)_\s$, since $U(1)$ with $2$ flavors, with $\CN=4$ $3d$ susy, is a balanced quiver, so the basic monopoles $\M^\pm$ have scaling dimension $1$ and sit in the same supermultiplet of the off diagonal currents of $SU(2)_\s$. The coupling to the $4d$ fields respect the $SU(2)_\a \times SU(2)_\s$ global symmetry.

The $3d$ theory itself displays $\CN=4$ supersymmetry and is the so called $TSU(2)$ theory, which is the S-duality for $4d$ $\CN=4$ SYM with gauge group $SU(2)$, but here we are coupling the $3d$ theory to a $\CN=2$ $4d$ (free) theory, so we expect that the full $3d-4d$ system has only $4$ supercharges. In particular the $SU(2) \times SU(2)$ $3d$ R-symmetry of the $TSU(2)$ theory is broken to $U(1)_R \times U(1)_\b$.

\subsection{Boundary three-point function: $U(1)$ with $4$ flavors and $\CW_{mon}$ coupled to $4$ free hypers}
The three point function  gives us the partition function for a $3d-4d$ coupled system
\be
\langle \Psi_{\beta_1}(0) \Psi_{\beta_2}(1)\Psi_{\beta_3}(\infty) \rangle^{\rm FZZT}_{{\mathcal D}, \s_1,\s_2,\s_3}=\CZ[\CB_{U(1), N_f=4}](\beta_1,\beta_2,\beta_3;\s_1,\s_2,\s_3)
\ee
can be found from $C^{\s_3 \s_2 \s_1}_{\b_3 \b_2 \b_1}$ in eq. (5.45) of \cite{Nakayama:2004vk}
\bea \label{ZNF4}\CZ[\CB_{U(1), N_f=4}] = \frac{\Gb(Q/2-\bt_1\pm \bt_2\pm\bt_3)S_b(Q/2 \pm\st_1-\st_3-\bt_3)S_b(Q/2 \pm\st_2+\st_3-\bt_2)}{\Gb(-2\bt_1)\Gb(-2\bt_2)\Gb(-2\bt_3)\Gb(Q)} \times \quad\\
  \times \int_{-i\infty}^{+i\infty}\!\!\!\!\!\! S_b(\frac{Q}{4}\!+\!x\!-\!\st_3\!\pm\!\bt_2)S_b(\frac{Q}{4}\!+\!x\!\pm\!\st_1\!-\!\bt_1)S_b(\frac{Q}{4}\!-\!x\!+\!\st_3\!+\!\bt_1\!\pm\bt_3)S_b(\frac{Q}{4}\!-\!x\!\pm\!\st_2)dx \nonumber \eea
 We shifted the integration variable $x$ by $Q/4-\s_2$ and defined $\bt_{i}=-Q/2+\b_{i}$, $\st_{i}=-Q/2+\s_{i}$.

The $4d$ hypers $X^{4d}_{I,J}$ $\Gb(Q/2-\bt_1\pm \bt_2\pm\bt_3)$ are half of the tri-fundamental of $SU(2)_{\bt_1} \times SU(2)_{\bt_2} \times SU(2)_{\bt_3}$. Only the four half-hypers with negative $\bt_1$ charge appear. The $4d$ fields appearing in the denominator $\Gb(-2\bt_1)\Gb(-2\bt_2)\Gb(-2\bt_3)$ become $4d$ $\CN=2$ vector multiplets upon gluing the boundary three-point function with a bulk-boundary propagator or with another copy of the boundary three-point function, they will not play a role in this subsection.

The $3d$ gauge theory is $U(1)$ with $4$ flavors of charge $+1$ and $4$ flavors of charge $-1$. It enjoys $\CN=2$ supersymmetry. The $8$ charged $3d$ fields can be organized in $4$ charged doublets:
\begin{itemize}
\item $S_b(Q/4 + x-\st_3\pm\bt_2)$ is a $SU(2)_{\bt_2}$-doublet $Q_{1,2}$
\item $S_b(Q/4 + x \pm \st_1-\bt_1)$ is a $SU(2)_{\st_1}$-doublet $Q_{3,4} = Q_{\a}$
\item $S_b(Q/4 -x+\st_3+\bt_1\pm\bt_3)$ is a $SU(2)_{\bt_3}$-doublet $\Qt_{1,2}$
\item $S_b(Q/4 -x \pm\st_2)$ is a $SU(2)_{\st_2}$-doublet $\Qt_{3,4} = \Qt_{\dot{\a}}$
\end{itemize}
and $8$ gauge singlets
\begin{itemize}
\item $S_b(Q/2 \pm \st_1-\st_3-\bt_3)$ is a $SU(2)_{\st_1}$-doublet $X^{3d}_{\a}$
\item $S_b(Q/2 \pm \st_2+\st_3-\bt_2)$ is a $SU(2)_{\st_2}$-doublet $\tilde{X}^{3d}_{\dot{\a}}$
\item $\Gb(Q/2-\bt_1\pm \bt_2\pm\bt_3)$ is a $SU(2)_{\bt_2} \times SU(2)_{\bt_3}$-bifundamental $Z^{4d}$
\end{itemize}
Notice that the $4d$ fields $Z^{4d}$ are not charged under the $\st_i$ fugacities. We use the notation $SU(2)_{\b_i}$ even if it will turn out that the $\b_i$-fugacities are associated to $U(1)$ symmetries.

As opposed to \ref{BB}, in the integral \ref{ZNF4} there is no FI term, moreover the fugacities of the $8$ charged fields sums up to zero. Together, these facts imply that both $U(1)$ topological and the $U(1)$ axial symmetries of the $U(1)$ with $N_f=4$ gauge theory are broken. The way to achieve this breaking is to include in the $3d$ superpotential two terms containing monopole operators $\CW_{mon} = \M^+ + \M^-$ (see \cite{Benini:2017dud} for a study of $U(N_c)$ gauge theories with such a superpotential). $\M^\pm$ is the basic supersymmetric monopole operator with topological charge $\pm 1$.

The $16$ gauge invariant quadratic mesons $Q_i \Qt_j$ of the $U(1)$ with $N_f=4$ gauge theory have the following global symmetry fugacities
\be \left( \begin{array}{c|cccc}
 & Q_1  & Q_2  & Q_3  & Q_4 \\ \hline
\Qt_1  & \bt_1 +\bt_2 + \bt_3 & \bt_1 -\bt_2 + \bt_3  & \st_1 + \st_3 + \bt_3 & -\st_1 + \st_3 + \bt_3 \\
\Qt_2  &  \bt_1 + \bt_2 - \bt_3 & \bt_1 -\bt_2 - \bt_3  & \st_1 + \st_3 - \bt_3  & -\st_1 + \st_3 - \bt_3  \\
\Qt_3  &   \st_2 -\st_3 + \bt_2 &  \st_2 -\st_3 - \bt_2  & \st_1 + \st_2 -\bt_1 & -\st_1 + \st_2 -\bt_1  \\
\Qt_4  &  -\st_2 -\st_3 + \bt_2 & -\st_2 -\st_3 - \bt_2  &  \st_1 - \st_2 -\bt_1 & -\st_1 - \st_2 -\bt_1  
 \end{array} \right)\ee

Comparing these charges with the charges of the $8$ gauge singlets, we find that the $3d$ superpotential compatible with the Cartan generators of all the global symmetries is given by the monopoles and $8$ flipping terms coupling the $8$ gauge-singlets to $8$ of the $16$ mesons:
\be \label{W4d3d} \CW = \M^+ + \M^- + \tilde{X}^{3d}_{\dot{\a}} Q_{1} \Qt^{\dot{\a}} + X^{3d}_\a Q^{\a} \Qt_1  + \sum_{I,J=1,2}Z^{4d}_{I,J} Q_{I} \Qt_{J} \ee
The non-Abelian $SU(2)_{\st_1} \times SU(2)_{\st_2}$ symmetry is respected by these interactions.

\begin{figure}\centering
\begin{tikzpicture}
\node[gauge] (B) at (0,0) {$U(1)^{3d}_x$};
\node[flavor] (C) at (3,-3) {$SU(2)^{3d}_{\st_1}$};
\node[flavor] (D) at (-3,3) {$SU(2)^{4d}_{\bt_2}$};
\node[flavor] (E) at (3,3) {$SU(2)^{4d}_{\bt_3}$};
\node[flavor] (F) at (-3,-3) {$SU(2)^{3d}_{\st_2}$};
\draw[thick, ->] (B)--(C) ;
\draw[thick, ->] (B)--(D) ;
\draw[thick, ->] (E)--(B) ;
\draw[thick, ->] (F)--(B) ;
\draw[ ->] (D)--(F) ;
\draw[ ->] (C)--(E) ;
\draw[ultra thick, ->] (D)--(E) ;
\node at (0,3.3) {$X^{4d}$} ;
\node at (3.4,0) {$X^{3d}_\a$} ;
\node at (-3.4,0) {$\tilde{X}^{3d}_{\dot{\a}}$} ;
\node at (2,-1.5) {$Q_\a$} ;
\node at (2,1.5) {$\Qt_{1,2}$} ;
\node at (-2,-1.5) {$\Qt_{\dot{\a}}$} ;
\node at (-2,1.5) {$Q_{1,2}$} ;
\end{tikzpicture}
\end{figure}

The above superpotential is enough to fix all the R-charges, as we now explain. There is a $\IZ_2$ charge-conjugation symmetry $Q_i \leftrightarrow \Qt_i \,, X^{3d} \leftrightarrow \tilde{X}^{3d}$, so $R[Q_i]=R[\Qt_i]$. The $SU(2)_{\st_1} \times SU(2)_{\st_2}$ symmetry imposes $R[Q_3]=R[Q_4]$. The R-charge of the basic monopoles in a $3d$ $\CN=2$ $U(1)$ gauge theory with $4$ flavors is given in terms of the R-charge of the elementary fields by the usual formula
\be \label{Rmon} R[\M^\pm] = \sum_{i=1}^4 (1-R[Q_i]) \ee
Let us consider for a moment the $3d$ theory $U(1)$ with $N_f=4$ with $\CW_{mon}$ and the $4$ flipping singlets $X,\tilde{X}$ in isolation, without coupling to $4d$ fields $Z^{4d}$. Performing $\CZ$ extermination, for the isolated $3d$ theory we find
\be R[Q_1]\simeq 0.7532 \,, \quad R[Q_2]\simeq 0.3215 \,, \quad R[Q_{3,4}]\simeq 0.4624 \, \quad R[X^{3d},\tilde{X}^{3d}]\simeq 0.7839 \ee
This implies that the model we are discussing is different from the models called $\CT_{(4,4)}$ in \cite{Dimofte:2012pd}, which is $U(1)$ with $N_F=4$ and $\CW=\M^++\M^-$, where the R-charges of the $Q$'s are all $\frac{1}{2}$.

Now instead we couple the $3d$ theory to $4$ free $4d$ fields $Z^{4d}$, which must have R-charge $1$.\footnote{It is also interesting to see what happens if we think of the $4$ fields $Z^{4d}$ as $3d$ gauge invariant singlets $Z^{3d}$. In this case we have to perform a Z-extremization on two variables. The result is 
\be R[Q_1]\simeq 0.771 \,, \quad R[Q_2]\simeq 0.511 \,, \quad R[Q_{3,4}]\simeq 0.359 \,.\ee  Notice that $R[Z^{3d}_{1,1}]=2-2R[Q_1]<\frac{1}{2}$, so the unitarity bound for $Z^{3d}_{1,1}$ would be violated.} This, because of the last term in \ref{W4d3d}, implies that $R[Q_1]=R[Q_2]=\frac{1}{2}$. Combining with eq. \ref{Rmon}, we conclude that  in the $3d-4d$ coupled system
\be R[Q_1,Q_2,Q_3,Q_4]=\frac{1}{2} \,,\qquad R[X^{3d},\tilde{X}^{3d}]=1 \ee

The chiral ring operators of the boundary condition have thus integral R-charges. At the lowest level, $R=\Delta=1$, there are $16$ operators: the $8$ mesons not appearing the superpotential and the $8$ gauge singlet fields $X,\tilde{X},Z$. The fugacities of these $16$ ops are given by
\be 
\left( \begin{array}{ccc}
  Z_{11} & Z_{12}  & X^{3d}_\a  \\
  Z_{21} & Z_{22}  & Q_\a\Qt_2  \\
  \tilde{X}^{3d}_{\dot{\a}} & Q_2\Qt_{\dot{\a}}  & Q_\a\Qt_{\dot{\a}}
 \end{array} \right)
\!: \left( \begin{array}{c|c|cc}
  -\bt_1 -\bt_2 - \bt_3 & -\bt_1 +\bt_2 - \bt_3  & -\st_1 - \st_3 - \bt_3 & \st_1 - \st_3 - \bt_3 \\ \hline
 -\bt_1 - \bt_2 + \bt_3 & -\bt_1 +\bt_2 + \bt_3  & \st_1 + \st_3 - \bt_3  & -\st_1 + \st_3 - \bt_3  \\ \hline
  - \st_2 +\st_3 - \bt_2 &  \st_2 -\st_3 - \bt_2  & \st_1 + \st_2 -\bt_1 & -\st_1 + \st_2 -\bt_1  \\
  \st_2 +\st_3 - \bt_2 & -\st_2 -\st_3 - \bt_2  &  \st_1 - \st_2 -\bt_1 & -\st_1 - \st_2 -\bt_1  
 \end{array} \right)
 \ee

Looking at this quantum numbers, we see that the operators $X^{3d}_\a$ and the operators $Q_\a\Qt_2$ can be organized in a $SU(2)_{\st_3}$ doublet. Similarly the operators $\tilde{X}^{3d}_{\dot{\a}}$ and the operators $Q_2\Qt_{\dot{\a}}$ form another $SU(2)_{\st_3}$ doublet. It is natural to expect that the $U(1)_{\s_3}$ symmetry is actually enhanced to $SU(2)_{\s_3}$. This enhancement should happen on a special point in the conformal manifold of the $3d-4d$ system. In order to prove this it should be useful to study non trivial dualities for the theory under consideration, as in \cite{Dimofte:2012pd}. The symmetry enhancement should also follow from the cyclic $\IZ_3$ symmetry of the $3$-points boundary function: all three symmetries associated to $\s_i$ are on equal footing and enhance to $SU(2)$. This cyclic symmetry is indeed non-trivial to prove, see \cite{Ponsot:2001ng}.

It was shown in \cite{Dimofte:2012pd} (see also \cite{Teschner:2012em}) that the theory $\CT_{4,4}$, $U(1)$ with $N_F=4$ and $\CW=\M^++\M^-$ (with $SU(4) \times SU(4)$ global symmetry) is part of a web of dual theories. $20=1+18+1$ dual phases are of the form $U(1)$ with $N_F=4$, $\CW=\M^++\M^-$, with $0$ or $8$ or $16$ gauge singlets fields flipping the mesons. There are also dual phases of the form $SU(2)$ with $6$ doublets, again with some gauge singlets flipping fields ($6$ or $10$). This set of mutually dual $3d$ theories was coupled in \cite{Dimofte:2012pd} to $16$ free $4d$ hypers to produce $SO(12)$-invariant boundary conditions. We can act with all these dualities also on our $3d$ theory $U(1)$ with $N_F=4$, $\CW=\M^++\M^-$ and $4$ gauge singlet flipping fields, producing many dual phases with a varying number of $3d$ flipping fields.

\subsection{Gluing the building blocks: gauge theory interpretation}

In this subsection we consider two simple examples of gluing procedure for the building blocks discussed above and their gauge theory interpretation. 
The gluing factor is given by the conformal block of CFT on the strip, which amounts to consider just one copy of Virasoro descendants. The corresponding gauge theory factor is one copy of Nekrasov instanton partition
function with real coupling. As we will show later, also the one-loop factors of the gauge theory change. Indeed, concerning the vector, one has to take into account only half of the modes on each side to reproduce the CFT OPE coefficients.
Moreover, the reality of the mass parameters implies a reduction of the flavor symmetry of the gauge theory from $SU(2)$s to $U(1)$s: this is reflected in the global symmetries of the boundary 3d gauge theories. 

We first consider the boundary four-point function. From the gauge theory viewpoint, this is obtained from $N_f=4$ $SU(2)$ gauge theory on $HS^4$
with real masses and gauge coupling, coupled to a 3d sector living on the $S^3$ boundary which is discussed below.
The four-point function is obtained by boundary gluing two disks with three boundary punctures as in the following 
figure
\begin{center}
\vspace{.5cm}
\includegraphics[scale=0.6]{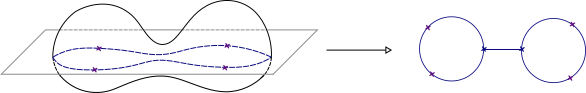} 
\vspace{.5cm}
\end{center}
The corresponding amplitude reads
\bea
<\Psi_{Q-\beta_4}^{\sigma_1\sigma_4}(x_4)\Psi_{\beta_3}^{\sigma_4\sigma_3}(x_3)\Psi_{\beta_2}^{\sigma_3\sigma_2}(x_2)\Psi_{\beta_1}^{\sigma_2\sigma_1}(x_1)>&=&\nonumber\\  \int d\beta C^{\sigma_4\sigma_2\sigma_1}_{Q-\beta_4\beta\beta_1}C^{\sigma_4\sigma_3\sigma_2}_{Q-\beta\beta_3\beta_2}F^s(\Delta_{\beta_i},\Delta_{\beta},x_1,x_2,x_3,x_4)
\eea
As we will see this Liuoville correlation function corresponds to ${\cal N}=2$ gauge theory on Hemi-$S^4$ with Neumann BCs and with certain matter content. We consider the one-loop part in this channel
\bea
&C^{\sigma_4\sigma_2\sigma_1}_{\beta_4\beta_1\beta}C^{\sigma_4\sigma_3\sigma_2}_{Q-\beta\beta_3\beta_2}=\nonumber\\
& \frac{\Gb(Q-\b\pm \bt_1\pm\bt_4)S_b(Q \pm\st_1-\s_3-\bt_4)S_b(\pm\st_2+\s_3-\bt_1)}{\Gb(Q-2\b)\Gb(-2\bt_1)\Gb(-2\bt_4)\Gb(Q)}
 \int \frac{S_b(x+Q/2 \pm \st_1-\b)S_b(x+Q/2-\s_3\pm\bt_1)}{S_b(x+3Q/2-\s_3-\b\pm\bt_4)S_b(x+Q/2\pm\st_2)}dx\nonumber\\
&  \frac{\Gb(Q+\b\pm \bt_2\pm\bt_3)S_b(Q \pm\st_1-\s_3-\bt_3)S_b(\pm\st_2+\s_3-\bt_2)}{\Gb(Q+2\b)\Gb(-2\bt_2)\Gb(-2\bt_3)\Gb(Q)}
 \int \frac{S_b(y+Q/2 \pm \st_1+\b)S_b(y+Q/2-\s_3\pm\bt_2)}{S_b(y+3Q/2-\s_3+\b\pm\bt_3)S_b(y+Q/2\pm\st_2)}dy\nonumber\\
\eea
where we have used the redefined fugacities  $-\frac{Q}{2}+\b_i \equiv \bt_i$. Note that 
\bea
\frac{1}{\Gb(Q-2\b)\Gb(-Q+2\b)}&=&\frac{1}{\Gb(-2\bt)\Gb(2\bt)}
\eea
after redefining $-\frac{Q}{2}+\b \equiv\bt$.
 For  $b=1$
\bea
\frac{1}{\Gb(-2\bt)\Gb(2\bt)}&=&\frac{1}{\Gamma_2(-2\bt)\Gamma_2(2\bt)}\nonumber\\
&=&G(-2\bt)G(2\bt)=\frac{G(2-2\bt)G(2\bt)}{\Gamma(2\bt)\Gamma(1-2\bt)}\nonumber\\
&=&\frac{1}{\Gamma_2(2-2\bt)\Gamma_2(2\bt)}\frac{1}{\Gamma(2\bt)\Gamma(1-2\bt)}\nonumber\\
&=&\frac{\Upsilon(2\bt)}{\Gamma(1-2\bt)\Gamma(2\bt)}
\eea
which is nothing but the vector one loop determinant  for the choice of  Neumann BCs on dynamical fields. This is so because $\bt$ denotes vev of the scalar of bulk vector multiplet and there is an overall integration over it. For the bulk hypers it is  clear that four half hypers with masses $\pm \bt_1\pm\bt_4$ are coupled to the half of vector multiplet with negative sign of $\bt$ and another four half hypers with masses $\pm \bt_2\pm\bt_3$ are coupled to the other half of vector with positive sign of $\bt$.

Therefore
\bea
&C^{\sigma_4\sigma_2\sigma_1}_{\frac{Q}{2}+\bt_4\frac{Q}{2}+\bt_1\frac{Q}{2}+\bt}C^{\sigma_4\sigma_3\sigma_2}_{\frac{Q}{2}-\bt\frac{Q}{2}+\bt_3\frac{Q}{2}+\bt_2}=\frac{\Upsilon(2\bt)}{\Gamma(1-2\bt)\Gamma(2\bt)}\nonumber\\
& \frac{\Gb(\frac{Q}{2}-\bt\pm \bt_1\pm\bt_4)S_b(Q \pm\st_1-\s_3-\bt_4)S_b(\pm\st_2+\s_3-\bt_1)}{\Gb(-2\bt_1)\Gb(-2\bt_4)\Gb(Q)}
 \int \frac{S_b(x \pm \st_1-\bt)S_b(x+Q/2-\s_3\pm\bt_1)}{S_b(x+Q-\s_3-\bt\pm\bt_4)S_b(x+Q/2\pm\st_2)}dx\nonumber\\
&\frac{\Gb(\frac{Q}{2}+\bt\pm \bt_2\pm\bt_3)S_b(Q \pm\st_1-\s_3-\bt_3)S_b(\pm\st_2+\s_3-\bt_2)}{\Gb(-2\bt_2)\Gb(-2\bt_3)\Gb(Q)}
 \int \frac{S_b(y \pm \st_1+\bt)S_b(y+Q/2-\s_3\pm\bt_2)}{S_b(y+Q-\s_3+\bt\pm\bt_3)S_b(y+Q/2\pm\st_2)}dy\nonumber\\
\eea
We now pass to consider the open gluing of a bulk-to-boundary disk amplitude in the following example 
\begin{center}
\vspace{.5cm}
\includegraphics[scale=0.6]{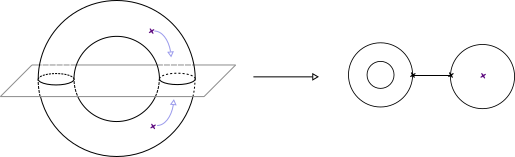} 
\vspace{.5cm}
\end{center}
The amplitude reads
\bea
&  BB(\at,Q-\beta,\sigma)C^{\s_3 \s_2 \s_1}_{\bt_3 \bt_2 \bt}=
\frac{\Gb( \pm 2 \at +\beta)\Gb(Q+\beta)^2 \Upsilon(\at)}{\Gb(Q)\Gb(-Q+2\beta)} 
S_b(\beta) \int e^{2 \pi i \sigma y} S_b(\pm y \pm \at -\beta/2+\frac{Q}{2})  dy\nonumber\\
& \frac{\Gb(Q-\b\pm \bt_2\pm\bt_3)S_b(Q \pm\st_1-\s_3-\bt_3)S_b(\pm\st_2+\s_3-\bt_2)}{\Gb(Q-2\b)\Gb(-2\bt_2)\Gb(-2\bt_3)\Gb(Q)}
 \int \frac{S_b(x+Q/2 \pm \st_1-\b)S_b(x+Q/2-\s_3\pm\bt_2)}{S_b(x+3Q/2-\s_3-\b\pm\bt_3)S_b(x+Q/2\pm\st_2)}dx\nonumber\\
\eea
Here again, after redefining the fugacities, we see the vector one loop for Neumann BCs as expected
\bea
&  BB(\at,\frac{Q}{2}-\bt,\sigma)C^{\s_3 \s_2 \s_1}_{\frac{Q}{2}+\bt_3 \frac{Q}{2}+\bt_2 \frac{Q}{2}+\bt}=\frac{\Upsilon(2\bt)}{\Gamma(1-2\bt)\Gamma(2\bt)}\nonumber\\
&\frac{\Gb( \pm 2 \at +\frac{Q}{2}+\bt)\Gb(\frac{3Q}{2}+\bt)^2 \Upsilon(\at)}{\Gb(Q)} 
S_b(\frac{Q}{2}+\bt) \int e^{2 \pi i \sigma y} S_b(\pm y \pm \at -\bt/2+\frac{Q}{4})  dy\nonumber\\
& \frac{\Gb(\frac{Q}{2}-\bt\pm \bt_2\pm\bt_3)S_b(Q \pm\st_1-\s_3-\bt_3)S_b(\pm\st_2+\s_3-\bt_2)}{\Gb(-2\bt_2)\Gb(-2\bt_3)\Gb(Q)}
 \int \frac{S_b(x \pm \st_1-\bt)S_b(x+Q/2-\s_3\pm\bt_2)}{S_b(x+Q-\s_3-\bt\pm\bt_3)S_b(x+Q/2\pm\st_2)}dx \, . \nonumber\\
\eea






\vspace{0.4cm}

\section{Open questions}

There is a number of open questions worth to be investigated. 

To start with, we did not discuss the  
interpretation and consequences of the $\mathbb{Z}_2$-quotient on the BPS spectrum and infrared properties of the supersymmetric gauge theories.
The Seiberg-Witten curve of the quotient gauge theory is expected to be given by a covering of the open/unoriented geometries on which the M5-branes
are wrapped. 
As noted in \cite{Tachikawa:2016xvs}, there is a non-trivial interplay between the involution and the S-duality properties of the BPS spectrum.

The localization computation we performed in Section 2 in the $\rpf$ case are valid for all gauge groups and thus amenable for a large N analysis.
One can easily obtain matrix models from the one-loop calculations presented in Section 2. 
This raises the question of a holographic dual description of these gauge theories.

Localization computations were performed on toric manifolds allowing for new computations in Donaldson theory \cite{Bershtein:2015xfa,Bershtein:2016mxz}. It would be interesting to investigate whether
our results could be used to extend Donaldson theory to un-orientable manifolds. 

As far as the AGT correspondence is concerned, the higher rank gauge theory computations performed in this paper should have a natural counterpart
in open/unoriented Toda CFT \cite{Wyllard:2009hg,Mironov:2009by,Bonelli:2009zp}.  More work is required to deepen our understanding of the correspondence in presence boundary punctures. Actually, the bulk-to-boundary overlap wavefunction should be related to the S-duality kernel for the relevant conformal block/gauge theory instanton sector, while boundary puncture degenerations should be related to the S-duality kernel associated to the disk three-point function.

 It would also be interesting to perform quotients of four-dimensional manifolds
other than $S^4$. A notable example is $S^2\times S^2$, whose two-dimensional CFT counterpart has been related to Liouville gravity \cite{Bawane:2014uka}. This should give access to a gauge theory description of the open sector of Liouville gravity, see \cite{Aleshkin:2017yty} for recent developments on this topic. 

One could also consider quotients of other six dimensional M5-brane geometries, such as $S^3\times M_3$, and investigate their consequences at the level of 3d-3d correspondence \cite{Dimofte:2010tz,Dimofte:2011ju}. 



\section*{ Acknowledgements}

We thank Y. Tachikawa for enlightening discussions at the early stage of this project and for sharing private notes on this topic. 
We thank E. Gava, S. Giacomelli, K. Hosomichi, B. Mares and K. Narain for useful discussions. We thank B. Le Floch for a careful reading of the manuscript and interesting comments.
This research was partly supported by the INFN Research Projects GAST and ST$\&$FI and by PRIN projects "Geometria delle variet\'a algebriche" and "Non-perturbative aspects of Gauge Theories And Strings". S. B. is supported by the MIUR-SIR grant RBSI1471GJ "Quantum Field Theories at Strong Coupling: Exact Computations and Applications".

\appendix 

\section*{APPENDICES}



\section{$SU(2)\times SU(2)$ harmonics}\label{app:harmonics}


Our choice of vielbeins is 
\begin{align}\label{eq:leftframe}
 e_L^1 &= -\frac{1}{2}\cos\psi\,\sin\rho\,d\theta -\frac{1}{2}\sin\theta\,\sin\psi\,\sin\rho\,d\phi, \nonumber\\
  e_L^2 &= +\frac{1}{2}\sin\psi\,\sin\rho\,d\theta -\frac{1}{2}\sin\theta\,\cos\psi\,\sin\rho\,d\phi,\nonumber\\
   e_L^3 &= -\frac{1}{2}\sin\rho\,d\psi -\frac{1}{2}\cos\theta\,\sin\rho\,d\phi,\nonumber \\
    e_L^4 &= d\rho.
\end{align}
The subscript will be explained shortly. The non-zero components of the spin connection derived from the above choice of vielbeins are
\begin{align*}
 \Omega_\psi^{12} &=-\frac{1}{2},\,\Omega_\psi^{34} =-\frac{1}{2}\cos\rho,\\
 \Omega_\theta^{13} &= -\frac{1}{2}\sin\psi,\, \Omega_\theta^{14}=-\frac{1}{2}\cos\rho\cos\psi,\,
  \Omega_\theta^{23} = -\frac{1}{2}\cos\psi,\,\Omega_\theta^{24}=-\frac{1}{2}\cos\rho\sin\psi,\\
  \Omega_\phi^{12}&= -\frac{1}{2}\cos\theta,\, \Omega_\phi^{13}=\frac{1}{2}\sin\theta\cos\psi, \,
  \Omega_\phi^{14}=-\frac{1}{2}\cos\rho\sin\psi\sin\theta,\\
  \Omega_\phi^{23}&=-\frac{1}{2}\sin\theta\sin\psi,\, \Omega_\phi^{24}=-\frac{1}{2}\cos\rho\cos\psi\sin\theta, \,\Omega_\phi^{34}=-\frac{1}{2}\cos\theta\cos\rho.
\end{align*}

Another choice of vielbeins is as follows:
\begin{equation}\label{eq:rightframe}
\begin{aligned}
  e_R^1 &= -\frac{1}{2}\cos\phi\,\sin\rho\,d\theta -\frac{1}{2}\sin\theta\,\sin\phi\,\sin\rho\,d\psi, \\
  e_R^2 &= -\frac{1}{2}\sin\phi\,\sin\rho\,d\theta +\frac{1}{2}\sin\theta\,\cos\phi\,\sin\rho\,d\psi,\\
   e_R^3 &= -\frac{1}{2}\sin\rho\,d\phi -\frac{1}{2}\cos\theta\,\sin\rho\,d\psi \\
    e_R^4 &= d\rho,
\end{aligned}
\end{equation}
which is essentially obtained by exchanging $\phi$ and $\psi$ in $e_L$. Call the above set the right-handed vielbeins, and those in Eq \ref{eq:leftframe} the left-handed vielbeins.

It can be verified that the six one-forms $\sin\rho\, e_L^a$ and $\sin\rho\, e_R^a$, where $a\in\{1,2,3\}$, satisfy the Killing equation $D_m v_n + D_n v_m =0$, and therefore correspond to generators of the isometry group of $\sfr$. The $SO(4)$ group is isomorphic to $SU(2)_L\times SU(2)_R$, where the subscripts on $SU(2)$ indicate which set of generators it is related to. In order to obtain the commutation relations of the two $SU(2)$ algebras in the canonical form, we rescale the generators (written as one-forms) as follows:
\begin{equation}
 \begin{array}{l}
  l^1 = \frac{i}{2}\sin\rho \,e^1_L,\\
  l^2 = -\frac{i}{2}\sin\rho \,e^2_L,\\
  l^3 = \frac{i}{2}\sin\rho \,e^3_L.
 \end{array}
\hspace{1cm}
 \begin{array}{c}
  r^1 = \frac{i}{2}\sin\rho \,e^1_R,\\
  r^2 = \frac{i}{2}\sin\rho \,e^2_R,\\
  r^3 = \frac{i}{2}\sin\rho \,e^3_R.
 \end{array}
\end{equation}
We shall denote the Lie derivatives with respect to $l^a$ and $r^a$ as $J_L^a$ and $J_R^a$. We also define $J_L^\pm = J_L^1 \pm i J_L^2$, which are raising and lowering operators for $SU(2)_L$ and co-incide with Lie derivatives with respect to $l^\pm = l^1\pm il^2$. Similar definitions hold for right-handed generators.

Let us first consider scalars. Lie derivative is the usual directional derivative: $J_L^a = l^{am}\partial_m$. Consider a basis in which $J^3_{L,R}$ are diagonal. Scalar functions belonging to this basis are given by $f(\theta,\rho)e^{i(q_L\psi+q_R\phi)}$. For such a function to be a highest weight function with respect to both $SU(2)$s, it has to be annihilated by $J^+_{L,R}$. (We do not have to consider $\rho$ dependence since none of the generators act along the $\rho$ direction. One can freely multiply an arbitrary function of $\rho$, and the resulting function would again be a highest weight function). It can be easily seen that for a function to be annihilated by both the raising operators, the highest weights with respect to the two $SU(2)$s, call them $j_L$ and $j_R$, must be equal. The highest weight function then takes the form
\begin{equation}
 \varPhi_{j_L} = \left(\cos\tfrac{\theta}{2}\right)^{2j_L}e^{ij_L(\psi+\phi)}
\end{equation}
up to a (possibly $\rho$-dependent) normalization factor. Acting $s$ times on $\varPhi_{j_L}$ by $J_R^-$, we get:
\begin{equation*}
 (J_R^-)^s\varPhi_{j_L} = e^{ij_L\psi+i(j_L-s)\phi}\left(\cos\tfrac{\theta}{2}\right)^{2j_L-s}\left(\sin\tfrac{\theta}{2}\right)^{s}
\end{equation*}
which is a highest weight state with respect to $SU(2)_L$ (therefore is an eigenstate of $J_L^3$ with eigenvalue $j_L$), but has $J_R^3$ eigenvalue $j_L-s$. To find the lowest weight state with respect to $SU(2)_R$, we act on the above state by $J_R^-$ once more and set the result equal to zero. Doing so, one finds that $(J_R^-)^s\varPhi_{j_L}$ is the lowest weight state when $s=2j_L$. Therefore, there are $2j_L$ states with $SU(2)_L$ weight equal to $j_L$. Similar result holds if we consider descendents with respect to the left lowering operator.

Let us now turn to one-forms. The Lie derivative of a one-form $\omega$, with respect to $l^a$ for instance, is given by:
\begin{equation*}
 (J^a_L\, \omega)_n=l^{am}(\partial_m\omega_n-\partial_n\omega_m)+\partial_n(l^{am}\omega_m).
\end{equation*}
We would like to repeat the same exercise and try to find one-forms that are simultaneously the highest weight states with respect to $SU(2)_R$ and $SU(2)_L$. The calculations are straightforward, but tedious, and we will therefore be sketchy in our description, emphasizing only the results that will be important later. Consider a one-form that is an eigenstate of $J^3_{L,R}$:
\begin{equation*}
 \omega_m(\psi,\theta,\phi) = e^{i(q_L\psi+q_R\phi)}\tilde\omega_m(\theta)
\end{equation*}
Apply the raising operators $J^+_{L,R}$ to the above one-form, and set the result equal to zero. One can divide the analysis into two cases:
\begin{itemize}
 \item $\tilde\omega_\rho(\theta)$ is not identically zero. In this case, simultaneous vanishing of $(J^+_L\,\omega)_\rho$ and $(J^+_R\,\omega)_\rho$ requires that the highest weights, which we denote $j_L$ and $j_R$, are equal. With this condition, the other components can be easily solved for, and we get the following highest weight one-form:
 \begin{equation*}
  \omega^0 \equiv e^{ij_L(\psi+\phi)}\left(\cos\tfrac{\theta}{2}\right)^{2j_L}\left(d\psi + d\phi+i\tan\tfrac{\theta}{2}\,d\theta +\alpha \,d\rho\right)
 \end{equation*}
  up to an overall normalization constant, and where $\alpha$ is some constant.
 \item $\tilde\omega_\rho(\theta)$ is identically zero. We may no longer conclude that $j_L = j_R$. The $\psi$ and $\phi$ components can be easily solved to give
 \begin{align*}
  \tilde\omega_\psi &= C_1\left(\cos\tfrac{\theta}{2}\right)^{j_L}\left(\sin\tfrac{\theta}{2}\right)^{-j_L}\left(\sin\theta\right)^{j_R}\\
  \tilde\omega_\phi &= C_2\left(\cos\tfrac{\theta}{2}\right)^{j_R}\left(\sin\tfrac{\theta}{2}\right)^{-j_R}\left(\sin\theta\right)^{j_L}
 \end{align*}
  The component $\tilde\omega_\theta$ can now be solved for in multiple ways, and the various solutions agree only when the combinations $C_1 (1+j_l-j_R)$ and $C_2 (-1+j_l-j_R)$ both vanish. This can happen if either $C_1=0,C_2\neq 0, j_L-j_R =1$ or $C_1\neq 0,C_2= 0, j_L-j_R =-1$. (If both $C_1$ and $C_2$ vanish, then we have a one-form that is identically zero). We have, then, the following two possibilities, up to normalization:
  \begin{align*}
   \omega^L &\equiv e^{ i(j_L-1)\phi + i\,j_L\psi}\left(\cos\tfrac{\theta}{2}\right)^{2j_L-2}\left(\sin\theta\, d\phi + i \,d\theta\right)\\
   \omega^R &\equiv e^{ i(j_L+1)\phi + i\,j_L\psi}\left(\cos\tfrac{\theta}{2}\right)^{2j_L}\left(\sin\theta\, d\psi + i \,d\theta\right)
  \end{align*}
where, in the first solution $j_L = j_R +1$, whereas in the second solution $j_L = j_R -1$. We have written $j_R$ everywhere in terms of $j_L$.
\end{itemize}
Let us now consider scalar combinations of the highest weight one-forms and the left Killing vectors $l^{am}$ (where $a\in \{1,2,3\}$). Let us first do this for the one-form with $j_L = j_R$:
\begin{equation}
 \omega^{0a}\equiv l^{am}\omega^0_m = e^{ij_L(\psi+\phi)}\left(\cos\tfrac{\theta}{2}\right)^{2j_L}\left(e^{-i\psi}\tan\tfrac{\theta}{2},ie^{-i\psi}\tan\tfrac{\theta}{2},-i \right).
\end{equation}
Further, consider the combinations $\omega^{0\pm} = \omega^{01}\pm i\omega^{02}$:
\begin{equation*}
 \omega^{0+} =0,\hspace{0.5cm}\omega^{0-} = 2 e^{ij_L(\psi+\phi)-i\psi}\left(\cos\tfrac{\theta}{2}\right)^{2j_L-1}\left(\sin\tfrac{\theta}{2}\right).
\end{equation*}
Note that the third component $\omega^{03}$ is already a scalar highest-weight function with weight $j_L$. The combination $\omega^{0+}$ is trivially belongs to a highest-weight representation, whereas $\omega^{0-}$ can be seen to be a level one descendent of a highest weight function with weight $j_L$:  $\omega^{0-} \propto J_L^- \varPhi_{j_L}$. Therefore, all of $\{\omega^{0+},\omega^{03},\omega^{0-}\}$ belong to the highest weight representation with weight $j_L$.

The same exercise carried out for $\omega^L$ gives:
\begin{equation*}
 (\omega^{L+},\omega^{L3},\omega^{L-}) = \left(0,0,2e^{i(j_L-1)(\phi+\psi)}\left(\cos\tfrac{\theta}{2}\right)^{j_L-1}\right),
\end{equation*}
which implies that all the above components belong to a scalar highest-weight representation with weight $j_L-1$. Finally, from $\omega^R$ we get:
\begin{equation*}
 (\omega^{R+},\omega^{R3},\omega^{R-}) \sim \left(\varPhi_{j_L+1}, J_L^-\varPhi_{j_L+1},(J_L^-)^2\varPhi_{j_L+1}\right)
\end{equation*}
(The $\sim$ sign indicates that each of the terms on the left is proportional to the corresponding term on the right, but the constant of proportionality, which is irrelevant to our purposes, may differ for each term.)

Let us summarize: If $\omega$ is a highest weight one-form with right weight $j_R$, then the scalar combinations of $\omega$ with $\{l^\pm,l^3\}$ belong to a scalar highest weight representation with highest weight $j_R$. This means that we can study one-form highest weight functions in terms of scalar highest weight functions and their descendents.

\section{Supersymmetry transformations}
\label{app:susy}
The supersymmetry transformation of a vector multiplet is given by \cite{Hama:2012bg}:
\begin{equation}\label{eq:SUSYVect}
\begin{aligned}
&\sq A_m =i\xi^A\sigma_m \bar\lambda_A - i \bar\xi^A \bar\sigma_m \lambda_A, \\
&\sq\phi =-i\xi^A\lambda_A, \\
&\sq\bar\phi =+i\bar\xi^A\bar\lambda_A, \\
&\sq\lambda_A =\tfrac{1}{2}\sigma^{mn}\xi_A(F_{mn}+8\bar\phi T_{mn})+2\sigma^m\bar\xi_A D_m\phi+\sigma^m D_m\bar\xi_A\phi+2i\xi_A[\phi,\bar\phi]+D_{AB}\xi^B, \\
&\sq\bar\lambda_A =\tfrac{1}{2}\bar\sigma^{mn}\bar\xi_A(F_{mn}+8\phi \bar T_{mn})+2\bar\sigma^m \xi_A D_m\bar\phi+\bar\sigma^m D_m\xi_A\bar\phi-2i\bar\xi_A[\phi,\bar\phi]+D_{AB}\bar\xi^B, \\
&\sq D_{AB} =-i\bar\xi_A\bar\sigma^mD_m\lambda_B
 -i\bar\xi_B\bar\sigma^mD_m\lambda_A
 +i\xi_A\sigma^mD_m\bar\lambda_B
 +i\xi_B\sigma^mD_m\bar\lambda_A \\ 
 &\hspace{1.6cm} -2[\phi,\bar\xi_A\bar\lambda_B+\bar\xi_B\bar\lambda_A] +2[\bar\phi,\xi_A\lambda_B+\xi_B\lambda_A].
\end{aligned}
\end{equation}
The square of the supersymmetry transformation is given by
\begin{equation}\label{eq:susysquare}
 \begin{aligned}
  &\sq^2 A_m = iv^nF_{nm}+D_m\Phi,\\
  &\sq^2\phi =iv^nD_n\phi +i[\Phi,\phi]+(w+2\Theta)\phi,\\
  &\sq^2\bar\phi =iv^nD_n\bar\phi +i[\Phi,\bar\phi]+(w-2\Theta)\bar\phi,\\
  &\sq^2\lambda_A =iv^nD_n\lambda_A +i[\Phi,\lambda_A]+\left(\tfrac{3}{2}w+\Theta\right)\lambda_A +\tfrac{i}{4}\sigma^{kl}\lambda_AD_kv_l+\Theta_{AB}\lambda^B,\\
  &\sq^2\bar\lambda_A =iv^nD_n\bar\lambda_A +i[\Phi,\bar\lambda_A]+\left(\tfrac{3}{2}w-\Theta\right)\bar\lambda_A +\tfrac{i}{4}\bar\sigma^{kl}\bar\lambda_AD_kv_l+\Theta_{AB}\bar\lambda^B,\\
  &\sq^2D_{AB} =iv^nD_nD_{AB} +i[\Phi,D_{AB}]+2wD_{AB}+\Theta_{AC}D^C_{\phantom{C}B}+\Theta_{BC}D^C_{\phantom{C}A},
 \end{aligned}
\end{equation}
where 
\begin{align*}
 v^m &= 2\bar\xi^A\bar\sigma^m\xi_A\\
 \Phi &= -2i\phi\bar\xi^A\bar\xi_A + 2i\bar\phi\xi^A\xi_A\\
 w&=-\tfrac{i}{2}\left( \xi^A\sigma^mD_m\bar\xi_A + D_m\xi^A\sigma^m\bar\xi_A\right)\\
 \Theta &= -\tfrac{i}{4}\left( \xi^A\sigma^mD_m\bar\xi_A - D_m\xi^A\sigma^m\bar\xi_A\right)\\
 \Theta_{AB} &= -i \xi_{(A}\sigma^mD_m\bar\xi_{B)} +i D_m\xi_{(A}\sigma^m\bar\xi_{B)}.
\end{align*}
In particular, it can be seen that the generator of gauge transformations is
\begin{equation}\label{eq:GaugeGenerator}
 \hat\Phi = \Phi - iv^nA_n.
\end{equation}
For the Killing spinors in Eq. \ref{eq:Kspinors}, we see that the generators of scaling and $U(1)_\mathcal{R}$ tranformations, $w$ and $\Theta$ respectively, are both zero. The other generators are
\begin{equation}\label{eq:BosonicGenerators}
 v^m\partial_m = 2\partial_\psi,\hspace{0.2cm}
 \hat\Phi = 2i\phi\sin^2\tfrac{\rho}{2} + 2i\bar\phi\cos^2\tfrac{\rho}{2} -2iA_\psi,\hspace{0.2cm}
 \Theta_{AB} =\left(\begin{array}{cc}
                0 & 1\\
                1 & 0
               \end{array}\right).
\end{equation}

We now describe the gauge-fixing procedure, more details regarding which may be found in \cite{Pestun:2007rz}. In order to perform gauge-fixing, one introduces a BRST-like operator $\sqb$, and a zoo of fields that satisfy
\begin{equation}\label{eq:BRSTGhost}
\begin{array}{llll}
  \sqb c = a_0 + icc, & \sqb\bar c = b, & \sqb\bar a_0 = \bar c_0, & \sqb b_0 = c_0, \\
  \sqb a_0=0, & \sqb b = [a_0,\bar c], & \sqb\bar c_0 = [a_0,\bar a_0], & \sqb c_0 = [a_0,b_0].
\end{array}
\end{equation}
To understand the role of each of these fields, consider the gauge-fixing action
\begin{equation}
 \sqb V_{GF,\xi}\equiv\sqb\mathrm{Tr}\left[\bar c (i\partial_m A^m) + \bar c b_0 + c\left(\bar a_0-\tfrac{\xi}{2}a_0\right)\right].
\end{equation}
(What goes into the action is of course $(\sq +\sqb)  V_{GF,\xi}$, but it can be shown that $\sq V_{GF,\xi}$ does not alter the partition function \cite{Pestun:2007rz}. Further, it will be shown below that $\xi$ drops out from the final expression, and therefore can be chosen to be zero. But we retain it temporarily to make the role of some of the gauge-fixing fields more transparent). Expanding the above expression we get
\begin{equation}
 \begin{split}
  &\mathrm{Tr}\left[b (i\partial_m A^m)- \bar c(i\partial_m D^mc)+bb_0-\bar cc_0-c\bar c_0\right]\\
  &+\mathrm{Tr}\left[\left(icc+a_0 \right)\left(\bar a_0-\tfrac{\xi}{2}a_0 \right)\right].
 \end{split}
\end{equation}
In the first line, the constant fields $b_0, c_0, \bar c_0$, on integrating over, absorb the zero modes of $b,\bar c,c$ respectively. After this is done, the integration over the remaining modes of $b,\bar c,c$ imposes the gauge condition $\partial_m A^m=0$ (via the term $ib\partial_m A^m$), and provides the Fadeev-Popov determinant (via the term $i\bar c\partial_m D^mc$). This successfully fixes the gauge.

The terms in the second line do not give any non-trivial contribution to the partition function on integration. To see this, rewrite it as
\begin{equation*}
 \mathrm{Tr}\left[ -\tfrac{\xi}{2}\left(a_0-\tfrac{1}{\xi}\bar a_0+ \tfrac{i}{2}cc\right)^2 + \tfrac{1}{2\xi}\left(\bar a_0 + \tfrac{i\xi}{2}cc\right)^2\right].
\end{equation*}
Integrating over $a_0$ on a real locus and then over $\bar a_0$ on an imaginary locus shows that $\xi$ drops out and there are no other non-trivial contributions.

To complete this section on gauge fixing, we also need to describe the action of $\sq$ on the gauge-fixing fields:
\begin{equation}\label{eq:SUSYGhost}
\begin{array}{llll}
  \sq c = -\hat\Phi, & \sq\bar c = 0, & \sq \bar a_0 = 0, & \sq b_0 = 0, \\
  \sq a_0=0, & \sq b = iv^m\partial_m\bar c, & \sq\bar c_0 = 0, & \sq c_0 = 0.
\end{array}
\end{equation}
and that of $\sqb$ on the vectormultiplet fields:
\begin{equation}\label{eq:BRSTVect}
 \sqb A_m = D_m c,\hspace{0.2cm} \sqb\, \mathrm{scalar} = i[c,\mathrm{scalar}],\hspace{0.2cm} \sqb\, \mathrm{fermion} = i\{c,\mathrm{fermion}\}.
\end{equation}
$\hat\Phi$ that appears in $\sq c$ was introduced in Eq. \ref{eq:GaugeGenerator}. This ensures that the square of $\sqh \equiv \sq + \sqb$ is, besides other bosonic transformations shown in Eq. \ref{eq:susysquare}, a constant gauge transformation by $a_0$ rather than a gauge transformation by $\hat\Phi$. In particular, for the case at hand,
\begin{equation*}
 \sqh^2 = i\,\mathrm{Lie}(v) + \mathrm{Gauge}(a_0)+SU(2)_\mathcal{R}(\Theta_{AB}),
\end{equation*}
where the generators in parentheses are given in Eq. \ref{eq:BosonicGenerators}.

Finally, we look at hypermultipets. These are multiplets that have had too much coffee.
%
\subsection{Hypermultiplet Lagrangian for ${\cal N}=2^*$}
To give  mass to a single hypermultiplet, we have to couple it to a background vector multiplet with abelian gauge group.
This abelian gauge group is taken as the subgroup of the commutant of the embedding of $SU(2)$ in $Sp(r)$. For this  coupling to background vector multiplet to be supersymmetric, the gaugino variation of this extra multiplet is set to zero and solved for the vacuum solution and finds the following 
\bea
\phi=\bar{\phi}=\phi_0,\quad D_{AB}^0=2\phi \omega_{AB}
\eea
where $\phi_0$ is a constant and $\omega_{AB}$ in the notation of \cite{Hama:2012bg} is given by
\bea
\omega_{AB}=\frac{4\xi_A\sigma^{mn}\xi_B(T_{mn}-S_{mn})}{\xi^A\xi_A}=-\frac{4\bar{\xi}_A\bar{\sigma}^{mn}\bar{\xi}_B(\bar{T}_{mn}-\bar{S}_{mn})}{\bar{\xi}_A\bar{\xi}^A}
\eea
Then we get the following Lagrangian for massive matter multiplet
 \bea
  L_{mat}&=&\frac{1}{2}D_mq^AD^mq_A-q^A\{\phi,\bar{\phi}\}q_A+\frac{\iota}{2}q^AD_{AB}q^B+(\frac{1} {8}(R+M+\{\phi_0,\phi_0\})\epsilon_{AB}+2\phi_0\omega_{AB})q^Aq^B\nonumber\\&-&\frac{\iota}{2}\bar{\psi}\bar{\sigma}^mD_m\psi
  -  \frac{1}{2}\psi\phi\psi+  \frac{1}{2}\bar{\psi}\bar{\phi}\bar{\psi} -  \frac{1}{2}\psi\phi_0\psi+  \frac{1}{2}\bar{\psi}\phi_0\bar{\psi}+\frac{\iota}{2}\psi\sigma^{kl}T_{kl}\psi-\frac{\iota}{2}\bar{\psi}\bar{\sigma}^{kl}\bar{T}_{kl}\bar{\psi}\nonumber\\&-&q^A\lambda_A\psi+\bar{\psi}\bar{\lambda}q^A-\frac{1}{2}F^AF_A.\nonumber\\
\eea
where it is understood that vector multiplet fields $\Phi$ carry $Sp(r)$ indices as $\Phi^I_J$ and importantly for the background vector multiplet the $I,J$ indices belong to the commutant of  gauge group in $Sp(r)$.
%

\section{Kernel equations}
\label{app:KE}

Before we delve into the details of the calculation, let us first make some useful definitions and observations. Analogous to the what was done in the section on highest-weight one-forms, define $A_a = l_a^{\phantom{a}m}A_m$ for $a\in\{1,2,3\}$. Further, define $A_\pm = A_1\pm i A_2$. Let us note the action of $\sqh^2$ on these new variables (and $A_\rho$) ):
\begin{align*}
 \sqh^2A_+ &= 2e^{i\psi}\partial_\psi\left(e^{-i\psi}A_+\right)\\
 \sqh^2A_- &= 2e^{-i\psi}\partial_\psi\left(e^{i\psi}A_-\right)\\
 \sqh^2A_3 &= 2\partial_\psi A_3\\
 \sqh^2A_\rho &= 2\partial_\psi A_\rho\nonumber\\
 \sqh^2 \phi_2 &= 2\partial_\psi \phi_2.
\end{align*}
Similarly, we define $\chi_\pm = \chi_1 \pm i\chi_2$. These transform under $\sqh^2$ (up to a gauge transformation) as
\begin{align*}
 \sqh^2\chi_+ &= 2e^{i\psi}\partial_\psi\left(e^{-i\psi}\chi_+\right)\\
  \sqh^2\chi_- &= 2e^{-i\psi}\partial_\psi\left(e^{i\psi}\chi_+\right)\\
  \sqh^2\chi_3 &= 2\partial_\psi \chi_3
  \nonumber\\
 \sqh^2 c &= 2\partial_\psi c
 \nonumber\\
 \sqh^2 \bar{c}_2 &= 2\partial_\psi \bar{c}_2.
\end{align*}

Our solution hinges on the following observations: Firstly, $\sqh$ squares to a Lie derivative over scalars 
along $l_3$ (with some shifts as in the cases of $A_\pm$ and $\chi_\pm$). Secondly, $D_{10}$ (like all $D_{ij}$) commutes with $\sqh^2$. These two facts together imply that we can expand the fields $(A_\pm, A_3,A_\rho,\phi_2)$ and $(\chi_\pm, \chi_3,c,\bar c)$ in $SU(2)_L \times SU(2)_R$ harmonics, and analyse $D_{10}$ over the Fourier modes. Further, these Fourier modes (with adequate shifts for $A_\pm$ and $\chi_\pm$) will have the same eigenvalue with respect to $\sqh^2$. Thirdly, as we will show explicitly later, the equations whose solutions give $\krd$ and $\ckrd$ can be written only in terms of the $J_L$s and without the $J_R$s. This implies that the $J^3_R$ eigenvalue will only serve to count the multiplicity of the solutions, if a solution exists (we will see this explicitly while computing $\ckrd$).
Concretely, one can expand the fields as follows
\begin{align*}
 A_+(\psi,\theta,\phi,\rho)&=\sum_{q_L,q_R}e^{-i(q_L-1)\psi}e^{-i q_R\phi}A_{+,(q_L,q_R)}(\theta,\rho),\\
A_-(\psi,\theta,\phi,\rho)&=\sum_{q_L,q_R}e^{-i(q_L+1)\psi}e^{-i q_R\phi}A_{-,(q_L,q_R)}(\theta,\rho),\\
A_3(\psi,\theta,\phi,\rho)&=\sum_{q_L,q_R}e^{-i(q_L)\psi}e^{-i q_R\phi}A_{3,(q_L,q_R)}(\theta,\rho),\\
A_r(\psi,\theta,\phi,\rho)&=\sum_{q_L,q_R}e^{-i(q_L)\psi}e^{-i q_R\phi}A_{\rho,(q_L,q_R)}(\theta,\rho),\\
\phi_2(\psi,\theta,\phi,\rho)&=\sum_{q_L,q_R}e^{-i(q_L)\psi}e^{-i q_R\phi}\phi_{2,(q_L,q_R)}(\theta,\rho),
\end{align*}
and 
\begin{align*}
\chi_+(\psi,\theta,\phi,\rho)&=\sum_{q_L,q_R}e^{i(q_L+1)\psi}e^{i q_R\phi}\chi_{+,(q_L,q_R)}(\theta,\rho),\\
\chi_-(\psi,\theta,\phi,\rho)&=\sum_{q_L,q_R}e^{i(q_L-1)\psi}e^{i q_R\phi}\chi_{-,(q_L,q_R)}(\theta,\rho),\\
\chi_3(\psi,\theta,\phi,\rho)&=\sum_{q_L,q_R}e^{i(q_L)\psi}e^{i q_R\phi}\chi_{3,(q_L,q_R)}(\theta,\rho),\\
c(\psi,\theta,\phi,\rho)&=\sum_{q_L,q_R}e^{i(q_L)\psi}e^{i q_R\phi}c_{(q_L,q_R)}(\theta,\rho)\\
\bar{c}(\psi,\theta,\phi,\rho)&=\sum_{q_L,q_R}e^{i(q_L)\psi}e^{i q_R\phi}\bar c_{(q_L,q_R)}(\theta,\rho),
\end{align*}
where the summation over $q_R$ runs from $-j_R$ to $j_R$ and the summation over $q_L$ runs from $-j_L -1$ to $j_L + 1$. We recall from appendix \ref{app:harmonics} that for scalar harmonics, $j_L = j_R$. We have also labelled the modes so that all modes with subscripts $(q_L,q_R)$ have the same $\sqh^2$ eigenvalue (not considering the gauge transformation, see below).

For simplicity, we take the gauge group to be abelian. This is analogous to what has been done in \cite{Pestun:2007rz} and \cite{Hama:2012bg}, where the authors first calculate the index for the abelian case, and then account for contribution due to a non-abelian gauge group to the eigenvalue with respect to $\sqh^2$. This contribution to the eigenvalue is $\sum_{\alpha \in \Delta}a_0\cdot \alpha$, where $\Delta$ represents all the roots of the Lie algebra of $G$.

Finally a few words on notation: We will suppress the subscript $L$ in the following while referring to $J_L$s, since we will work almost exclusively with the left-handed generators. As noted in appendix \ref{app:harmonics}, the $\theta$-dependence of a scalar harmonic is given by $Y^{j_L,q_L.q_R}(\theta)$ (which are related to the Jacobi polynomials), while the $\rho$ dependence is unrestricted by the $SU(2)_L\times SU(2)_R$ algebra. Therefore, a Fourier mode such as $\chi_{3,(q_L,q_R)}(\theta,\rho)$ can be decomposed as
\begin{equation*}
 \chi_{3,(q_L,q_R)}(\theta,\rho) = Y^{j_L,q_L.q_R}(\theta)\chi_{3,(q_L,q_R)}(\rho)
\end{equation*}
 We are therefore using the same symbol for the field, its Fourier mode, and a part of that mode that depends only on $\rho$. We hope no confusion will occur.

\subsection{The kernel equations for the vector multiplet}
The space $\krd$ consists of the solutions of five differential equations that are obtained as co-efficients of the fermionic fields $(\chi_a,c,\bar c)$ in Eq. \ref{eq:D10}. One of those equations is
\begin{align*}
 0=&\tfrac{1}{2} \sin \theta  \sin \rho \left(-4 \partial_\theta^2\phi_{p}-\sin^2\rho\, \partial_\rho^2\phi_{p}-4 \cot \theta \,
   \partial_{\theta}\phi_{p}\right)\nonumber\\
   &-3 \sin \rho \cos \rho \, \partial_\rho\phi_{p}-4 \csc ^2\theta \, \phi_{p} \left(q_L^2-2
   q_L q_R \cos \theta +q_R^2\right).
\end{align*}
The right-hand side of the above equation is proportional to the Laplacian of $\phi_{p}$, and therefore only has a constant solution. We presently focus only on non-constant solutions that correspond to $j_L >0$. The (non-zero) constant solutions will be discussed later. Therefore, we set $\phi_{p}=0$ for now.

With this condition, the other differential equations that yield $\krd$, in terms of the Fourier modes, are $\mathcal{E}_i = 0$, $i=1,...,4$, where
\begin{align*}
 \mathcal{E}_1 &= q_L \cos\rho A_- - \tfrac{1}{2}\sin\rho \partial_\rho A_- + \sec\rho J^-A_3 + \tfrac{1}{2}\sin\rho J^-A_\rho\\
 \mathcal{E}_2 &= q_L \cos\rho A_+ + \tfrac{1}{2}\sin\rho \partial_\rho A_+ + \sec\rho J^+A_3 - \tfrac{1}{2}\sin\rho J^+A_\rho\\
 \mathcal{E}_3 &= J^-A_+ - J^+A_- + (1+\sec^2\rho) \tan\rho\partial_\rho A_3 + q_L\sin\rho\cos\rho A_\rho\\
 \mathcal{E}_4 &= J^-A_+ + J^+A_- - 2q_LA_3 - \tfrac{3}{2}\sin\rho\cos\rho A_\rho-\tfrac{1}{2}\sin^2\rho\partial_\rho A_\rho.
\end{align*}
Some notation has been abused above in favour of readability: $A_+$, for example, stands for $A_{+,(q_L,q_R)}(\theta,\rho)$, and similarly others. We will not employ this notation anywhere else. As advertised earlier, the equations yielding the kernel are written entirely in terms of $J_L$s.

For clarity, we inform the reader right at the beginning that the result of the following analysis is that the \emph{kernel is empty}. We divide the analysis into the following cases
\begin{itemize}
 \item $|q_L| = j_L +1$:
Let us consider the case $q_L = j_L +1$.

The expressions $\mathcal{E}_{1,3,4}$ vanish identically on imposing $q_L = j_L +1$, and the only non-trivial equation arises $\mathcal{E}_2 = 0$:
\begin{equation*}
2(1+j_L)\cos\rho\, A_+^{(j_L,-j_L,-q_R)}(\rho)+\sin\rho\, \partial_\rho A_+^{(j_L,-j_L,-q_R)}(\rho)=0,
  \end{equation*}
solving which gives
\begin{equation*}
 A_+^{(j_L,-j_L,-q_R)}(\rho)=A^0_+(\sin\rho)^{-2(1+j_L)}.
\end{equation*}
This solution is clearly singular at the two poles $\rho = 0,\pi$, forcing us to set $A^0_+=0$. Therefore, we do not have a non-trivial solution.

The case $q_L = -(j_L +1)$ is almost identical. In this case, only the equation $\mathcal{E}_1 = 0$ is non-trivial, but the only solution that is non-singular at the poles is the trivial one.

\item $|q_L| = j_L$:
Again, we only provide details for the case  $q_L = j_L$. The case $q_L = -j_L$ is similar. The expression $\mathcal{E}_1$ identically vanishes. The remaining equations give the following system of equations:
\begin{equation*}
\partial_\rho \left(\begin{array}{c}
                      A_+\\
                      A_3\\
                      A_\rho
                    \end{array}\right)=
\left(\begin{array}{lll}
       -2j_L\cot\rho & -4j_L\sec\rho\csc\rho & 2j_L\\
       -\cot\rho & -\cot\rho(1+\sec^2\rho) & -j_L\cos^2\rho\\
       2\csc^2\rho & -4j_L\csc^2\rho &-3\cot\rho
      \end{array}\right)
\left(\begin{array}{c}
                     A_+\\
                     A_3\\
                     A_\rho
                    \end{array}\right)
\end{equation*}
where for brevity we have used the shorthand $A_+ \equiv A_+^{(j_L,1-j_L,-q_R)}(\rho)$, $ A_3\equiv  A_3^{(j_L,-j_L,-q_R)}(\rho)$ and $A_\rho \equiv A_\rho^{(j_L,-j_L,-q_R)}(\rho)$.
One can proceed by eliminating $ A_+^{(j_L,1-j_L,-q_R)}(\rho)$ between the second and third equations, and then integrating for $A_3^{(j_L,-j_L,-q_R)}(\rho)$. One gets
\begin{equation*}
 A_3^{(j_L,-j_L,-q_R)}(\rho) = -\tfrac{1}{2}\cos\rho\sin\rho A_\rho^{(j_L,-j_L,-q_R)}(\rho) + C\cos\rho(\sin\rho)^{-2(j_L +1)}.
\end{equation*}
Regularity of the solution at the poles forces $C=0$. One can finally use the first equation (along with the information obtained above) to solve for $A_\rho^{(j_L,-j_L,-q_R)}(\rho)$:
\begin{equation*}
 A_\rho^{(j_L,-j_L,-q_R)}(\rho) = (\sin\rho)^{-3-j_L}\left(C_1 \cos\rho - C_2 P_\frac{1}{2}^{(-\frac{1}{2},-2j_L-1)}\left(-\cos(2\rho)\right)\right)
\end{equation*}
where the special function $P_n^{(\alpha,\beta)}(x)$ in the second term is the Jacobi polynomial introduced in the appendix \ref{app:harmonics}. For the values of the parameters given here, it turns out to be a polynomial in $\cos^2\rho$ of degree $2j_L$, or equivalently, a linear combination of $\cos(2m\rho)$ where $m=0,...,2j_L$. The only values of $C_1$ and $C_2$ for which the above function is regular is $C_1 = C_2 = 0$. Therefore, $A_+^{(j_L,1-j_L,-q_R)}(\rho)$ and $A_3^{(j_L,-j_L,-q_R)}(\rho)$ are also zero, and we have no non-trivial solution.
\item $|q_L| < j_L$. Let us first deal with the case $q_L = 0$. One can solve $\mathcal{E}_3,4 = 0$ for $A_\pm^{(j_L,\pm1,-q_R)}$ in terms of the other fields and their derivatives, and then use this to solve the second-order ODE for $A_\rho^{(j_L,0,-q_R)}(\rho)$, for instance:
\begin{equation*}
 A_\rho^{(j_L,0,-q_R)}(\rho) = C_1\left[\tan\left(\tfrac{\rho}{2}\right)\right]^{1+2j_L} + C_2\left[\cot\left(\tfrac{\rho}{2}\right)\right]^{1+2j_L}\csc^2\rho,
\end{equation*}
whose regularity at the poles requires both constants to be zero. Therefore, there is again no non-trivial solution in this case.

Let us finally deal with the case $0\neq|q_L| < j_L$. Rather than directly integrating the equations and showing that the solutions are not smooth, we will show that the existence of a smooth solution leads to a contradiction.
\end{itemize}
Putting it all together, we see that $\krd$ is indeed empty.
\subsection{The cokernel equations for the vector multiplet}\label{sec:Cokernel}
To find the equations whose solutions yield the space $\ckrd$, we integrate the relevant part of the localizing action by parts and substitute the Fourier expansion. The relevant equations are $\mathcal{CE}_i=0$, $i=1,...,5$ where
\begin{align*}
 \mathcal{CE}_1&= 2(1-q_L)\cos\rho\,\chi_-+\sin\rho\,\partial_\rho\chi_-+2J^-\chi_3 + 2iJ^-\tilde c_p,\\
 \mathcal{CE}_2&= 2(1+q_L)\cos\rho\,\chi_++\sin\rho\,\partial_\rho\chi_++2J^+\chi_3 + 2iJ^+\tilde c_p,\\
 \mathcal{CE}_3&= -J^-\chi_++J^+\chi_- +\sin\rho\,\partial_\rho\chi_3+ 2iq_L\cos\rho\tilde c_p,\\
 \mathcal{CE}_4&= J^-\chi_++J^+\chi_- + 2q_L\cos\rho\chi_3+i\sin\rho\,\partial_\rho\tilde c_p,\\
 \mathcal{CE}_5&= 3\chi_3 + 2\csc^2\rho \left[J^+J^- + J^-J^+ + 2J_3^2 \right]c-3\cot\rho\,\partial_\rho c-\partial_\rho^2 c-2iq_L\tilde c_p.
\end{align*}
where we have used similar shorthand as we did while writing the kernel equations in terms of $J_L$s: $\chi_3$, for examples, in a shorthand for $\chi_{3,(q_L,q_R)}(\theta,\rho)$, and similarly others.
We now begin the analysis of the above equations, paralleling what we did for the kernel equations. However, in this case, will find that the cokernel is not empty.
\begin{itemize}
 \item $|q_L| = j_L +1$: 
  Let us first consider the case $q_L = j_L +1$. The equations $\mathcal{CE}_i= 0$ are satisfied identically for $i\in\{2,...,5\}$. The only non-trivial constraint comes from the equation $\mathcal{CE}_1= 0$, which gives
  \begin{equation*}
   \sin\rho\,\partial_\rho\chi_-^{(j_L,j_L,q_R)}(\rho) = 2j_L\cos\rho\,\chi_-^{(j_L,j_L,q_R)}(\rho)
  \end{equation*}
  which yields
  \begin{equation*}
   \chi_-^{(j_L,j_L,q_R)}(\rho)= \chi_-^0 (\sin \rho)^{2 j_L}.
  \end{equation*}
  All the other fermionic fields $(\chi_+,\chi_3,c,\tilde c_p)$ are identically zero. The multiplicity of this solution is $2j_L + 1$. The eigenvalue of $\sqh^2$ corresponding to the above solution is $2(j_L + 1)$. Denoting this eigenvalue as $n$, we see that the multiplicity corresponding to this eigenvalue is $n-1$.
  For the case $q_L = -j_L -1$, the only non-trivial constraint comes from the equation $\mathcal{CE}_2= 0$, which gives
  \begin{equation*}
   \chi_+^{(j_L,-j_L,q_R)}(\rho)= \chi_+^0 (\sin \rho)^{2 j_L}.
  \end{equation*}
  This is again the only non-vanishing field. The multiplicity of this solution is $2j_L + 1$. The eigenvalue of $\sqh^2$ corresponding to the above solution is $n\equiv -2(j_L + 1)$, so that the multiplicity corresponding to this eigenvalue is $|n|-1$.

 \item $|q_L| = j_L$. Let us first consider $q_L = j_L$. The equation $\mathcal{CE}_2=0$ vanishes identically (as does the field $\chi_+$). The equations $\mathcal{CE}_{1,3,4}=0$ give the system
 \begin{equation*}
\sin\rho\,\partial_\rho \left(\begin{array}{c}
                      \chi_-\\
                      \chi_3\\
                      \tilde c_p
                    \end{array}\right)=
\left(\begin{array}{lll}
       2(j_L-1)\cos\rho & -2 & -2i\\
       -2j_L & 0 & -2ij_L\cos\rho\\
       2ij_L & 2ij_L\cos\rho &0
      \end{array}\right)
\left(\begin{array}{c}
                     \chi_-\\
                      \chi_3\\
                      \tilde c_p
                    \end{array}\right)
\end{equation*}
where $ \chi_-, \chi_3, \tilde c_p$ stand for $ \chi_-^{(j_L,j_L-1,q_R)}(\rho), \chi_3^{(j_L,j_L,q_R)}(\rho), \tilde c_p^{(j_L,j_L,q_R)}(\rho)$ respectively. Besides this, $\mathcal{CE}_5=0$ gives the following second-order ODE:
\begin{equation*}
 \sin^2\rho\,\partial_\rho^2c + 3\cos\rho\sin\rho\,\partial_\rho c-4j_L(j_L+1)c + 2ij_L\sin^2\rho\tilde c_p-3\sin^2\rho\chi_3 = 0.
\end{equation*}
where $c$ stands for $c^{(j_L,j_L,q_R)}(\rho)$, and similarly $\tilde c_p$ and $\chi_3$.

We can proceed by eliminating $\chi_-^{(j_L,j_L-1,q_R)}(\rho)$ between the second and third equations of the first-order system. On solving the resulting differential equation we get
\begin{equation*}
 \chi_3^{(j_L,j_L,q_R)}(\rho) - i\tilde c_p^{(j_L,j_L,q_R)}(\rho) = C\left[\sin\rho \right]^{2j_L}.
\end{equation*}
To fully solve for $\chi_3^{(j_L,j_L,q_R)}(\rho)$ and $\tilde c_p^{(j_L,j_L,q_R)}(\rho)$, introduce the ansatz
\begin{equation*}
 \tilde c_p^{(j_L,j_L,q_R)}(\rho) =\tfrac{i}{2}C\left[\sin\rho \right]^{2j_L}+g(\rho)
\end{equation*}
into the first first-order ODE (while also eliminating $\chi_-^{(j_L,j_L-1,q_R)}(\rho)$ using the other equations). The ODE we get is
\begin{equation*}
 \sin^2\rho\,\partial_\rho^2g(\rho) + \tfrac{3}{2}\cos2\rho\,\partial_\rho g(\rho)-j_L\left(2j_L +7 + (2j_L-3)\cos2\rho \right)=0.
\end{equation*}
We will argue that $g(\rho)$ vanishes identically if it is required to be smooth. In order to see this, multiply the above equation by $\sin\rho g(\rho)$ and consider the integral of the resulting expression over $\rho \in (0,\pi)$. Integrating-by-parts the term with $\partial_\rho^2g(\rho)$ gives
\begin{equation*}
  \int_0^\pi\left[(-\sin\rho)\left(\sin^2\rho(\partial_\rho g)^2 + j_L(2j_L+7 + (2j_L-3)\cos 2\rho)g^2 \right)\right]d\rho \stackrel{!}=0.
\end{equation*}
where the boundary terms vanish due to the assumed regularity of $g(\rho)$ at the poles. Note that $g(\rho)$ is a real function. The co-efficient of $g^2$ above is easily verified to be positive for all allowed values of $j_L$. Therefore, the only way the above integral vanishes is if $g(\rho)=0$ identically. Using this solution we can also immediately see that $\chi_-^{(j_L,j_L-1,q_R)}(\rho)$ vanishes, and the second-order ODE for $c^{(j_L,j_L,q_R)}(\rho)$ gives us essentially the inhomogeneous Laplace equation:
\begin{equation*}
\sin^2\rho \partial_\rho^2c + \tfrac{3}{2}\sin2\rho\partial_\rho c - 4j_L(j_L+1)c = C\left(\tfrac{3}{2}+ j_L\right)\left[\sin\rho \right]^{2j_L+2}
\end{equation*}
where $c\equiv c^{(j_L,j_L,q_R)}(\rho)$. Since the homogenous Laplace equation does not have any non-trivial smooth solution on the sphere, the following solution is unique:
\begin{equation*}
 c^{(j_L,j_L,q_R)}(\rho) = -\tfrac{1}{4j_L}C\left[\sin\rho \right]^{2j_L}
\end{equation*}
To summarize, the solution is:
\begin{equation*}
 \begin{array}{ll}
  c^{(j_L,j_L,q_R)}(\rho) = -\tfrac{1}{4j_L}C\left[\sin\rho \right]^{2j_L}, & \tilde c_p^{(j_L,j_L,q_R)}(\rho) =\tfrac{i}{2}C\left[\sin\rho \right]^{2j_L}, \\
  \chi_3^{(j_L,j_L,q_R)}(\rho) =  \tfrac{1}{2}C\left[\sin\rho \right]^{2j_L}, & 
 \end{array}
\end{equation*}
with all other fields vanishing.
The eigenvalue of this solution is $2j_L$. Referring to this as $n$, the multiplicity is $n+1$.

The case $q_L = -j_L$ is similar: Following the same steps as above, we find that the non-vanishing fields of the solution set are
\begin{equation*}
 \begin{array}{ll}
  c^{(j_L,j_L,q_R)}(\rho) = -\tfrac{1}{4j_L}C\left[\sin\rho \right]^{2j_L}, & \tilde c_p^{(j_L,j_L,q_R)}(\rho) =-\tfrac{i}{2}C\left[\sin\rho \right]^{2j_L}, \\
  \chi_3^{(j_L,j_L,q_R)}(\rho) =  \tfrac{1}{2}C\left[\sin\rho \right]^{2j_L}, & 
 \end{array}
\end{equation*}
The eigenvalue of this solution is $n=-2j_L$ and its the multiplicity is $|n|+1$.
\item $|q_L| < j_L$: The analysis of this case is similar to the case $|q_L| < j_L$ for kernel equations, and therefore will not be detailed. The result is that we have no non-trivial solutions in this case.
\end{itemize}
We see that the total multiplicity for any integer eigenvalue $n\in \intz$ is $(|n| -1)+(|n| +1) = 2|n|$.


To the eigenvalues we found above, we need to add the contribution due to the non-abelian gauge group. After doing so, we see that the ratio of the determinants in Eq. \ref{eq:Detratio} is equal to the unregularized product
\begin{equation}\label{eq:unregS4}
  \prod_{\alpha \in \Delta}\prod_{n\geq 1}(n + ia\cdot \alpha)^{2n}(n - ia\cdot \alpha)^{2n}
\end{equation}
The partition function, which is the square-root of the above expression, on regularizing becomes
\begin{equation*}
  \prod_{\alpha \in \Delta_+} \frac{\Ups(ia\cdot \alpha)\Ups(-ia\cdot \alpha)}{(ia\cdot \alpha)^2}.
\end{equation*}
The above expression enters the expression of the total partition function, integrated of the Lie algebra of the gauge group. When the integral over the entire Lie algebra is reduced to an integral over the Cartan subalgebra, the integration measure contains a Vandermonde factor, which precisely cancels the $(ia\cdot \alpha)^2$ in the denominator. The total partition function then is
\begin{equation}
 Z_{\sfr}^\mathrm{vec} = \int \mathrm{d}a \,e^{-\frac{8\pi^2}{g^2}\mathrm{Tr}(a^2)}|\zins|^2\prod_{\alpha \in \Delta_+} \Ups(ia\cdot \alpha)\Ups(-ia\cdot \alpha)
\end{equation}
\subsection{Including matter}
We now introduce matter in the form of four hypernultiplets in the fundamental representation. In this case we will find that the cokernel is empty while the kernel is not.


Including the contributions due to the hypermultiplets, and specializing the formulae for $SU(2)$ gauge group, the complete expression for the one-loop contribution is
\begin{equation}\label{eq:TotalOneLoopS4}
\frac{\Ups(-2ia)}{\prod_{\pm\pm}\Ups(Q/2 + ia \pm im_1 \pm
im_2)}\frac{\Ups(2ia)}{\prod_{\pm\pm}\Ups(Q/2 - ia \pm im_3 \pm im_4)}
\end{equation}
\bibliographystyle{JHEP}
\bibliography{draftbib}

\providecommand{\href}[2]{#2}\begingroup\raggedright\begin{thebibliography}{10}

\bibitem{Nekrasov:2002qd}
N.~A. Nekrasov, {\it {Seiberg-Witten prepotential from instanton counting}},
  {\em Adv. Theor. Math. Phys.} {\bf 7} (2003), no.~5 831--864,
  [\href{http://xxx.lanl.gov/abs/hep-th/0206161}{{\tt hep-th/0206161}}].

\bibitem{Pestun:2016zxk}
V.~Pestun {\em et~al.}, {\it {Localization techniques in quantum field
  theories}},  \href{http://xxx.lanl.gov/abs/1608.02952}{{\tt 1608.02952}}.

\bibitem{Gava:2016oep}
E.~Gava, K.~S. Narain, M.~N. Muteeb, and V.~I. Giraldo-Rivera, {\it {$N = 2$
  gauge theories on the hemisphere $HS^4$}},  {\em Nucl. Phys.} {\bf B920}
  (2017) 256--297, [\href{http://xxx.lanl.gov/abs/1611.04804}{{\tt
  1611.04804}}].

\bibitem{Alday:2009aq}
L.~F. Alday, D.~Gaiotto, and Y.~Tachikawa, {\it {Liouville Correlation
  Functions from Four-dimensional Gauge Theories}},  {\em Lett. Math. Phys.}
  {\bf 91} (2010) 167--197, [\href{http://xxx.lanl.gov/abs/0906.3219}{{\tt
  0906.3219}}].

\bibitem{Gaiotto:2009we}
D.~Gaiotto, {\it {N=2 dualities}},  {\em JHEP} {\bf 08} (2012) 034,
  [\href{http://xxx.lanl.gov/abs/0904.2715}{{\tt 0904.2715}}].

\bibitem{Witten:1997sc}
E.~Witten, {\it {Solutions of four-dimensional field theories via M theory}},
  {\em Nucl. Phys.} {\bf B500} (1997) 3--42,
  [\href{http://xxx.lanl.gov/abs/hep-th/9703166}{{\tt hep-th/9703166}}].

\bibitem{Belavin:2011pp}
V.~Belavin and B.~Feigin, {\it {Super Liouville conformal blocks from N=2 SU(2)
  quiver gauge theories}},  {\em JHEP} {\bf 07} (2011) 079,
  [\href{http://xxx.lanl.gov/abs/1105.5800}{{\tt 1105.5800}}].

\bibitem{Nishioka:2011jk}
T.~Nishioka and Y.~Tachikawa, {\it {Central charges of para-Liouville and Toda
  theories from M-5-branes}},  {\em Phys. Rev.} {\bf D84} (2011) 046009,
  [\href{http://xxx.lanl.gov/abs/1106.1172}{{\tt 1106.1172}}].

\bibitem{BMT1}
G.~Bonelli, K.~Maruyoshi, and A.~Tanzini, {\it {Instantons on ALE spaces and
  Super Liouville Conformal Field Theories}},  {\em JHEP} {\bf 08} (2011) 056,
  [\href{http://xxx.lanl.gov/abs/1106.2505}{{\tt 1106.2505}}].

\bibitem{Belavin:2011sw}
A.~A. Belavin, M.~A. Bershtein, B.~L. Feigin, A.~V. Litvinov, and G.~M.
  Tarnopolsky, {\it {Instanton moduli spaces and bases in coset conformal field
  theory}},  {\em Commun. Math. Phys.} {\bf 319} (2013) 269--301,
  [\href{http://xxx.lanl.gov/abs/1111.2803}{{\tt 1111.2803}}].

\bibitem{BMT2}
G.~Bonelli, K.~Maruyoshi, and A.~Tanzini, {\it {Gauge Theories on ALE Space and
  Super Liouville Correlation Functions}},  {\em Lett. Math. Phys.} {\bf 101}
  (2012) 103--124, [\href{http://xxx.lanl.gov/abs/1107.4609}{{\tt 1107.4609}}].

\bibitem{Fateev:2000ik}
V.~Fateev, A.~B. Zamolodchikov, and A.~B. Zamolodchikov, {\it {Boundary
  Liouville field theory. 1. Boundary state and boundary two point function}},
  \href{http://xxx.lanl.gov/abs/hep-th/0001012}{{\tt hep-th/0001012}}.

\bibitem{Teschner:2000md}
J.~Teschner, {\it {Remarks on Liouville theory with boundary}},
  \href{http://xxx.lanl.gov/abs/hep-th/0009138}{{\tt hep-th/0009138}}.
  [PoStmr2000,041(2000)].

\bibitem{Zamolodchikov:2001ah}
A.~B. Zamolodchikov and A.~B. Zamolodchikov, {\it {Liouville field theory on a
  pseudosphere}},  \href{http://xxx.lanl.gov/abs/hep-th/0101152}{{\tt
  hep-th/0101152}}.

\bibitem{LeFloch:2017lbt}
B.~Le~Floch and G.~J. Turiaci, {\it {AGT/Z$_2$}},
  \href{http://xxx.lanl.gov/abs/1708.04631}{{\tt 1708.04631}}.

\bibitem{Gupta:2012cy}
R.~K. Gupta and S.~Murthy, {\it {All solutions of the localization equations
  for N=2 quantum black hole entropy}},  {\em JHEP} {\bf 02} (2013) 141,
  [\href{http://xxx.lanl.gov/abs/1208.6221}{{\tt 1208.6221}}].

\bibitem{Hama:2012bg}
N.~Hama and K.~Hosomichi, {\it {Seiberg-Witten Theories on Ellipsoids}},  {\em
  JHEP} {\bf 09} (2012) 033, [\href{http://xxx.lanl.gov/abs/1206.6359}{{\tt
  1206.6359}}]. [Addendum: JHEP10,051(2012)].

\bibitem{Klare:2013dka}
C.~Klare and A.~Zaffaroni, {\it {Extended Supersymmetry on Curved Spaces}},
  {\em JHEP} {\bf 10} (2013) 218,
  [\href{http://xxx.lanl.gov/abs/1308.1102}{{\tt 1308.1102}}].

\bibitem{Bawane:2014uka}
A.~Bawane, G.~Bonelli, M.~Ronzani, and A.~Tanzini, {\it {$\mathcal{N}=2$
  supersymmetric gauge theories on $S^2\times S^2$ and Liouville Gravity}},
  {\em JHEP} {\bf 07} (2015) 054,
  [\href{http://xxx.lanl.gov/abs/1411.2762}{{\tt 1411.2762}}].

\bibitem{Pestun:2007rz}
V.~Pestun, {\it {Localization of gauge theory on a four-sphere and
  supersymmetric Wilson loops}},  {\em Commun. Math. Phys.} {\bf 313} (2012)
  71--129, [\href{http://xxx.lanl.gov/abs/0712.2824}{{\tt 0712.2824}}].

\bibitem{Vafa:1994tf}
C.~Vafa and E.~Witten, {\it {A Strong coupling test of S duality}},  {\em Nucl.
  Phys.} {\bf B431} (1994) 3--77,
  [\href{http://xxx.lanl.gov/abs/hep-th/9408074}{{\tt hep-th/9408074}}].

\bibitem{Bruzzo:2002xf}
U.~Bruzzo, F.~Fucito, J.~F. Morales, and A.~Tanzini, {\it {Multiinstanton
  calculus and equivariant cohomology}},  {\em JHEP} {\bf 05} (2003) 054,
  [\href{http://xxx.lanl.gov/abs/hep-th/0211108}{{\tt hep-th/0211108}}].

\bibitem{Okuda:2010ke}
T.~Okuda and V.~Pestun, {\it {On the instantons and the hypermultiplet mass of
  N=2* super Yang-Mills on $S^{4}$}},  {\em JHEP} {\bf 03} (2012) 017,
  [\href{http://xxx.lanl.gov/abs/1004.1222}{{\tt 1004.1222}}].

\bibitem{Tachikawa:unpub}
Y.~Tachikawa, {\it Unpublished notes}, .

\bibitem{Nakayama:2004vk}
Y.~Nakayama, {\it {Liouville field theory: A Decade after the revolution}},
  {\em Int. J. Mod. Phys.} {\bf A19} (2004) 2771--2930,
  [\href{http://xxx.lanl.gov/abs/hep-th/0402009}{{\tt hep-th/0402009}}].

\bibitem{Hikida:2002bt}
Y.~Hikida, {\it {Liouville field theory on a unoriented surface}},  {\em JHEP}
  {\bf 05} (2003) 002, [\href{http://xxx.lanl.gov/abs/hep-th/0210305}{{\tt
  hep-th/0210305}}].

\bibitem{Dimofte:2012pd}
T.~Dimofte and D.~Gaiotto, {\it {An E7 Surprise}},  {\em JHEP} {\bf 10} (2012)
  129, [\href{http://xxx.lanl.gov/abs/1209.1404}{{\tt 1209.1404}}].

\bibitem{Dimofte:2011ju}
T.~Dimofte, D.~Gaiotto, and S.~Gukov, {\it {Gauge Theories Labelled by
  Three-Manifolds}},  {\em Commun. Math. Phys.} {\bf 325} (2014) 367--419,
  [\href{http://xxx.lanl.gov/abs/1108.4389}{{\tt 1108.4389}}].

\bibitem{Dimofte:2011py}
T.~Dimofte, D.~Gaiotto, and S.~Gukov, {\it {3-Manifolds and 3d Indices}},  {\em
  Adv. Theor. Math. Phys.} {\bf 17} (2013), no.~5 975--1076,
  [\href{http://xxx.lanl.gov/abs/1112.5179}{{\tt 1112.5179}}].

\bibitem{Dimofte:2013lba}
T.~Dimofte, D.~Gaiotto, and R.~van~der Veen, {\it {RG Domain Walls and Hybrid
  Triangulations}},  {\em Adv. Theor. Math. Phys.} {\bf 19} (2015) 137--276,
  [\href{http://xxx.lanl.gov/abs/1304.6721}{{\tt 1304.6721}}].

\bibitem{Benini:2017dud}
F.~Benini, S.~Benvenuti, and S.~Pasquetti, {\it {SUSY monopole potentials in
  2+1 dimensions}},  {\em JHEP} {\bf 08} (2017) 086,
  [\href{http://xxx.lanl.gov/abs/1703.08460}{{\tt 1703.08460}}].

\bibitem{Ponsot:2001ng}
B.~Ponsot and J.~Teschner, {\it {Boundary Liouville field theory: Boundary
  three point function}},  {\em Nucl. Phys.} {\bf B622} (2002) 309--327,
  [\href{http://xxx.lanl.gov/abs/hep-th/0110244}{{\tt hep-th/0110244}}].

\bibitem{Teschner:2012em}
J.~Teschner and G.~Vartanov, {\it {6j symbols for the modular double, quantum
  hyperbolic geometry, and supersymmetric gauge theories}},  {\em Lett. Math.
  Phys.} {\bf 104} (2014) 527--551,
  [\href{http://xxx.lanl.gov/abs/1202.4698}{{\tt 1202.4698}}].

\bibitem{Tachikawa:2016xvs}
Y.~Tachikawa and K.~Yonekura, {\it {Gauge interactions and topological phases
  of matter}},  {\em PTEP} {\bf 2016} (2016), no.~9 093B07,
  [\href{http://xxx.lanl.gov/abs/1604.06184}{{\tt 1604.06184}}].

\bibitem{Bershtein:2015xfa}
M.~Bershtein, G.~Bonelli, M.~Ronzani, and A.~Tanzini, {\it {Exact results for $
  \mathcal{N} $ = 2 supersymmetric gauge theories on compact toric manifolds
  and equivariant Donaldson invariants}},  {\em JHEP} {\bf 07} (2016) 023,
  [\href{http://xxx.lanl.gov/abs/1509.00267}{{\tt 1509.00267}}].

\bibitem{Bershtein:2016mxz}
M.~Bershtein, G.~Bonelli, M.~Ronzani, and A.~Tanzini, {\it {Gauge theories on
  compact toric surfaces, conformal field theories and equivariant Donaldson
  invariants}},  {\em J. Geom. Phys.} {\bf 118} (2017) 40--50,
  [\href{http://xxx.lanl.gov/abs/1606.07148}{{\tt 1606.07148}}].

\bibitem{Wyllard:2009hg}
N.~Wyllard, {\it {A(N-1) conformal Toda field theory correlation functions from
  conformal N = 2 SU(N) quiver gauge theories}},  {\em JHEP} {\bf 11} (2009)
  002, [\href{http://xxx.lanl.gov/abs/0907.2189}{{\tt 0907.2189}}].

\bibitem{Mironov:2009by}
A.~Mironov and A.~Morozov, {\it {On AGT relation in the case of U(3)}},  {\em
  Nucl. Phys.} {\bf B825} (2010) 1--37,
  [\href{http://xxx.lanl.gov/abs/0908.2569}{{\tt 0908.2569}}].

\bibitem{Bonelli:2009zp}
G.~Bonelli and A.~Tanzini, {\it {Hitchin systems, N=2 gauge theories and
  W-gravity}},  {\em Phys. Lett.} {\bf B691} (2010) 111--115,
  [\href{http://xxx.lanl.gov/abs/0909.4031}{{\tt 0909.4031}}].

\bibitem{Aleshkin:2017yty}
K.~Aleshkin, V.~Belavin, and C.~Rim, {\it {Minimal gravity and Frobenius
  manifolds: bulk correlation on sphere and disk}},
  \href{http://xxx.lanl.gov/abs/1708.06380}{{\tt 1708.06380}}.

\bibitem{Dimofte:2010tz}
T.~Dimofte, S.~Gukov, and L.~Hollands, {\it {Vortex Counting and Lagrangian
  3-manifolds}},  {\em Lett. Math. Phys.} {\bf 98} (2011) 225--287,
  [\href{http://xxx.lanl.gov/abs/1006.0977}{{\tt 1006.0977}}].

\end{thebibliography}\endgroup
\end{document}